\documentclass[11pt,a4paper]{article}
\usepackage[english]{babel}
\usepackage[mathscr]{eucal}
\usepackage{amsmath,amsbsy,amsthm,amsfonts,amssymb,makeidx,enumerate,bm,wasysym}
\usepackage{courier,vmargin,graphicx,wrapfig,float}
\usepackage{caption,subcaption}
\usepackage{epstopdf}

\theoremstyle{plain}

\theoremstyle{remark}

\def\2F1{{}_{2}F_{1}}
\def\aa{\alpha}
\def\AA{\mathcal{A}}
\def\ac{\mathrm{a}}
\def\af{a}
\def\AF{\mathscr{A}}
\def\al{\alpha}
\def\am{\mathfrak{a}}
\def\at{b}
\def\atan{\tan^{-1}}
\def\be{\beta}
\def\cu{\xi}
\def\dcu{\dot{\cu}}
\def\dd{\displaystyle}
\def\de{\partial}
\def\decu{\delta\cu}
\def\dsX{ds}
\def\ee{\varepsilon}
\def\EE{\mathcal{E}}
\def\EEf{\EE_f}
\def\EEg{\EE_g}
\def\EFf{\mathfrak{E}_f}
\def\EFg{\mathfrak{E}_g}
\def\ellm{\lambda}
\def\ellp{\Lambda}
\def\En{\mbox{E}}
\def\eps{\varepsilon}
\def\FF{\mathcal{F}}
\def\Ga{\Gamma}
\def\geqs{\geqslant}
\def\GG{\mathcal{G}}
\def\gmn{\mathtt{g}}
\def\gr{gr}
\def\gt{\mbox{g}_{\earth}}
\def\HH{\mathcal{H}}
\def\Hk{\mathfrak{H}}
\def\Hr{\mathscr{H}}
\def\II{\mathscr{I}}
\def\Jd{\mathfrak{h}}
\def\JJ{\mathfrak{J}}
\def\jmn{\mathfrak{j}}
\def\l{\left}
\def\leqs{\leqslant}
\def\L0{\mathfrak{L}}
\def\Mink{\mathfrak{M}}
\def\MM{\mathscr{M}}
\def\muh{\al}
\def\nuh{\be}
\def\pacu{\partial \cu}
\def\parn{\par \noindent}
\def\pf{\omega}
\def\PP{\mathcal{P}}
\def\pt{\gamma}
\def\q0{\QQ_0}
\def\QQ{\mathcal{Q}}
\def\r{\right}
\def\R{\mathbb{R}}
\def\RF{\mathfrak{R}}
\def\Ric{\mbox{Ric}}
\def\Riem{\mbox{Riem}}
\def\RJ{\mathfrak{K}}
\def\RL{\mathscr{L}}
\def\Rm{\mbox{R}}
\def\rr{\varrho}
\def\RR{\mathcal{K}}
\def\Rs{\mbox{R}}
\def\Rsc{\mathcal{R}}
\def\rv{r}
\def\si{\sigma}
\def\Si{\Sigma}
\def\ss{\varsigma}
\def\SS{\mathscr{S}}
\def\ta{\tau}
\def\Tar{\mathfrak{T}}
\def\Te{\mbox{T}}
\def\Th{\Theta}
\def\To{\mathbf{T}}
\def\Top{\To_{\ellp}}
\def\Tom{\To_{\ellm}}
\def\Tm{\mbox{T}}
\def\tt{\mathrm{w}}
\def\TT{\mathcal{T}}
\def\uu{\mathrm{u}}
\def\vfi{\varphi}
\def\Xc{\Xi}
\def\xf{\varpi}
\def\xr{\mathrm{v}}
\def\XX{\mathcal{X}}
\def\yr{\mathit{y}}
\def\Z{\mathbb{Z}}
\def\ze{\zeta}
\def\dsOS{d\tilde{s}}
\def\FOS{\tilde{F}}
\def\XOS{\tilde{\XX}}
\def\HOS{\tilde{\HH}}
\def\aOS{\tilde{a}}
\def\bOS{\tilde{b}}
\def\dOS{\tilde{\ell}}
\def\ROS{\tilde{R}}
\def\kOS{\tilde{k}}
\def\qOS{\tilde{q}}

\numberwithin{equation}{section}

\author{Davide Fermi}

\setmarginsrb{2.2 cm}{1.3 cm}%
             {2.2 cm}{1.3 cm}%
             {1.2 cm}{1.2 cm}%
             {1.2 cm}{1.2 cm}

\begin{document}
\begin{titlepage}
\vskip-1.cm
\noindent
\begin{center}
{\LARGE{\textbf{A time machine for free fall into the past}}}\vspace{0.12cm}\\
\end{center}
\begin{center}
{\large
Davide Fermi$\,{}^a$, Livio Pizzocchero$\,{}^b$} \\
\vspace{0.5cm}
${}^a$,${}^b$ Dipartimento di Matematica, Universit\`a di Milano\\
Via C. Saldini 50, I-20133 Milano, Italy\\
and Istituto Nazionale di Fisica Nucleare, Sezione di Milano, Italy \vspace{0.2cm}\\
${}^a$ e--mail: davide.fermi@unimi.it \\
${}^b$ e--mail: livio.pizzocchero@unimi.it
\vspace{0.2cm}
\end{center}
\begin{abstract}
Inspired by some recent works of Tippett-Tsang and Mallary-Khanna-Price, we present a new spacetime model containing
closed timelike curves (CTCs). This model is obtained postulating
an \textsl{ad hoc} Lorentzian metric on $\R^4$, which differs from
the Minkowski metric only inside a spacetime region bounded by two
concentric tori. The resulting spacetime is topologically trivial,
free of curvature singularities and is both time and space
orientable; besides, the inner region enclosed by the smaller
torus is flat and displays geodesic CTCs. Our model shares some
similarities with the time machine of Ori and Soen but it has the
advantage of a higher symmetry in the metric, allowing for the
explicit computation of a class of geodesics. The most remarkable
feature emerging from this computation is the presence of
future-oriented timelike geodesics starting from a point in the
outer Minkowskian region, moving to the inner spacetime region
with CTCs, and then returning to the initial spatial position at
an earlier time; this means that time travel to the past can be
performed by free fall across our time machine. The amount of time
travelled into the past is determined quantitatively; this amount
can be made arbitrarily large keeping non-large the proper
duration of the travel.
An important drawback of the model is the violation of the classical energy conditions,
a common feature of many time machines. Other problems emerge from our computations
of the required (negative) energy densities and of the tidal accelerations;
these are small only if the time machine is gigantic.
\end{abstract}
\vspace{0.2cm} \noindent
\hfill \parn
\par \vspace{0.3truecm} \noindent \textbf{AMS Subject classifications:} 83C05, 83C20\,.
\par \vspace{0.3truecm} \noindent \textbf{PACS}: 04.20.Cv, 04.20.Gz, 04.90.+e\,.
\end{titlepage}

\vfill\eject\noindent
$\phantom{a}$
\vskip-1.7cm
\noindent
\section{Introduction}
The construction of spacetime geometries admitting time travels is a recurrent subject
in general relativity; within this framework, a time travel is usually described in terms
of a closed timelike curve (CTC).
\parn
Excellent surveys on the subject were written by Thorne \cite{ThoRev} and Lobo \cite{Lobo}.
A threefold classification of the existing literature has been proposed in a recent work
of Tippett and Tsang \cite{sing}; we will integrate the scheme of these authors
with the addition of a fourth class, which leads to the following description.
\vspace{0.05cm}\parn
\textsl{First class}:
exact solutions of the Einstein equations (typically with high symmetry, in many cases
with strong angular momentum). This streamline originated from G\"odel's solution \cite{God},
describing a stationary and homogeneous universe filled with rotating dust
(and with a fine-tuned cosmological constant);
in this cosmological model, each event belongs to a CTC. Prior to G\"odel's work, van Stockum \cite{VanS}
had solved the Einstein equations in presence of a rigidly rotating, infinite cylinder of dust
(and with zero cosmological constant); the existence of CTCs in this spacetime was noted
much later by Tipler \cite{Tip}.
To proceed, let us recall the Taub-NUT spacetime \cite{NUT,Taub}, a spatially homogeneous,
vacuum solution with topology $\R \times S^3$; the existence of CTCs in this model was
pointed out by Misner \cite{Mis}, who also proposed a two-dimensional analogue of this
geometry \cite{Mis2} (see also the detailed analysis given by Thorne in \cite{ThoM}).
\parn
The Kerr rotating black hole \cite{Kerr} also possesses CTCs which, however, are hidden
behind an event horizon; indeed, these curves appear in the maximal extension of Kerr's
solution, near the ring singularity (see, e.g., \cite{Hawk}).
Tippett and Tsang also mention the Tomimatsu-Sato rotating, vacuum spacetime \cite{Tom35}
(a generalization of Kerr's model) and some spacetimes with moving cosmic strings.
Especially, they refer to the papers of Gott \cite{Gott} and of Deser, Jackiw, 't Hooft \cite{Des};
in addition to these references, we would also cite a paper of Grant \cite{Grant}
(considering a generalized version of Misner space, closely related to Gott's model)
and the very recent work of Mallary, Khanna and Price \cite{sing}
(who examine the existence of CTCs in a spacetime containing naked line singularities).
\vspace{0.05cm}\parn
\textsl{Second class:} \textsl{ad hoc} built spacetimes, i.e., geometries specifically projected
to produce CTCs. In these cases, the metric is given and the stress-energy tensor is derived
\textsl{a posteriori} from the right-hand side of the Einstein equations;
the undesired, exotic features emerging from this construction (typically, the violation
of the standard energy conditions) are regarded as secondary issues.
Probably, the most influential papers in this class are those of Ori and Soen
\cite{Ori1, Ori2, Ori3, Ori4}. In particular, \cite{Ori3,Ori4} present a time machine with
a toroidal spatial core, surrounded by a region where the spacetime metric is conformally flat.
CTCs are developed inside the toroidal region when the external time coordinate reaches a specific
value, and violations of the energy conditions appear simultaneously.
\parn
The cited work of Tippett and Tsang \cite{tardis} also belongs to
this class. Therein, two flat spacetime geometries are connected
via a curved transition region, where the energy conditions are
violated; the inner flat region contains CTCs. The model of
\cite{tardis} is simpler than the Ori-Soen spacetime in many
aspects, but a price must be paid for this: this spacetime is not
time-orientable, and naked curvature singularities appear in the
transition region. 
\vspace{0.05cm}\parn
\textsl{Third class:} \textsl{ad hoc} geometries originally designed to allow hyperfast
space travel, which have natural variants possessing CTCs. Let us
mention the celebrated Alcubierre's warp drive \cite{Alcu},
allowing for superluminal motion of a spaceship, and the
improvements of this model suggested by Krasnikov, Everett and
Roman \cite{Eve1,Kras}; some of these authors also indicated how
to use warp drives to produce CTCs \cite{Eve2,Kras}, developing a
qualitative idea of Hawking \cite{Hawk2}. As for the second class,
these geometries are generally postulated and the stress-energy
tensor is subsequently obtained from the Einstein equations; the
standard energy conditions are violated, a fact that was proven to
be unavoidable by Olum \cite{Olum} if one adopts a specific definition
of superluminal travel arrangement, proposed by this author.
\vfill\eject\noindent
The third class also contains wormhole-type geometries (which, again, violate the energy conditions);
notably, the static wormhole of Ellis, Morris and Thorne \cite{Ell,MT1} can be converted
into a time machine by accelerating one of its two mouths, as shown by Morris, Thorne and
Yurtsever \cite{MT2}. (Let us also mention a related paper by Echeverria, Klinkhammer and Thorne
\cite{Eche}, mentioned later in connection with the paradoxes of time travels.)
\parn
\textsl{Fourth class.} This is formed by just one model by Ori \cite{Ori2007},
which has connections with the first two classes but fits none of them exactly.
A striking feature of this model is that it presents no violation
of the energy conditions, since its matter content consists
only of ordinary dust (or vacuum). The spacetime of \cite{Ori2007} is the union of
three regions $\MM_0, \MM_1$ and $\MM_2$. $\MM_0$ is an internal, toroidal vacuum core
containing CTCs, with a ``pseudo-Schwarzschild'' metric obtained from the usual
Schwarzschild line element performing a Wick rotation on the polar angle $\theta$.
$\MM_2$ is an external, asymptotically flat vacuum region with the usual Schwarzschild
metric outside a sphere. $\MM_1$ is an intermediate region, called the envelope,
matching $\MM_0$ and $\MM_2$; this is filled with dust of non-negative density.
We have pointed out that the spacetime metric is given a priori in $\MM_0$ and $\MM_2$,
so there is a partial resemblance with the time machines of the second class;
the situation is very different in the envelope $\MM_1$, where the metric is described
as the solution of the Einstein equations with suitable Cauchy data.
The above solution is not explicitly known, so the appearance of pathological
structures (including black holes) cannot be excluded; a numerical investigation of these
issues was indicated in \cite{Ori2007} as a goal for future works, and is still pending
to the best of our knowledge.
\parn
The possibility of time travels originates well known paradoxes, which were analysed
by Friedman \textsl{et al.} \cite{MNT} and by Frolov and Novikov \cite{NovB}
(see Ch.\! 16 and the literature cited therein). Echeverria, Klinkhammer and Thorne
\cite{Eche} considered the Cauchy problem for a billiard ball in two exemplifying
spacetimes with CTCs generated by wormholes; due to the interaction of
the ball with copies of itself emerging from time travels, certain initial
data for the Cauchy problem produce infinitely many solutions
(against the conventional expectation of one solution at most). \parn
Another paradoxical aspect of time machines is the appearing of
divergences in the observables of semiclassical or quantized field theories.
A result of this kind was obtained by Krasnikov \cite{Krad}
and can be described as follows: under precise technical conditions,
a spacetime describing the creation of a time machine contains
``almost closed'' null geodesics,
returning again and again to an arbitrarily small region where they
are perceived as a ``multi-photon bundle'' of arbitrarily large energy.
The problem of infinities was discussed by Hawking \cite{Hawk2}
for quantum field theories on spacetimes with CTCs; here
the author considered a case study in which divergences exist
even after renormalization, and he suggested that
this should happen typically. To overcome the problem, Hawking formulated the famous
\textsl{chronology protection conjecture}: the laws of (quantum) physics forbid CTCs.
The viewpoint of Hawking has been discussed elsewhere, and even questioned (see, in particular,
a counterexample suggested by Li-Xin Li \cite{LiX}).
\parn
Dealing with the above mentioned paradoxes and problems is not among our purposes;
here, we just propose to enrich the second class of spacetimes with time travels,
introducing a new model (which violates the energy conditions).
In setting up this model, we were mainly stimulated by the paper of Tippett-Tsang \cite{tardis}
(the work of Mallary, Khanna and Price \cite{sing} also gave us some general motivation
to consider this subject); later on we realized that our construction has a closer contact
with the model of Ori-Soen \cite{Ori2,Ori3,Ori4}.
\parn
Our model is topologically trivial, possesses no curvature singularity, and is both space and time orientable;
it consists of a toroidal ``time machine'', which contains CTCs and is surrounded by flat
Minkowski space. These features resemble the Ori-Soen geometry, where a toroidal machine is
surrounded by a \textsl{conformally} flat spacetime region; however, there are relevant
differences between that model and ours.
\parn
Let us first point out the technical differences in the metric structure; later on, we will
emphasize their physical implications. Differently from the Ori-Soen model (and similarly
to the one of Tippett-Tsang), our metric is non-flat only in a transition region individuated
by two concentric tori $\Tom, \Top$, with the same major radius $R$ and minor radii $\ellm,\ellp$;
moreover, our metric exhibits more manifest symmetries, which allow us to reduce to quadratures
the Lagrange equations for a class of geodesic motions (both outside and inside the time machine).
Using these exact solutions, we can prove the following: a \textsl{freely falling} observer
(with suitable initial velocity) can start a trip at a spacetime point $(t_0=0, {\bf{x}}_0)$
in the outer Minkowskian region, fall into the time machine, re-emerge from it and finally return
to its initial space position ${\bf{x}}_0$ at a time $t_2 < 0$ (where time and position
are measured with respect to an inertial frame for the outer Minkowskian region;
the subscript ${}_{2}$ to indicate the end of the travel is naturally suggested by our
computations in Section \ref{freefallSec}). Obviously enough,
one can reinterpret this in terms of a CTC crossing the time machine (to close the observer's
worldline, it suffices to add a segment corresponding to the observers' permanence at
${\bf{x}}_0$ from time $t_2 < 0$ to time $t_0 = 0$).
\parn
Independently of the last remark, we think that time travel from $(t_0 = 0, {\bf{x}}_0)$
to $(t_2 < 0, {\bf{x}}_0)$ via free fall across the time machine is the most interesting
aspect of this model. We do not know whether such a time travel is possible in the Ori-Soen model,
since these authors just pointed out the existence of CTCs \textsl{inside} their time machine;
let us also mention that the explicit computation of geodesic motions in the Ori-Soen metric
is a non-trival affair, so it is difficult to ascertain the possibility of a time travel similar
to ours via free fall.
\parn
Let us illustrate other features of the time travel from $(t_0 = 0, {\bf{x}}_0)$ to
$(t_2 < 0, {\bf{x}}_0)$ in our model. First of all, adjusting suitably the initial
velocity one can make $|t_2|$ arbitrarily large: in other terms, the observer can go
back into the past as far as he/she wants. Moreover, the duration $\tau_2$ of the time
travel according to the observer's clock (i.e., the proper time along the observer's world line)
can be made arbitrarily small: to this purpose, the observer must start his/her trip with
a sufficiently large Lorentz factor with respect to the outer Minkowski frame. In this
way the observer can go back into his/her past, say, of one billion years while his/her
clock indicates a duration of only one year for the trip.
\parn
To conclude the analysis of our model, we estimate the tidal acceleration
experienced by a freely falling extended body when it crosses the transition region
inside the time machine. We also determine the energy densities measured by two kinds of observers:
some suitably defined, ``fundamental'' observers and, alternatively, the freely falling observers
performing a time travel (to this purpose, we adopt the previously mentioned idea to derive the
stress-energy tensor from the metric via the Einstein equations).
As expected, the tidal accelerations and the energy densities (of both kinds)
vanish identically in the outer Minkowskian region and in the flat region inside the machine;
in the transition region between $\Tom$ and $\Top$, the tidal accelerations are
non-zero and negative densities appear, yielding violations of the standard
energy conditions. \parn
Both the tidal accelerations and the energy densities are inversely proportional to the
square of the major radius $R$ (if the minor radii $\ellm,\ellp$ are comparable with it);
so, these quantities are small if the time machine is gigantic.
Indeed, our analysis is fully quantitative and exemplified by tables with numerical values.
If the radius $R$ is astronomical (say, 100 light years), the tidal acceleration
is sustainable for a human being, even for certain ultra-relativistic motions, and the energy
density according to a fundamental observer has an absolute value much smaller than $1\gr/cm^3$
(in units where $c = 1$); the energy density measured during free fall is below
$1\gr/cm^3$ even for some ultra-relativistic initial speeds
({\footnote{These considerations about the energy density are not related to the
evaluation of the effects produced on a freely falling observer by the exotic
matter in the transition region between the tori $\Tom,\Top$. As a matter of fact,
free fall is possible only if no direct interaction is assumed between the observer and this exotic matter.}}).
\vskip 0.1cm
\noindent
Let us describe the organization of the present work.
In Section \ref{secMod} we introduce the spacetime $\Tar$ describing our time machine;
this is the base manifold $\R^4$, equipped with a suitably defined Lorentzian metric $g$.
The main characters of this section are a set of ``cylindrical'' spacetime coordinates
$(t,\vfi,\rho,z)$, the previously mentioned pair of concentric tori $\Tom,\Top$ and
a sufficiently regular shape function $\XX$ that equals $1$ inside $\Tom$ and vanishes
identically outside $\Top$; explicit choices for $\XX$ are proposed in Appendix \ref{App1}.
\parn
In Section \ref{SecOr} we exhibit an orthonormal tetrad $(E_{(\muh)})_{\muh \in \{0,1,2,3\}}$
for $\Tar$, and use it to induce time and space orientations.
Contextually we introduce the ``fundamental observers'', whose worldlines are
the integral curves of the timelike vector field $E_{(0)}$.
In Section \ref{secKil} we discuss some evident symmetries of $\Tar$ and their physical
implications.
\parn
In Section \ref{secGeo} we consider the Lagrangian formalism for the causal
geodesics in $\Tar$; in particular, using the previously mentioned symmetries
we reduce to quadratures the computation of causal geodesics in the
plane $\{z = 0\}$.
Section \ref{freefallSec} is the core of the present work: here, we use the previously
established results to prove the existence of a timelike geodesic (with fine-tuned
initial velocity) which starts at any point in the outer Minkowskian region at time $t_0 = 0$,
crosses the tori $\Top,\Tom$ and eventually returns to its initial spatial position
at time $t_2 < 0$. Certain integrals in the quadrature formulas for the geodesics are
analysed in Appendix \ref{App2}.
\parn
In Section \ref{secacc} and in the related Appendix \ref{App3} we discuss the
tidal forces experienced by an extended body whose particles fall freely along geodesics
as in Section \ref{freefallSec}.
Section \ref{secen} and the related Appendix \ref{App4} deal with the energy densities
measured by fundamental and freely falling observers, and point out the violation
of the classical energy conditions.
\vskip 0.05cm \noindent
Some of the results presented in this paper have been derived
using the software \verb"Mathematica" for both symbolic
and numerical computations.
\vspace{-0.1cm}

\section{Description of the model}\label{secMod}
To begin with, let us consider the $4$-dimensional Minkowski spacetime $\Mink = (\R^4,\eta)$,
where $\eta$ denotes the usual, flat Lorentzian metric on the
base manifold $\R^4$. We fix the units of measure so that the speed of light is $c = 1$
({\footnote{Let us stress that we \textsl{do not} set the universal
gravitational constant to be $G = 1$.}});
moreover, we introduce on $\R^4 = \R \times \R^3$ a set of coordinates $(t,\vfi,\rho,z)$
where $t$ is the natural coordinate on $\R$ and $(\vfi,\rho,z) \in \R/(2\pi\Z) \times
(0,+\infty)\times \R$ are standard cylindrical coordinates on $\R^3$,
so that the line element $ds_0^2$ corresponding to the Minkowski metric $\eta$ reads
\begin{equation}
ds_0^2 \,=\, -\,dt^2 + \rho^2 d\vfi^2 + d\rho^2 + dz^2 ~. \label{ds0def}
\end{equation}
Next, let us fix $\ellm,\ellp,R \in (0,+\infty)$ such that $\ellm < \ellp < R$ and
consider in $\R^3$ the pair of concentric tori
\begin{equation}
\To_{\ell} \,:=\, \big\{\,\sqrt{(\rho-R)^2 + z^2}\, =\, \ell\,\big\} \qquad~
\big(\ell = \ellm,\ellp\big) \label{Todef}
\end{equation}
(see Figure 1 for a graphical representation of these tori, having a common major
radius $R$ and minor radius $\ellm$ or $\ellp$).
\begin{figure}[t!]
    \centering
        \begin{subfigure}[b]{0.475\textwidth}
                \includegraphics[width=\textwidth]{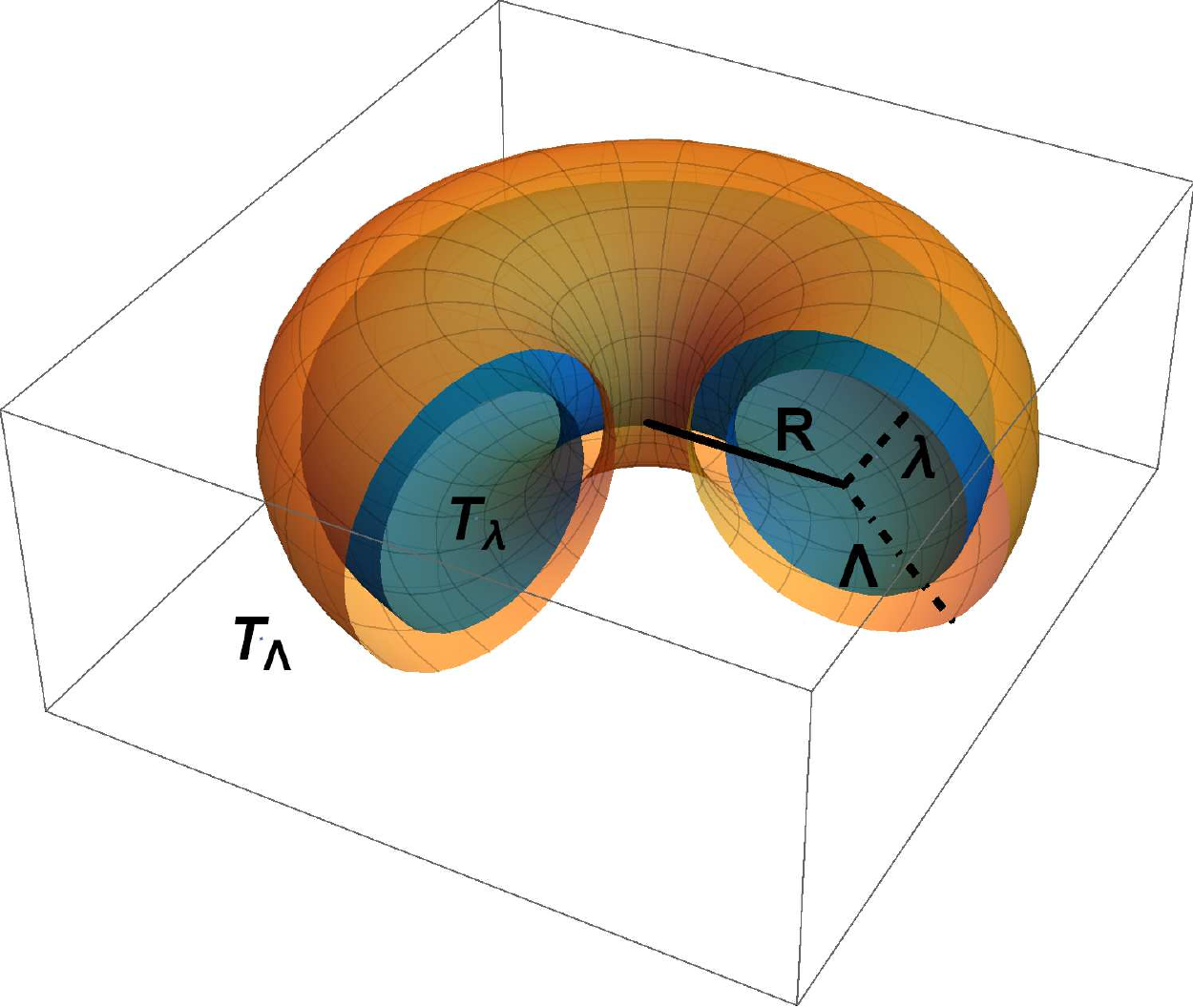}
                \caption*{Figure 1}
        \end{subfigure}
        \hspace{0.4cm}
        \begin{subfigure}[b]{0.475\textwidth}
                \includegraphics[width=\textwidth]{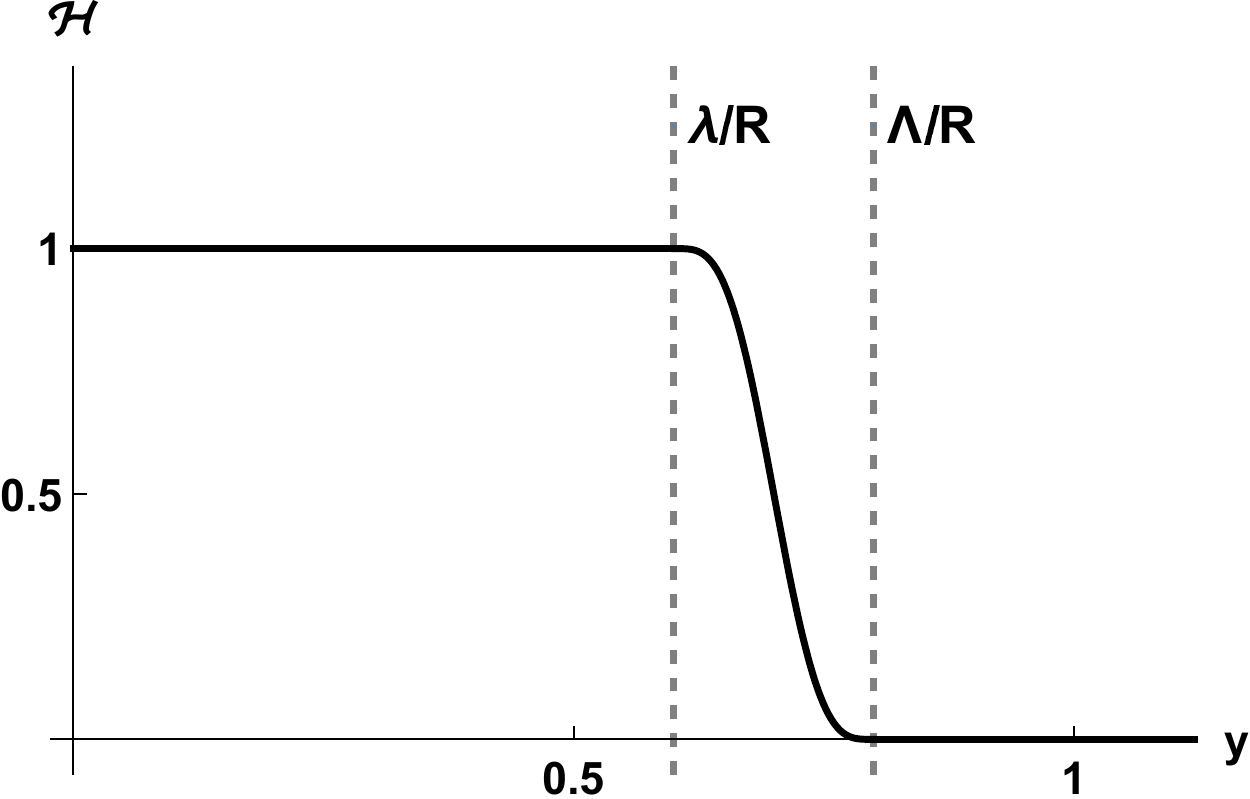}
                \caption*{Figure 2}
        \end{subfigure}
       \caption*{{\small Fig. 1: the concentric tori $\Tom$ (in blue) and $\Top$ (in orange);
       here we have fixed $\ellm/R = 3/5$, $\ellp/R = 4/5$. \\
       Fig. 2: plot of a possible shape function $\HH$; here we have considered the function
       $\HH = \HH_{(k)}$ defined as in Eq.s \eqref{HkExp} \eqref{Fkdef} of Appendix \ref{App1},
       with $k = 3$.}} \label{fig:ToriHk}
\end{figure}
\parn
In addition, we introduce a regular function that equals $1$ in the region inside $\Tom$
and vanishes identically in the region outside $\Top$
({\footnote{Obviously enough, these two regions are defined, respectively, by the inequalities
$\sqrt{(\rho-R)^2+z^2} < \ellm$ and $\sqrt{(\rho-R)^2+z^2} > \ellp$.}});
more precisely, we set
\begin{equation}
\XX(\rho,z) \,:=\, \HH\!\l(\sqrt{(\rho/R-1)^2+(z/R)^2}\;\r) \,, \label{Xidef}
\end{equation}
where $\HH \in C^k([0,+\infty);\R)$ ($k \in \{2,3,...,\infty\}$) is an assigned shape function, such that
\begin{equation}
\HH(\yr) = 1 ~~ \mbox{for $\yr \in [0,\ellm/R]$} ~, \qquad\quad
\HH(\yr) = 0 ~~ \mbox{for $\yr \in [\ellp/R,+\infty)$}
\label{chidef}
\end{equation}
(see Figure 2; Appendix \ref{App1} suggests a possible choice $\HH = \HH_{(k)}$). \parn
Inspired by the ``interpolation strategies'' of Alcubierre \cite{Alcu}, Krasnikov
\cite{Kras}, Tippett-Tsang \cite{tardis} and other authors,
we use the function $\XX$ of Eq. \eqref{Xidef} to introduce on $\R^4$ the quadratic form
\begin{equation}\begin{array}{c}
\dd{\dsX^2 \,:= } \vspace{0.1cm} \\
\dd{-\,\big[\big(1\!-\!\XX(\rho,z)\big)dt + \XX(\rho,z)\,\af\,R\,d\vfi\big]^2\!
+ \big[\big(1\!-\!\XX(\rho,z)\big)\rho\,d\vfi - \XX(\rho,z)\,\at\,dt\big]^2\!
+ d\rho^2 + dz^2 ~.} \label{dsXidef}
\end{array}\end{equation}
Here, $\af,\at \in \R \setminus \{0\}$ are two parameters which are dimensionless
in natural units with $c = 1$; the physical meaning of these parameters will be clarified
by the following analysis. \parn
Let us stress that, outside the larger torus $\Top$, the quadratic form $\dsX^2$
defined in Eq. \eqref{dsXidef} coincides with the Minkowskian line element $ds_{0}^2$
written in Eq. \eqref{ds0def}. On the other hand, inside the smaller torus $\Tom$,
$\dsX^2$ reduces to the flat line element
\begin{equation}
ds_1^2 \,=\, -\,\af^2 R^2 \,d\vfi^2 + \at^2\,dt^2 + d\rho^2 + dz^2 ~;
\label{ds1}
\end{equation}
this shows that, contrary to what happens in the Minkowskian region outside $\Top$,
within $\Tom$ the variable $\vfi \in \R/(2\pi\Z)$ plays the role of a time coordinate,
while $t \in \R$ is a spatial coordinate.
These considerations make evident that inside $\Tom$ there are closed timelike curves (CTCs):
these are naturally parametrised by the periodic coordinate $\vfi\,$
(see the beginning of Section \ref{secGeo} for further information on this topic). \parn
Anywhere on $\R^4$, Eq. \eqref{dsXidef} describes
$\dsX^2$ as an interpolation
of $ds_0^2$ and $ds_1^2$ based on $\XX$. We write $g$ for the symmetric bilinear form
associated to $\dsX^2$, and $(g_{\mu\nu})$ for its coefficients in a coordinate system $(x^\mu)$.
We claim that, when the parameters $\af,\at \in \R \setminus \{0\}$ fulfil the constraint
\begin{equation}
\af \, \at > 0 ~, \label{assa}
\end{equation}
the symmetric bilinear form $g$ determined by
$\dsX^2$ is indeed a Lorentzian metric of class $C^k$ ($k \geqs 2$) on $\R^4$.
\parn
To prove the above claim let us first point out that in the coordinate system
$(x^\mu) = (t,\vfi,\rho,z)$ we have
\begin{equation}
\det (g_{\mu\nu}) \,=\, -\,\Big[\rho\,\big(1\!-\!\XX(\rho,z)\big)^2
+ \af\,\at\, R \, \big(\XX(\rho,z)\big)^2\Big]^2 ~. \label{detg}
\end{equation}
Notice that, under the assumption \eqref{assa}, the two addenda within the square brackets
on the right-hand side of Eq. \eqref{detg} are both non-negative and vanish
simultaneously only for $\rho = 0$, thus $\det(g_{\mu\nu}) < 0$ whenever $\rho > 0$.
On the other hand, the axis $\{\rho = 0\}$ corresponds simply to a singularity of the
cylindrical coordinate system $(\vfi,\rho,z)$;
this singularity disappears if one uses a set of standard Cartesian coordinates in
a neighborhood of the said axis, where $g$ coincides with the Minkowski metric $\eta$.
The above considerations allow us to infer that $g$ is everywhere non-degenerate.
In addition, by direct inspection of the explicit expression \eqref{dsXidef} we can infer
that $1$ is an eigenvalue of $(g_{\mu\nu})$ with multiplicity two
({\footnote{To prove this statement, it suffices to notice that the components $g_{\mu\nu}$
with respect to the coordinates $(t,\vfi,\rho,z)$ form a block matrix, whose
only non-vanishing components are $g_{00}$, $g_{01} = g_{10}$, $g_{11}$ and $g_{22} = g_{33} = 1$.}});
this constrains the remaining two eigenvalues to have opposite signs in order to accomplish
the previously established condition $\det (g_{\mu\nu}) < 0$. The latter remarks prove that
$(g_{\mu\nu})$ has three positive eigenvalues and a negative one, i.e., that $g$ has signature $(3,1)$.
In order to prove that the metric $g$ is of class $C^k$ it suffices to analyse the
expressions of the coefficients $(g_{\mu\nu})$ (depending on $\XX \in C^k((0,+\infty)\times \R)$)
in the coordinate system $(t,\vfi,\rho,z)$, and to recall that $g$ coincides with $\eta$
in a neighborhood of the axis $\{\rho = 0\}$.
\parn
Of course, the $C^k$ nature of $g$ implies that the Riemann curvature tensor
(along with all the associated curvature invariants) is of class $C^{k-2}$, hence
free of singularities.
\parn
Summing up, the modified line element \eqref{dsXidef} determines a new spacetime
\begin{equation}
\Tar \,:=\, (\R^4,g) ~,
\end{equation}
which is of course topologically trivial, contains CTCs and possesses no curvature singularity at all.
In the forthcoming sections, we analyse in greater detail some interesting and non-trivial
features of $\Tar$. To perform this analysis, from now on we implement the condition \eqref{assa}
assuming that
\begin{equation}
\af > 0 \qquad \mbox{and} \qquad \at > 0 ~; \label{apbn}
\end{equation}
this causes no loss of generality since the complementary case where $\af < 0$ and $\at < 0$
can be straightforwardly recovered via the change of coordinate $t \to -t$
(or, alternatively, $\vfi \to -\vfi$).
Furthermore, we restrict the attention to cases where the shape function
\begin{equation}
\mbox{$\yr \mapsto \HH(\yr)$ is strictly decreasing for $\yr \in (\ellm/R,\ellp/R)$} ~,
\label{HHcond}
\end{equation}
which grants in particular that
\begin{equation}
\mbox{$0 < \XX(\rho,z) < 1$~~ for~ $\ellm < \sqrt{(\rho - R)^2 + z^2} < \ellp$} ~.
\end{equation}
The above requirements on $\HH$ and $\XX$ are not strictly necessary;
however, they do in fact allow to largely simplify some steps of the forthcoming analysis.
Note that all the realizations $\HH = \HH_{(k)}$ presented in Appendix \ref{App1}
fulfil the condition \eqref{HHcond}.

\subsection{A comparison with the Ori-Soen model}
For their time machine, Ori and Soen postulate in \cite{Ori2} a line element $\dsOS^2$ on $\R^4$,
using for the latter a coordinate system $(t,\vfi,\rho,z)$ like ours;
this line element depends on some parameters $\aOS,\bOS,\kOS > 0$, $\qOS \in \R$ and $0 < \dOS < \ROS$
and reads
\begin{equation}\begin{array}{c}
\dd{\dsOS^2 =} \vspace{0.05cm}\\
\hspace{-3cm}\dd{\FOS(t)\, \Big[-dt^2\! + 2\, \XOS(\rho,z)\, \Big(\aOS\,t\,dt - \bOS\,\big((\rho - \ROS)\,d\rho + z\,dz\big)\Big)\,\rho\,d\vfi\; + } \vspace{0.1cm} \\
\hspace{4cm}\dd{+ \,\Big(1 + \XOS(\rho,z)^2\big(\,\bOS^2 ((\rho-\ROS)^2 + z^2) - \aOS^2 t^2\,\big)\Big)\, \rho^2\, d\vfi^2 + d\rho^2\! + dz^2 \Big]\,;}
\end{array}\end{equation}
$$ \FOS(t) := 1 + \qOS \l(\!t- {1 \over \aOS}\r)\! - \kOS \l(\!t - {1 \over \aOS}\r)^{\!\!2}\,, \qquad
\XOS(\rho,z) := \HOS\Big(\sqrt{\big(\rho/\dOS - \ROS/\dOS\,\big)^2 + \big(z/\dOS\,\big)^2}\;\Big)\;, $$
$$ \HOS(\yr) := \l\{\!\begin{array}{ll}
\dd{\big(1-\yr^4\big)^3} & \dd{0 < \yr < 1\,,} \vspace{0.05cm}\\
\dd{0} & \dd{\yr \geqs 1\,.}
\end{array} \r. $$
Like the function $\HOS$, the metric is of class $C^2$; thus, no curvature singularity
occurs. The line element $\dsOS^2$ is \textsl{conformally}
flat outside the toroidal region bounded by $\To_{\dOS} := \{(\rho-\ROS)^2\! + \!z^2 = \dOS^2\}$;
inside $\To_{\dOS}$ the metric is not flat (not even conformally) and CTCs appear.
\parn
Differently from ours, the line element $\dsOS^2$ depends on the coordinate $t$;
CTCs and violations of the energy conditions appear only for $t > 1/\aOS$,
i.e., there is an activation time for the machine.
The price to pay for this $t$-dependence is that the Ori-Soen metric exhibits less
symmetries than ours; due to this the explicit computation of its geodesics is problematic,
as already mentioned in the Introduction.
In the forthcoming Sections \ref{secKil}-\ref{freefallSec} we will emphasize the
symmetries of our metric and use them to reduce to quadratures a class of geodesics,
describing time travel by free fall. It is not clear whether these calculations
would be possible in the Ori-Soen model; in any case, no attempt was ever done
in this direction.

\section{A tetrad. Time and space orientations, fundamental observers}\label{SecOr}
Let us proceed to determine for our model and orthonormal tetrad, consisting of four orthonormal
vector fields $(E_{(\muh)})_{\muh \in \{0,1,2,3\}}$ of class $C^k$ ($k \geqs 2$).
The facts stated hereafter can be readily inferred by direct inspection of the line
element $\dsX^2$ defined in Eq. \eqref{dsXidef}; for this reason, we will not dwell
too much on the related, elementary computations.\parn
Taking into account the explicit expression \eqref{dsXidef}, it is natural to consider
the set of $1$-forms
\begin{equation}\begin{array}{c}
\dd{e^{(0)} :=\, \big(1\!-\!\XX(\rho,z)\big)\,dt + \XX(\rho,z)\,\af\, R\,d\vfi ~, \qquad
e^{(1)} :=\, \big(1\!-\!\XX(\rho,z)\big)\,\rho\,d\vfi -\,\XX(\rho,z)\,\at\,dt ~,} \vspace{0.15cm} \\
\dd{e^{(2)} :=\, d\rho ~, \qquad e^{(3)} :=\, dz ~;}
\end{array}\end{equation}
these form a basis fulfilling
\begin{equation}
g \,=\, \eta_{\al\be} ~ e^{\al} \otimes e^{\be} ~, \label{gea}
\end{equation}
where $(\eta_{\muh \nuh}) := \mbox{diag}(-1,1,1,1)$\,.\parn
Let us now consider the dual vector fields $E_{(\nuh)}$, defined by
$\langle e^{(\muh)},E_{(\nuh)}\rangle = \delta^{\muh}_{\;\nuh}$; due to
Eq. \eqref{gea}, we have
\begin{equation}
g(E_{(\muh)},E_{(\nuh)}) \,=\, \eta_{\muh\nuh} ~. \label{normE}
\end{equation}
These vector fields are automatically granted to be of class $C^k$
and have the explicit expressions
({\footnote{\label{footTetra}Here we are implicitly making reference to Cartan's formalism
(see, e.g., Chapter 9 of \cite{Pleb}); in particular, due to Eq. \eqref{gea} (or \eqref{normE})
we have the component identity
$$ E_{(\muh)}^{\mu} = \eta_{\muh \nuh}\; g^{\mu\nu}\, e^{(\nuh)}_\nu
\qquad\qquad (\mu,\nu,\muh,\nuh \in \{0,1,2,3\}) ~, $$
which can be used to infer the explicit expressions in Eq. \eqref{EComp}.}})
\begin{equation}\begin{array}{c}
\dd{E_{(0)} =\, {\big(1\!-\!\XX(\rho,z)\big)\,\rho\,\de_t
+ \XX(\rho,z)\,\at\,\de_\vfi \over \rho\,\big(1\!-\!\XX(\rho,z)\big)^2\!
+ \af\,\at\,R\,\XX(\rho,z)^2} ~, \qquad
E_{(1)} =\, {\big(1\!-\!\XX(\rho,z)\big)\,\de_\vfi -\,\XX(\rho,z)\,\af\,R\,\de_t
\over \rho\,\big(1\!-\!\XX(\rho,z)\big)^2\! + \af\,\at\,R \,\XX(\rho,z)^2} ~,} \vspace{0.15cm} \\
\dd{E_{(2)} =\, \de_\rho ~, \qquad E_{(3)} \,= \de_z ~.} \label{EComp}
\end{array}\end{equation}
Eq. \eqref{normE} shows that $E_{(0)}$ and $E_{(i)}$ ($i \in \{1,2,3\}$)
are, respectively, timelike and spacelike everywhere.
Therefore, we can use these vector fields to establish both time and space
orientations for the spacetime $\Tar$.
In the following, we spend a few more words about the latter structures.
\parn
Let us consider the expression for $E_{(0)}$ given in Eq. \eqref{EComp}
and notice that, in the Minkowskian region outside $\Top$ (where $\XX = 0$),
this reduces to $E_{(0)} = \de_t$; this indicates, amongst else, that $E_{(0)}$
makes sense even at points where $\rho = 0$.
On account of these facts, it is natural to define the future as the time orientation
containing $E_{(0)}$. Besides, from the said expression in Eq. \eqref{EComp} it follows
that $E_{(0)} = 1/(\af R)\,\de_{\vfi}$ inside the region delimited by $\Tom$
(where $\XX = 1$); considering the previously established convention on time orientation,
this means that the coordinate vector field $\de_{\vfi}$ is timelike and future-oriented inside $\Tom$.
\parn
As usual, in this paper the term \textsl{observer} is employed as a synonym of
the expression ``timelike worldline'' (with an obvious interpretation attached to it).
In particular, any integral curve of the tetrad vector field $E_{(0)}$ will be called
a \textsl{fundamental observer}. At any point of such a worldline (with $\rho > 0$),
$E_{(1)},E_{(2)},E_{(3)}$ span the orthogonal complement $E_{(0)}^{\perp}$ which is the
linear subspace of infinitesimal simultaneity corresponding to this observer.
However, it can be easily checked that $E_{(0)}^{\perp}$ is not closed with respect
to the commutators of vector fields which, by Frobenius theorem, means that there
does not exists a foliation of $\Tar$ into spacelike (hyper-)surfaces orthogonal to
the family of fundamental observers mentioned above.
\parn
Finally, let us discuss the possibility to define an orientation on the orthogonal complement
$E_{(0)}^{\perp}$, which could be understandably referred to as a ``space orientation''.
To this purpose let us first remark that in the region outside $\Top$ we have
$E_{(1)} = \rho^{-1}\de_{\vfi}$, $E_{(2)} = \de_\rho$ and $E_{(3)} = \de_z$, which indicates,
amongst else, that $E_{(1)}$ and $E_{(2)}$ are ill defined at $\rho = 0$. Keeping in mind this
fact, at all spacetime points with $\rho > 0$ we equip $E_{(0)}^{\perp}$ with the
orientation induced by the ordered triplet $(E_{(2)},E_{(1)},E_{(3)})$.
To go on we note that, in the region outside $\Top$ (where $E_{(0)} = \de_t$), $E_{(0)}^{\perp}$
is spanned as well by the vectors $\tilde{E}_{(i)} = \de_{x^i}$ ($i = 1,2,3$), defined starting
from the coordinate system $x^1\! = \rho \cos\vfi$, $x^2\! = \rho \sin\vfi$, $x^3 = z$.
It can be easily checked that the triplets $(E_{(2)},E_{(1)},E_{(3)})$ and
$(\tilde{E}_{(1)},\tilde{E}_{(2)},\tilde{E}_{(3)})$ are equi-oriented at all spacetime
points outside $\Top$ with $\rho > 0$
({\footnote{The peculiar ordering $(E_{(2)},E_{(1)},E_{(3)})$ is used just to ensure
this result of equi-orientation.}});
moreover, since the vectors $\tilde{E}_{(i)}$ also make
sense at $\rho = 0$, we can use the triple $(\tilde{E}_{(1)},\tilde{E}_{(2)},\tilde{E}_{(3)})$
to define a coherent orientation of $E_{(0)}^{\perp}$ at these points.

\section{Symmetries of the model}\label{secKil}
First of all, let us remark that none of the tetrad vector fields $E_{(\muh)}$ $(\muh \in \{0,1,2,3\})$
considered in the previous section is a generator of isometries for $\Tar$, since none of them
fulfils the Killing equation $\mathcal{L}_{E_{(\muh)}} g = 0$ ($\mathcal{L}$ denotes the Lie derivative).
Nevertheless, the spacetime $\Tar$ does in fact possesses a number of self-evident symmetries,
both discrete and continuous, which we are going to discuss separately in the
following paragraphs. \vspace{-0.2cm}

\paragraph{Discrete symmetries.} On the one hand, it can be easily checked by direct
inspection that the transformation with coordinate representation
\begin{equation}
(t,\vfi,\rho,z) \to (-\,t,-\,\vfi,\rho,z)
\end{equation}
preserves the line element $ds^2$ of Eq. \eqref{dsXidef}, thus describing a (discrete) symmetry of $\Tar$.
Let us notice that under this transformation the vector fields
$E_{(0)}$ and $E_{(1)}$ are mapped, respectively, to $-\,E_{(0)}$ and $-\,E_{(1)}$;
on the contrary, $E_{(2)}$ and $E_{(3)}$ are left unchanged. Recalling that the tetrad
$(E_{(\muh)})_{\muh \in \{0,1,2,3\}}$ determines the time and space orientations of $\Tar$,
we can say that the spacetime $\Tar$ is in fact invariant
under the simultaneous reversal of the time and space orientations. \parn
On the other hand, due to the specific choice \eqref{Xidef} of the shape function $\XX$,
it appears that $ds^2$ is also invariant under the transformation
\begin{equation}
(t,\vfi,\rho,z) \to (t,\vfi,\rho,-\,z) ~, \label{zz}
\end{equation}
i.e., under reflection across the plane $\{z = 0\}$.
By direct inspection of the explicit expressions in Eq. \eqref{EComp} it can be readily inferred
that, under the transformation \eqref{zz}, the vector fields $E_{(0)},E_{(1)},E_{(2)}$
are left unchanged while $E_{(3)}$ is mapped to $-E_{(3)}$. Thus, $\Tar$ is invariant
under reversal of the space orientation. \parn
Summing up, the previous arguments show that $\Tar$ is invariant both under the sole reversal
of space orientation and under the simultaneous reversal of space and time orientations.
Therefore, $\Tar$ is also invariant under the sole reversal of the time orientation.
\vspace{-0.3cm}

\paragraph{Killing vector fields and stationary limit surfaces.}
Let us now pass to the analysis of the continuous symmetries of $\Tar$. \parn
First of all let us repeat that, outside the larger torus $\Top$,
the metric $g$ of the spacetime under analysis coincides with that of flat Minkowski spacetime;
therefore, it can be readily inferred that this region admits a maximal, $10$-dimensional
algebra of Killing vector fields. The same conclusion can be drawn for the region inside
the smaller torus $\Tom$, since therein the metric $g$ is flat as well. \parn
Next, let us pass to the analysis of global continuous symmetries. Since the metric coefficients
do not depend on the coordinates $t$ and $\vfi$, it can be inferred straightforwardly that both
\begin{equation}
K_{(0)} \,:=\, \de_t \qquad \mbox{and} \qquad K_{(1)} \,:=\, \de_\vfi \label{K0K1Kill}
\end{equation}
are Killing vector fields. \parn
It can be checked by elementary computations that $K_{(0)}$ is timelike on $\Th_{(0)}^{-}$,
null on $\Si_{(0)}$ and spacelike on $\Th_{(0)}^{+}$, where
\begin{equation}
\Th_{(0)}^{\pm} := \big\{\,\XX(\rho,z) \gtrless (1+\at)^{-1} \big\} ~, \qquad
\Si_{(0)} := \big\{\,\XX(\rho,z) = (1+\at)^{-1} \big\} ~.  \label{Th0}
\end{equation}
$\Th_{(0)}^{-}$ is the spacetime region in which the orbits of $K_{(0)}$, being
timelike, can be interpreted as observers; this contains the region outside $\Top$
(where $\XX(\rho,z) = 0$). $\Si_{(0)}$ is the boundary of the region $\Th_{(0)}^{-}$
and so, in the language of \cite{Carr,Hawk}, it is a \textsl{stationary limit surface}
for $K_{(0)}$. \parn
A similar analysis can be performed for the other Killing vector field $K_{(1)}$;
this is timelike on $\Th_{(1)}^{-}$, null on $\Si_{(1)}$ and spacelike on $\Th_{(1)}^{+}$, where
\begin{equation}
\Th_{(1)}^{\pm} := \big\{\,\XX(\rho,z) \lessgtr (1+\af\,R/\rho)^{-1} \big\} ~, \qquad
\Si_{(0)} := \big\{\,\XX(\rho,z) = (1+\af\,R/\rho)^{-1} \big\} ~.  \label{Th1}
\end{equation}
$\Th_{(1)}^{-}$ contains the region inside $\Tom$ (where $\XX(\rho,z) = 1$);
its boundary $\Si_{(1)}$ is a stationary limit surface for $K_{(1)}$. \parn
With our assumptions on the shape function $\HH$, it can be easily checked that both
$\Si_{(0)}$ and $\Si_{(1)}$ are \textsl{timelike} hypersurfaces
({\footnote{As an example, let us account for this statement in the case of $\Si_{(0)}$.
To this purpose, it should be recalled that in the present work $\HH(\yr)$ is assumed
to be strictly decreasing for $\yr \in (\ellm/R,\ellp/R)$ (see the comments at the end of Section \ref{secMod}).
In consequence of this assumption, the surface $\Si_{(0)}$ can be described as
$$ \Si_{(0)} \,=\, \big\{\,F(\rho,z) \,:=\,
(\rho-R)^2+z^2 - \big[\ellp - (\ellp- \ellm)\,
\HH^{-1}\big(1/(1+\at)\big)\big]^2 =\, 0 \,\big\} ~, $$
where $\HH^{-1}$ denotes the local inverse of $\HH$ in the interval $(\ellm/R,\ellp/R)$
(notice that $0 < 1/(1+\at) < 1$, since $\at > 0$). Then, considering the vector field
$n_F \equiv (n_F^\mu) = (g^{\mu\nu} (dF)_\nu)$ normal to $\Si_{(0)}$, it can be inferred
by elementary computations that
$$ g(n_F,n_F) \,=\, 4\,\big[\ellp - (\ellp- \ellm)\, \HH^{-1}\big(1/(1+\at)\big)\big]^2 \,>\, 0 ~. $$
The above relation proves that $n_F$ is spacelike, which by definition is equivalent
to say that $\Si_{(0)}$ is timelike (see, e.g., Section 2.7 of \cite{Hawk}).}});
so, in particular, neither of them is a Killing horizon for $K_{(0)}$ or $K_{(1)}$.
Besides, it appears that the positions and the shapes of $\Si_{(0)}$
and $\Si_{(1)}$ depend strongly on the particular choices of the radii $\ellm,\ellp,R$
and of the parameters $\af,\at > 0$. In particular, recalling that we are assuming the
shape function $\HH(\yr)$ to be strictly decreasing for $\yr \in (\ellm/R,\ellp/R)$,
it can be checked by direct inspection of Eq.s \eqref{Th0} \eqref{Th1} that for $a/b < 1 - \ellp/R$
(compare with the forthcoming Eq. \eqref{impl}) there is a region of spacetime where
a test particle cannot remain at rest neither with respect to observers in the outer
Minkowskian region nor with respect to observers in the innermost region delimited by $\Tom$.

\section{Results on causal geodesics}\label{secGeo}
Here and in the rest of the paper, a geodesic in $\Tar$ will always
be represented in terms of an affine parametrization
$\cu : \tau \mapsto \cu(\tau)$; we will write $\dot{~}$ for
the derivative with respect to $\tau$. \parn
Let us first remark that, in the region outside $\Top$, all
geodesics do in fact coincide with those of flat Minkowski
spacetime; in particular, the orbits of the Killing vector field
$K_{(0)}$ are timelike geodesics in this region.
Similar considerations hold for the flat spacetime region inside
the smaller torus $\Tom$. Notably, as anticipated in Section
\ref{secMod}, we have CTCs\, $\cu : \tau \mapsto \cu(\tau)$
with the following representation in coordinates $(x^\mu) := (t,\varphi,\rho,z)$\,:
\begin{equation}
(\cu^\mu(\ta)) = \big(t_0,\vfi_0 + \Omega\,\ta\,(\mbox{mod $2\pi$}),\rho_0,z_0\big),
\qquad 0 \leqs \ta \leqs {2 \pi/\Omega} ~, \label{CTC}
\end{equation}
where $\Omega \!>\! 0$, $t_0 \!\in\! \R$, $\vfi_0 \!\in\! \R/(2\pi \Z)$
and $\rho_0 \!>\!0$, $z_0 \in \R$ are such that
$0 \!\leqs \!\sqrt{(\rho_0/R \!-\! 1)^2\!+\!(z_0/R)^2}$ $< \ellm/R$\,; such a
curve is future-oriented
({\footnote{The metric $g$ of $\Tar$ has constant coefficients in coordinates
$(t,\vfi,\rho,z)$ in the region inside $\Tom$ (where $\XX = 1$ and $\dsX^2$
has the form \eqref{ds1}); $\cu$ is represented in these coordinates by an affine
function of $\ta$, so it is a geodesic. For $0 \leqs \tau \leqs 2\pi/\Omega$
we have $\dcu(\ta) = \Omega\,\de_{\vfi}$; this is a timelike vector, we now
discuss its orientation. It is easily checked that
$g\big(E_{(0)}\big(\cu(\ta)\big),\dcu(\ta)\big) = -\,\af\,R\,\Omega < 0$
(recall our assumption \eqref{apbn} and Eq. \eqref{EComp});
since we have chosen $E_{(0)}$ to be everywhere future-oriented
(see Section \ref{SecOr}), the latter identity shows that $\dcu(\ta)$
is future-oriented as well.}}).
In passing, let us also remark that the above curves coincide with the orbits
of the Killing vector field $K_{(1)}$ inside $\Tom$.
\vspace{0.2cm} \parn
Let us now pass to the study of different causal geodesics, not necessarily confined
outside $\Top$ or inside $\Tom$. As a matter of fact we are going to show in the subsequent
Section \ref{freefallSec} that, at least for suitable choices of the
parameters $\af,\at$ and of the shape function $\HH$, there exist timelike geodesics
which start from the region outside $\Top$, cross both $\Top$ and $\Tom$ and
return outside $\Top$; this fact is not self-evident a priori and has non-trivial
consequences to be discussed later on.
\parn
To this purpose, let us first recall that any (affinely parametrised) geodesic
can be characterized as a solution $\cu$ of the Euler-Lagrange equations associated
to the Lagrangian function
({\footnote{Of course, if $X$ is tangent to $\Tar$ at a point $p$, $g(X,X)$ stands for
$g_p(X,X)$; the notation $L(X)$ understands the dependence on $p$.
Similar remarks will never be repeated in the remainder of this paper.}})
\begin{equation}
L: T\Tar \to \R\;, \qquad L(X) \,:=\, {1 \over 2}\; g\big(X,X\big)
\end{equation}
whose representation in our usual coordinates is
\begin{equation}\begin{array}{c}
L(x^\mu, \dot{x}^\mu) = \vspace{0.1cm} \\
\dd{-{1 \over 2} \,\big[\,\big(1\!-\!\XX(\rho,z)\big) \dot{t} + \XX(\rho,z)\,\af\,R\,\dot{\vfi}\,\big]^2\!
+ {1 \over 2} \,\big[\,\big(1\!-\!\XX(\rho,z)\big)\rho\,\dot{\vfi} - \XX(\rho,z)\,\at\,\dot{t}\,\big]^2\!
+ {1 \over 2} \,\dot{\rho}^2 + {1 \over 2} \,\dot{z}^2 ~.} \label{lagcoord}
\end{array}\end{equation}
For simplicity, from now on we restrict the attention to the plane $\{z = 0\}$ of
$\Tar$, that we equip with the coordinates $(x^A)_{A \in \{0,1,2\}} := (t,\varphi,\rho)$.
Our considerations involve the dimensionless variable
\begin{equation}
\rv \,:=\, \rho/R \,\in\, (0,+\infty) \label{rvdef}
\end{equation}
and the function (see Eq.s \eqref{Xidef} \eqref{chidef})
\begin{equation}
\rho \in (0,+\infty) \,\mapsto\, \XX(\rho,0) \,=\, \Hr(\rho/R)~, \qquad\;
\Hr(\rv) \,:=\, \HH\big(\,|\rv-1|\,\big) ~, \label{Hrdef}
\end{equation}
which fulfils, in particular,
\begin{equation}\begin{array}{c}
\dd{\Hr(\rv) = 0 ~~ \mbox{for ~$\rv \in (0, 1 - \ellp/R] \cup [1 + \ellp/R, + \infty)$}~,}
\vspace{0.2cm} \\
\dd{\Hr(\rv) = 1 ~~ \mbox{for ~$\rv \in [1 - \ellm/R, 1 + \ellm/R]$} ~.}
\end{array}\end{equation}
Let us write $\L0$ for the Lagrangian $L$ restricted to
the (tangent bundle of the hyper-) plane $\{ z = 0 \}$. The coordinate
representation of $\L0$ is obtained setting $z=0,\dot{z}=0$ in
Eq. \eqref{lagcoord}, and can be written as follows:
\begin{equation}\begin{array}{c}
\dd{\L0\big(x^A,\dot{x}^A\big) \,:=} \vspace{0.1cm}\\
\dd{- {1 \over 2}\,\big[\,\big(1\!-\!\Hr(\rho/R)\big)\,\dot{t} + \af\,R\;\Hr(\rho/R)\,\dot{\vfi}\,\big]^2\!
+ {1 \over 2}\, \big[\,\rho\,\big(1\!-\!\Hr(\rho/R)\big)\,\dot{\vfi} - \at\,\Hr(\rho/R)\,\dot{t}\,\big]^2\!
+ {1 \over 2}\, \dot{\rho}^2  .}\label{RedLagDef}
\end{array}\end{equation}
It is readily checked that there are geodesics $\cu$ of $\Tar$ lying in the
plane $\{ z = 0 \}$, and that such geodesics coincide with the
solutions of the Euler-Lagrange equations induced by $\L0$
({\footnote{We have just stated that the geodesics in $\Tar$ are
characterized by the Euler-Lagrange equations for $L$.
One checks by elementary means that the equation
$(d/d \ta) (\de L/\de \dot{z}) - \de L/\de z = 0$ is fulfilled setting $z(\ta) := 0$;
this statement depends crucially on the fact that $(\de_z \XX)(\rho,0) = 0$
(see Eq. \eqref{Xidef}). The remaining Euler-Lagrange equations induced by $L$ coincide,
if $z(\ta) = 0$, with those associated to $\L0$.}})
({\footnote{Notice that $\L0$ could be seen as the Lagrangian associated to
a $3$-dimensional spacetime, obtained from $\Tar$ by suppression of
the coordinate $z$; because of this, all the considerations that follow
could be interpreted in terms of generic geodesics in this $3$-dimensional space-time.}}).
\parn
The Lagrange equations induced by $\L0$ possess a maximal
number of first integrals, and can be solved by quadratures. The said first integrals are the energy
and two conserved momenta; let us give more details on this subject. \parn
The energy function, defined via the general theory
of Lagrangian systems, coincides with $\L0$ due to the purely ``kinetic''
nature of this Lagrangian. Thus $\L0(\dcu(\tau)) = {1 \over 2}\, g(\dcu(\tau),
\dcu(\tau)) $ = const. along any solution $\cu$ of the Lagrange
equations; this corresponds to the well known conservation law
for the norm of the velocity of any geodesic.
\parn
We are mainly interested in causal geodesics, i.e., in null or timelike geodesics.
In the null case, we obviously have
\begin{equation}
\L0(\dcu) \,=\, 0 ~. \label{m2EqL}
\end{equation}
In the timelike case, after possibly rescaling $\ta$ by a constant factor, we can arrange things so that
$g(\dcu,\dcu) = -1$ i.e.
\begin{equation}
\L0(\dcu) \,=\, -\,{1 \over 2} ~, \label{m2EqT}
\end{equation}
which is equivalent to saying that the parameter $\ta$ is proper time. \parn
To go on, let us notice that the explicit expression \eqref{RedLagDef} for $\L0$
does not depend explicitly on $t,\vfi$; so, the system admits as conserved quantities
the canonical momenta
({\footnote{In passing, let us remark that $p_t$ and $p_{\vfi}$ are
strictly related to the Killing vector fields
$K_{(0)}$ and $K_{(1)}$, defined in Eq. \eqref{K0K1Kill}. More precisely,
for each vector $X$ tangent to $\{z = 0 \}$ one has
$$ p_t(X) \,=\,g(K_{(0)},X) ~, \qquad p_{\vfi}(X) \,=\, g(K_{(1)},X) ~. \vspace{-0.2cm} $$
This is readily checked expressing $p_t(X)$, $p_\vfi(X)$ in terms
of the components $\dot{t}, \dot{\vfi}, \dot{\rho}$ of $X$,
and comparing with the coordinate expressions of $g(K_{(0)},X)$,
$g(K_{(1)},X)$.}})
\begin{equation}
p_t \,:=\, {\de \L0 \over \de \dot{t}}~, \qquad
p_{\vfi} \,:=\, {\de \L0 \over \de \dot{\vfi}} ~.
\end{equation}
\vfill\eject\noindent
$\phantom{a}$
\vskip-1.5cm
\noindent
In hindsight, it is convenient to replace the momenta $p_t, p_{\vfi}$ with the related quantities
\begin{equation}\begin{array}{c}
\dd{\pt \,\equiv\, \pt(\rho,\dot{t}, \dot{\vfi}) \,:=\, -\,p_t\, =} \vspace{0.1cm}\\
\dd{\big[\,\big(1\!-\!\Hr(\rho/R)\big)^2 - \at^2\,\Hr(\rho/R)^2 \,\big]\,\dot{t}\,
+\,\big(\af + \at\,\rho/R\big)\,\Hr(\rho/R)\big(1\!-\!\Hr(\rho/R)\big)\,R\,\dot{\vfi} ~,}
\label{ptdef}
\end{array}\end{equation}
\begin{equation}\begin{array}{c}
\dd{\pf \,\equiv\, \pf(\rho,\dot{t}, \dot{\vfi})\,:=\,-\;{p_\vfi \over \pt\,R}  \,=} \vspace{0.05cm}\\
\dd{{1 \over \pt}\,\big(\af + \at\,\rho/R\big)\,\Hr(\rho/R)\big(1\!-\!\Hr(\rho/R)\big)\,\dot{t}\,
-\,{1 \over \pt}\,\big[\,(\rho/R)^2 \big(1\!-\!\Hr(\rho/R)\big)^2\! - \af^2\,\Hr(\rho/R)^2\,\big]\,R\,\dot{\vfi} ~.}
\label{pfdef}
\end{array}\end{equation}
Note that both $\pt$ and $\pf$ are dimensionless; hereafter we show that $\pt > 0$ in the
situation in which we are mainly interested.
To be precise, let us consider the case of a future-oriented causal curve passing through
the Minkowskian region outside $\Top$ (where $\rho > R + \ellp$ and $\Hr = 0$); then Eq. \eqref{ptdef} gives
\begin{equation}
\pt(\rho,\dot{t}, \dot{\vfi}) \,=\, \dot{t} \label{68}
\end{equation}
along the curve, implying that $\pt > 0$. In particular, let us consider the case of
a future-oriented timelike curve parametrized by proper time $\ta$; in this case we have
$\pt = \dot{t} \equiv dt/d \ta$ along the curve, indicating that $\pt$ is the familiar
``Lorentz factor'' of relativity in Minkowski spacetime.
Thus $\pt \geqs 1$ and the limits $\pt \to 1^{+}$, $\pt \to + \infty$ correspond, respectively,
to non-relativistic and ultra-relativistic motions with respect to the coordinate frame $(t,\vfi,\rho,z)$. \parn
Eq.s \eqref{ptdef} \eqref{pfdef} are easily solved for $\dot{t},\dot{\vfi}$ in terms
of $\rho,\pt,\pf$; this gives
({\footnote{Notice that the denominators in Eq.s \eqref{tp} \eqref{fp} coincide, apart
from overall multiplicative constant factors ($-1$ and $-R$, respectively), with the
metric determinant $\det (g_{\mu \nu})$ (see Eq. \eqref{detg}) at $(x^\mu) = (t,\vfi,\rho,0)$;
besides, recall that $\det (g_{\mu \nu}) < 0$ (see the considerations reported below Eq. \eqref{detg}).
These facts suffice to infer that the expressions in the cited equations are well defined whenever $\rho > 0$.}})
\begin{equation}\begin{array}{c}
\dd{\dot{t}(\rho,\pt,\pf) \,=\,} \vspace{0.1cm}\\
\dd{\pt\;{\big[(\rho/R)^2 \big(1\!-\!\Hr(\rho/R)\big)^2\! - \af^2\,\Hr(\rho/R)^2\big]
+ \big(\af\!+\!\at\,\rho/R\big)\,\Hr(\rho/R)\big(1\!-\!\Hr(\rho/R)\big)\,\pf \over
\big[(\rho/R) \big(1\!-\!\Hr(\rho/R)\big)^2\! + \af\,\at\,\Hr(\rho/R)^2\big]^2} ~,} \label{tp}
\end{array}\end{equation}
\begin{equation}\begin{array}{c}
\dd{\dot{\vfi}(\rho,\pt,\pf) \;=} \vspace{0.1cm}\\
\dd{\pt\; {\big(\af\!+\!\at\,\rho/R\big)\,\Hr(\rho/R)\big(1\!-\!\Hr(\rho/R)\big)\,
- \big[\big(1\!-\!\Hr(\rho/R)\big)^2 - \at^2\,\Hr(\rho/R)^2\big]\,\pf \over
R\,\big[(\rho/R) \big(1\!-\!\Hr(\rho/R)\big)^2\! + \af\,\at\,\Hr(\rho/R)^2\big]^2} ~.} \label{fp}
\end{array}\end{equation}
In particular, the above relations give
\begin{equation}
\dot{t}(\rho,\pt,\pf) \,=\, \pt\,, \quad \dot{\vfi}(\rho,\pt,\pf) \,=\,
-\,\pt\, {\pf\,R \over \rho^2} \qquad
\mbox{for~ $\rho \in (0, R - \ellp] \cup [R + \ellp, + \infty)$}~, \label{ppWX0}
\end{equation}
\begin{equation}
\dot{t}(\rho,\pt,\pf) \,=\; -\,\pt\;{1 \over \at^2}\,, \quad
\dot{\vfi}(\rho,\pt,\pf) \,=\, \pt\;{\pf \over \af^2 R}
\qquad\mbox{for~ $\rho \in [R - \ellm, R + \ellm]$}~. \label{ppWX1}
\end{equation}
Notably, the first identity in Eq. \eqref{ppWX1} shows that $\dot{t}$ is (constant and)
negative inside the smaller torus $\Tom$; in consequence of this, the coordinate
time $t$ decreases along future-oriented causal geodesics in the region inside $\Tom$.
This fact is crucial for the possibility of time travels to the past,
a topic to be discussed in more detail in the following Section \ref{freefallSec}. \parn
Now, let us consider the reduced Lagrangian
\begin{equation}
\RL_{\pt,\pf}(\rho,\dot{\rho}) \,:=\,
\Big[ \L0(x^\alpha,\dot{x}^{\alpha}) - \big((-\pt)\,\dot{t} + (-\,\pt\,\pf\,R)\,\dot{\vfi}\big)
\Big]_{\dot{t} \,=\, \dot{t}(\rho,\pt,\pf),~\dot{\vfi}\,=\, \dot{\vfi}(\rho,\pt,\pf)}
\end{equation}
(recall that $-\,\pt = p_t$ and $-\,\pt\,\pf\,R = p_{\vfi}$); by direct computation, this can be expressed as
\begin{equation}
\RL_{\pt,\pf}(\rho,\dot{\rho}) \,=\, {1 \over 2}\;\dot{\rho}^2 - V_{\pt,\pf}(\rho) ~,
\end{equation}
where we have introduced the effective potential
\begin{equation}\begin{array}{c}
\dd{V_{\pt,\pf}(\rho) \,:=\,} \vspace{0.1cm}\\
\dd{\l({\pt^2 \over 2}\r) {\big[\af\,\Hr(\rho/R) - \big(1\!-\!\Hr(\rho/R)\big)\,\pf\big]^2
- \big[(\rho/R)\,\big(1\!-\!\Hr(\rho/R)\big) + \at\,\Hr(\rho/R)\,\pf\big]^2 \over
\big[(\rho/R)\,\big(1-\Hr(\rho/R)\big)^2 + \af\,\at\;\Hr(\rho/R)^2\big]^2} ~.} \label{Vdef}
\end{array}\end{equation}
Let us remark that $V_{\pt,\pf}$ depends on the radial coordinate $\rho$
only through the dimensionless ratio $\rv := \rho/R$
(we have taken this fact into account in Fig.s 3a-3b, showing the graphs of
$V_{\pt,\pf}$ as a function of $\rv$ for some choices of the parameters).
Moreover,
\begin{equation}
V_{\pt,\pf}(\rho) \,=\, {\pt^2 \over 2} \l({R^2\,\pf^2 \over \rho^2} - 1 \r) \qquad
\mbox{for~ $\rho \in (0, R - \ellp] \cup [R + \ellp, + \infty)$}~, \label{VXRed0}
\end{equation}
\begin{equation}
V_{\pt,\pf}(\rho) \,=~ \mbox{const.}~ =\, {\pt^2 \over 2\,\af^2}\l({\af^2 \over \at^2} - \pf^2\r)
\qquad \mbox{for~ $\rho \in [R - \ellm, R + \ellm]$}~. \label{VXRed1}
\end{equation}
\begin{figure}[]
    \centering
        \begin{subfigure}[b]{0.47\textwidth}
                \includegraphics[width=\textwidth]{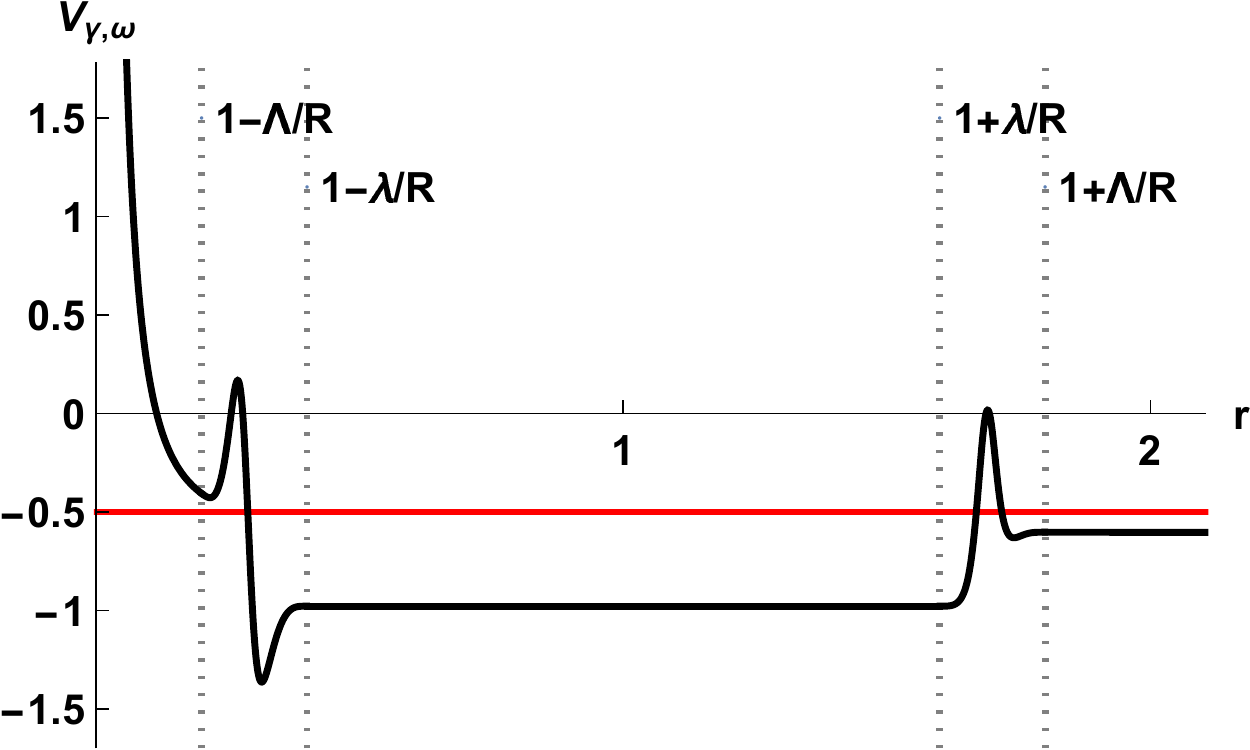}
                \caption*{Figure 3a: $\pf = -\,0.08$.} \label{ga}
                \vspace{0.cm}
        \end{subfigure}
        \hspace{0.3cm}
        \begin{subfigure}[b]{0.47\textwidth}
                \includegraphics[width=\textwidth]{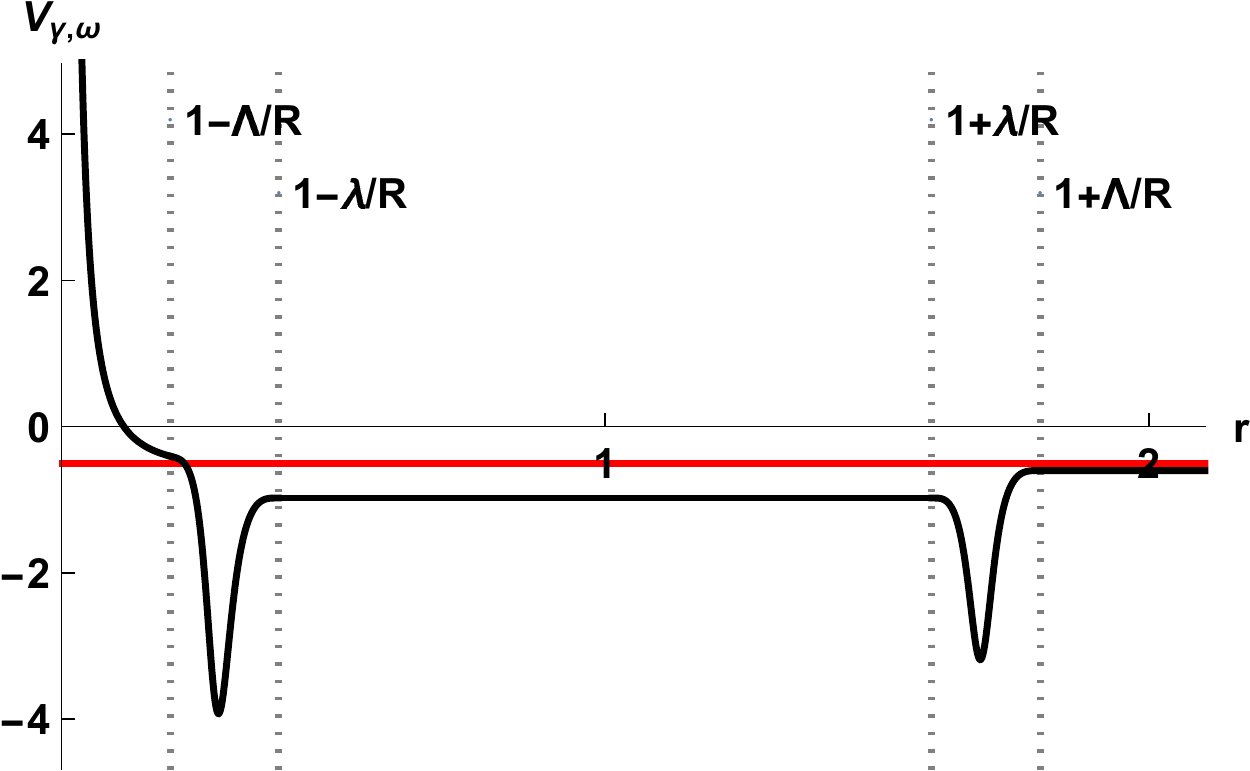}
                \caption*{Figure 3b: $\pf = 0.08$.} \label{gb}
                \vspace{0.cm}
        \end{subfigure}
        \caption*{{\small Fig.s 3a-3b: the potential $V_{\pt,\pf}$ (black line)
        as a function of the dimensionless variable $\rv := \rho/R$, for $\ellm/R = 3/5$,
        $\ellp/R = 4/5$, $\af = 9/100$, $\at = 10$, $\pt = 1.1$ and for two opposite
        choices of $\pf$. In both cases, the shape function $\HH = \HH_{(k)}$ is the one
        given in Eq.s \eqref{HkExp} \eqref{Fkdef} of Appendix \ref{App1}, with $k = 3$. The red
        lines correspond to the energy value $\En = -1/2$.}}
        \label{fig:Potential}
\end{figure}
\par\noindent
Clearly, $\RL_{\pt,\pf}$ can be interpreted as the Lagrangian function
associated to a classical point particle moving along the half-line $(0,+\infty)$,
in presence of the potential $V_{\pt,\pf}$\,. The total energy of this one-dimensional
system, a conserved quantity, is
\begin{equation}
\En \,:=\, {1 \over 2}\,\dot{\rho}^2 + V_{\pt,\pf}(\rho) ~;
\end{equation}
this is found to coincide with the Lagrangian $\L0$ and this result, along with
Eq.s \eqref{m2EqL} \eqref{m2EqT}, gives
\begin{equation}
\En \,=\, \l\{\!\begin{array}{ll}
\dd{0} & \quad \mbox{for null geodesics} ~, \vspace{0.1cm} \\
\dd{-1/2} & \quad \mbox{for timelike geodesics} ~.
\end{array} \r.
\end{equation}
Of course, for each solution of the Euler-Lagrange equations, $\rho(\ta)$ ($\ta\!\in\!\R$)
is confined within a connected component of the region $\{ \rho \in (0,+\infty) ~|~
V_{\pt,\pf}(\rho) \leqs \En \}$ and conservation of the total energy can be used
to reduce to quadratures the computation of $\rho(\ta)$. \parn
Let us consider an interval $[\ta_i,\ta_f] \subset \R$ and assume
\begin{equation}
\mbox{sign}\, \dot{\rho}(\ta) = \si \in \{\pm 1\} \quad \mbox{for $\ta_i < \ta < \ta_f$}~,
\qquad~ \rho(\ta_i) = \rho_i,~~\rho(\ta_f) = \rho_f ~. \label{assum}
\end{equation}
Then, on the said interval we have
\begin{equation}
\dot{\rho} \equiv \dot{\rho}(\rho,\En,\pt,\pf)\,=\, \si\, \sqrt{2 \big(\En - V_{\pt,\pf}(\rho)\big)} ~, \label{dacit}
\end{equation}
whence
\begin{equation}
\ta_f - \ta_i \,=\, \si \int_{\rho_i}^{\rho_f}
{d \rho \over \sqrt{2 \big(\En - V_{\pt,\pf}(\rho)\big)}}~. \label{tafi}
\end{equation}
In the case of a timelike geodesic ($\En=-1/2$), the above equation gives the variation
of the proper time along this part of the geodesic. \parn
Keeping the assumptions \eqref{assum}, let
$$
t(\ta_h) \,=\, t_h~, \quad \vfi(\ta_h) \,=\, \vfi_h \qquad (h \in \{i,f\})
$$
and consider the maps $[\rho_i, \rho_f] \ni \rho \mapsto t(\rho),\,\vfi(\rho)$
obtained composing the functions $[\ta_i, \ta_f] \ni \ta \mapsto t(\ta),\,\vfi(\ta)$
with the inverse function $\rho \mapsto \ta(\rho)$ of the map $[\ta_i, \ta_f] \ni \ta
\mapsto \rho(\ta)$. Then, using the notation $' \equiv d/d \rho$, we have
$$
t'(\rho) \,=\, {\dot{t}(\rho, \pt, \pf) \over \dot{\rho}(\rho, \En,\pt,\pf) }~, \quad
\vfi'(\rho) \,=\, { \dot{\vfi}(\rho, \pt, \pf) \over \dot{\rho}(\rho, \En,\pt,\pf) }~,
$$
where $\dot{t}$, $\dot{\vfi}$, $\dot{\rho}$ are as in Eq.s \eqref{tp}, \eqref{fp}, \eqref{dacit}.
Using the explicit expression for $\dot{\rho}$ and integrating, we get
\begin{equation}
t_f - t_i \,=\, \si \!\int_{\rho_i}^{\rho_f}\! d\rho \;
{\dot{t}(\rho, \pt, \pf) \over \sqrt{2 \big(\En - V_{\pt, \pf}(\rho)\big)}}~, \label{rulet}
\end{equation}
\begin{equation}
\vfi_f - \vfi_i \,=\, \si \!\int_{\rho_i}^{\rho_f}\! d\rho \;
{\dot{\vfi}(\rho, \pt, \pf) \over \sqrt{2 \big(\En - V_{\pt, \pf}(\rho)\big)}} \label{rulefi}
\end{equation}
(the last equality being understood mod $2\pi$). \parn
In the subsequent sections we will confine the attention to the case of
timelike geodesics ($\En = -1/2$) and use the present results to show
that an observer freely falling in the plane $\{z = 0\}$ can travel
backwards in time.
No further discussion will be performed on null geodesics ($\En = 0$);
we plan to return to this subject in future works, where the present analysis
of null geodesics will be used to discuss the light signals emitted
by the time traveller towards the outer Minkowskian region.

\section{Free fall and time travel into the past}\label{freefallSec}
\subsection{Free fall under special assumptions}\label{subspec}
Let us consider a massive test particle freely falling in the plane $\{z=0\}$.
The worldline of such a particle is a timelike geodesic and can be analysed following
the framework described in the previous section with $\En = -1/2$, i.e., using a proper time
parametrization $\ta \mapsto \cu(\ta)$. \parn
We make the following assumptions (i)(ii), involving the initial conditions $\cu(0)$, $\dcu(0)$
and the dimensionless parameters $\pt$, $\pf$ associated to $\cu$ (see Eq.s \eqref{ptdef}
\eqref{pfdef}): \vspace{-0.1cm}
\begin{itemize}
\item[(i)] We have
\begin{equation}
t(0) = 0\,,~ \dot{t}(0) >0\,, \qquad \vfi(0) = 0\,, \qquad
\rho(0) = \rho_0 > R + \ellp\,, ~\dot{\rho}(0) < 0\,. \label{assum0}
\end{equation}
The choices of $t(0)$ and $\vfi(0)$ are conventional, and imply no loss of generality.
The condition $\dot{t}(0) > 0$ indicates that $\dcu(0)$ is future-oriented:
by continuity, $\dcu(\ta)$ will be future-oriented for all $\ta$.
The conditions on $\rho(0)$ and $\dot{\rho}(0)$ mean that the particle is initially in the
Minkowskian region outside the larger torus $\Top$, with radial velocity pointing towards $\Top$.
\vfill\eject\noindent
$\phantom{a}$
\vskip-1.5cm
\noindent
\item[(ii)] The effective potential $V_{\pt, \pf}$ defined in Eq. \eqref{Vdef}
fulfils
\begin{equation}
V_{\pt, \pf}(\rho) \l\{\!\begin{array}{ll}
> -1/2 & ~\mbox{for $\rho \in (0, \rho_1)$}~, \\
= -1/2 & ~\mbox{for $\rho = \rho_1$}~, \\
< -1/2 & ~\mbox{for $\rho \in (\rho_1, + \infty)$}
\end{array} \r.
\qquad \mbox{for some ~$\rho_1 \in (0, R - \ellp)$} \label{condv}
\end{equation}
(this holds, e.g., in the case of Fig. 3b). Together with (i), this ensures that,
for $\ta \geqs 0$, the radial coordinate $\rho = \rho(\ta)$ of the particle will decrease
until a minimum value $\rho_1$ and then it will increase (we shall return on this later).
\end{itemize}

\subsection{Implications of (i)(ii) on the parameters $\bm{\af,\at,\pt,\pf}$}
Firstly let us recall that, for a motion like the one under analysis, $\pt$ is the familiar
Lorentz factor of special relativity (a fact already mentioned after Eq. \eqref{68}).
In particular, we have the lower bound $\pt \geqs 1$;
moreover, the equality $\pt=1$ cannot be realized in the particular case that we are considering
for this would imply that the three-velocity of the particle vanishes at $\ta=0$, against
the assumption $\dot{\rho} <0$ of Eq. \eqref{assum0}. In conclusion, we have
\begin{equation}
\pt > 1 ~. \label{imp0}
\end{equation}
Secondly we remark that, according to (ii), for suitable $\ta > 0$ we have
$\rho(\ta) \in [R - \ellm,R + \ellm]$; for the same values of $\ta$,
Eq. \eqref{EComp} (with $\XX(\rho,z) = 1$) and Eq. \eqref{ppWX1} give
$E_{(0)}(\cu(\ta)) = 1/(\af R)\,\de_{\vfi}$
and $\dot{\vfi}(\ta) = \pt\,\pf/(\af^2 R)$, which implies
$g\big(E_{(0)}(\cu(\ta)), \dcu(\ta)\big) = -\,\pt\,\pf/\af$\,. On the other hand,
$E_{(0)}(\cu(\ta))$ and $\dcu(\ta)$ have the same time orientation
(indeed, they are both future-oriented); thus, $g\big(E_{(0)}(\cu(\ta)), \dcu(\ta)\big) < 0$.
Recalling that we are assuming $\af > 0$ (see Eq. \eqref{apbn}), the facts pointed
out above give
\begin{equation}
\pf > 0 ~. \label{imp3}
\end{equation}
Let us recall that, according to Eq.\! \eqref{ppWX0}, we have $\dot{\vfi}(0) = -\,\pt\,\pf R/\rho^2_0$;
so, Eq. \eqref{imp3} implies $\dot{\vfi}(0) < 0$.\parn
To go on, let us consider the point $\rho_1$ mentioned in Eq. \eqref{condv}. This can be readily determined
solving the equation $V_{\pt,\pf}(\rho_1)= -1/2$ with the expression \eqref{VXRed0}
for $V_{\pt,\pf}$, which  gives
\begin{equation}
\rho_1 \,= {R\;\pf \over \sqrt{1 - 1/\pt^2}} ~. \label{ro1}
\end{equation}
The right-hand side of the above equation must belong to the interval $(0, R - \ellp)$,
so we are forced to assume that
\begin{equation}
{\pf \over \sqrt{1 - 1/\pt^2}} \,<\, 1 - {\ellp \over R} ~. \label{imp1}
\end{equation}
Finally the condition $V_{\pt,\pf}(\rho) < -1/2$ in Eq. \eqref{condv}, required to hold for all
$\rho \in (\rho_1, + \infty)$, must be fulfilled in particular for
$\rho \in [R - \ellm, R + \ellm]$ where $V_{\pt,\pf}(\rho)$ has the
constant value indicated in Eq. \eqref{VXRed1}; this yields the inequality
\begin{equation}
{\pt^2 \over 2\,\af^2} \l({\af^2 \over \at^2} - \pf^2\r) < -\,{1 \over 2} ~. \label{imp2}
\end{equation}
By elementary manipulations, one finds that the constraints \eqref{imp0}, \eqref{imp3},
\eqref{imp1} and \eqref{imp2} imply
\begin{equation}\begin{array}{c}
\dd{{\af \over \at} \,<\, 1 - \,{\ellp \over R} ~, \qquad ~
\pt \,>\, \sqrt{{(1 - \ellp/R)^2 + \af^2 \over (1 - \ellp/R)^2 - \af^2/\at^2} }~,}
\vspace{0.1cm}\\
\dd{{\af \over \at} \,\sqrt{1 + {\at^2 \over \pt^2}} \,<\, \pf \,<\,
\l(1 - {\ellp \over R} \r) \sqrt{1 - {1 \over \pt^2}}} \label{impl}
\end{array}\end{equation}
(note that the inequalities in the first line are equivalent to the relation
${\af \over \at} \sqrt{1\! + \!{\at^2 \over \pt^2}} <
\l(1\! -\! {\ellp \over R} \r) \sqrt{1 \!-\! {1 \over \pt^2}}$\,).\parn
The above arguments show that the assumptions (i)(ii) imply \eqref{impl}.
Investigating the validity of the converse implication \eqref{impl} $\Rightarrow$ (i)(ii)
is a non trivial task, since the shape function $\HH$ is implicitly involved in
the condition (ii); in any case, this problem is not relevant for our purposes.
\vspace{-0.5cm}
\newpage
$\phantom{a}$
\vskip-1.5cm
\noindent

\subsection{The claimed time travel: qualitative features}\label{subqual}
The essential qualitative features of a timelike geodesic motion $\cu$ under the
assumptions (i)(ii) of subsection \ref{subspec} have been sketched in the
accompanying comments.
To be more precise, the cited assumptions ensure that:\vspace{-0.1cm}
\begin{itemize}
\item[(a)] for $\ta \geqs 0$, the coordinate $\rho(\ta)$ of the particle will decrease
until reaching the minimum value $\rho_1$ at a certain proper time $\ta_1$; \vspace{-0.2cm}
\item[(b)] after this, the coordinate $\rho(\ta)$ will increase and return to its
initial value $\rho_0$ at a proper time $\ta_2$:
\begin{equation}
\rho(\ta_2) \,=\, \rho_0~. \label{cro0}
\vspace{-0.2cm}
\end{equation}
\end{itemize}
To proceed, let us put
\begin{equation}
t_2 \,:=\, t(\ta_2)~, \qquad \vfi_2 \,:=\, \vfi(\ta_2)~;
\end{equation}
we claim that we can choose $\af,\at,\rho_0,\pt,\pf$ (and $\HH$) so that
\begin{equation}
t_2 \,<\, 0 ~, \label{ct2}
\end{equation}
\begin{equation}
\vfi_2 \,=\, 0~ (\mbox{mod}\, 2 \pi)~. \label{cfi2}
\end{equation}
Eq.s \eqref{cro0} \eqref{cfi2} indicate that the final space position of the particle,
as measured in the coordinate frame $(t,\vfi,\rho,z)$, coincides with the initial position.
Taking this into account, the inequality \eqref{ct2} means that the event $\cu(\ta_2)$
(the end of the travel) is \textsl{in the past} of the initial event $\cu(0)$ with respect
to the chronological structure of the Minkowskian region outside $\Top$.
Note that $t_2$ is the time indicated at the end of the travel by a clock initially set
to zero and kept at $\vfi = 0$, $\rho = \rho_0$, $z = 0$ during the whole travel of the
freely falling particle. On the other hand, $\ta_2$ is the final time indicated
by a clock initially set to zero, which has travelled with the particle. \parn
All the above claims will be proved by the forthcoming quantitative analysis of the geodesic $\cu$.
We also make a stronger claim: choosing appropriately $\af,\at,\pt,\pf,\HH$ we can make
$|t_2|$ arbitrarily large, i.e., go arbitrarily far in the past keeping $\ta_2$
(the proper duration of the trip) small with respect to $|t_2|$.

\subsection{The claimed time travel: quantitative analysis}
For the moment, we consider any choice of $\af,\at,\rho_0,\pt,\pf$ (and $\HH$) fulfilling (i)(ii).
Let us combine the general rules \eqref{tafi} \eqref{rulet} and \eqref{rulefi} for the
variations of $\ta$, $t$ and $\vfi$ with the qualitative features (a)(b) of the geodesic $\cu$
under analysis; these imply, in particular, that $\si = -1$ on $(0,\ta_1)$ and $\si = 1$ on $(\ta_1,\ta_2)$.
Therefore, via the elementary identity $-\!\int_{\rho_0}^{\rho_1} + \int_{\rho_1}^{\rho_0} = 2 \int_{\rho_1}^{\rho_0}$,
we get
\begin{equation}\begin{array}{c}
\dd{t_2 \,=\, 2 \int_{\rho_1}^{\rho_0}\!\!
{d \rho \; \dot{t}(\rho, \pt, \pf) \over \sqrt{-1  - 2 V_{\pt, \pf}(\rho)}} ~,
\qquad \vfi_2 \,=\, 2 \int_{\rho_1}^{\rho_0}\!\!
{d \rho \;\dot{\vfi}(\rho, \pt, \pf) \over \sqrt{-1  - 2 V_{\pt, \pf}(\rho)}} ~,}
\vspace{0.1cm} \\
\dd{\ta_2 \,=\, 2 \int_{\rho_1}^{\rho_0} \!\!
{d \rho \over \sqrt{-1 - 2V_{\pt,\pf}(\rho)}}~. } \label{integrals}
\end{array}\end{equation}
(To be precise, the symbol ``$\vfi_2$'' in Eq. \eqref{integrals} stands for a determination
of the angle $\vfi_2$\,.)\parn
Next, we write
$$
\int_{\rho_1}^{\rho_0} \,=\, \int_{\rho_1}^{R - \ellp} \,+\, \int_{R - \ellp}^{R - \ellm}
\,+\, \int_{R - \ellm}^{R + \ellm} \,+\, \int_{R + \ellm}^{R + \ellp} \,+\,
\int_{R + \ellp}^{\rho_0}
$$
and use this decomposition for the integrals in Eq. \eqref{integrals}, together
with the following indications. \vspace{-0.15cm}
\begin{enumerate}[-]
\item In the intervals $[\rho_1, R - \ellp]$ and $[R + \ellp, \rho_0]$ we have
for $\dot{t}(\rho, \pt, \pf)$, $\dot{\vfi}(\rho, \pt, \pf)$ and $V_{\pt, \pf}(\rho)$
the simple expressions \eqref{ppWX0} \eqref{VXRed0}, which allow us to calculate
explicitly the corresponding integrals; moreover, we can use for $\rho_1$ the explicit
expression \eqref{ro1}.
\vfill\eject\noindent
$\phantom{a}$
\vskip-1.5cm
\noindent
\item In the interval $[R - \ellm,R + \ellm]$, $\dot{t}(\rho, \pt, \pf)$,
$\dot{\vfi}(\rho, \pt, \pf)$ and $V_{\pt, \pf}(\rho)$ have the constant values
\eqref{ppWX1} \eqref{VXRed1}, so the evaluation of the corresponding integrals is a trivial task.
\vspace{-0.15cm}
\item The integrals $\int_{R - \ellp}^{R - \ellm}$ and $\int_{R + \ellm}^{R + \ellp}$
must be written using for $\dot{t}(\rho, \pt, \pf)$, $\dot{\vfi}(\rho, \pt, \pf)$ and
$V_{\pt,\pf}(\rho)$ the full expressions \eqref{tp} \eqref{fp} \eqref{Vdef}, involving
the function $\Hr(\rho/R) = \HH(|\rho/R-1|)$. It is convenient to re-express these
integrals in terms of the dimensionless variable $\rv := \rho/R$.
\vspace{-0.cm}
\end{enumerate}
In this way, we obtain\parn
\noindent \vbox{
\begin{equation}
{t_2 \over R} \,=\,-\;{4\,\af \over \at^2}\;{\ellm \over R}
\l[\,\pf^2 - \,{\af^2 \over \at^2} \l(1 + {\at^2 \over \pt^2}\r)\r]^{-1/2} +
\label{eqt2} \vspace{-0.1cm}
\end{equation}
$$ + \;{2 \over 1\!-\!1/\pt^2}\!
\l[ \sqrt{\!\l(\!1\!-\!{1 \over \pt^2}\!\r)\!\!\l(\!1\!-\!{\ellp\over R}\r)^{\!\!2}\!-\pf^2}\,
+ \sqrt{\!\l(\!1\!-\!{1 \over \pt^2}\!\r)\!\!\l({\rho_0 \over R}\r)^{\!2}\!-\pf^2}\,
- \sqrt{\!\l(\!1\!-\!{1 \over \pt^2}\!\r)\!\!\l(\!1\!+\!{\ellp \over R}\r)^{\!\!2}\!-\pf^2}\; \r]\! +
\vspace{-0.2cm} $$
$$ + \;2 \l(\int_{1- \ellp/R}^{1- \ellm/R} + \int_{1+\ellm/R}^{1+\ellp/R} \r) d\rv\;
{\big[\,\rv^2 \big(1\!-\!\Hr(\rv)\big)^{2}\! - \af^2\Hr(\rv)^2\,\big]
+ (\af\!+\!\at\,\rv)\,\Hr(\rv)\,\big(1\!-\!\Hr(\rv)\big)\, \pf \over
\rv\,\big(1\!-\!\Hr(\rv)\big)^2 + \af\,\at\,\Hr(\rv)^2} \; \times
\vspace{-0.15cm} $$
$$ \!\! \times\! \l[\big[\,\rv \big(1\!-\!\Hr(\rv)\big)\!+\!\at\,\Hr(\rv)\pf\,\big]^2
\!-\! \big[\,\af\,\Hr(\rv)\!-\!\big(1\!-\!\Hr(\rv)\big)\pf\,\big]^2
\!- {1 \over \pt^2}\,\big[\,\rv\big(1\!-\!\Hr(\rv)\big)^2\!+\!\af\,\at\,\Hr(\rv)^2\,\big]^2\r]^{\!-1/2}\!;
\vspace{0.25cm} $$}
\parn \vbox{
\begin{equation}
\vfi_2 \,=\, {4\,\pf \over \af}\;{\ellm \over R}
\l[\,\pf^2 - \,{\af^2 \over \at^2} \l(1 + {\at^2 \over \pt^2}\r)\r]^{-1/2} +
\vspace{-0.1cm} \label{eqfi2}
\end{equation}
$$ \begin{array}{c}
\hspace{-7.5cm}\dd{-\;2\l[\atan\!\l({1 \over \pf}\,
\sqrt{\!\l(\!1\!-\!{1 \over \pt^2}\!\r)\!\l(\!1\!-\!{\ellp \over R}\r)^{\!\!2}\!-\pf^2}\,\r) + \r.}\vspace{0.08cm} \\
\hspace{0.8cm} \dd{\l. +\,\atan\!\l({1 \over \pf}\,
\sqrt{\!\l(\!1\!-\!{1 \over \pt^2}\!\r)\!\l({\rho_0 \over R}\r)^{\!2}\!-\pf^2}\,\r)
- \,\atan\!\l({1 \over \pf}\,
\sqrt{\!\l(\!1\!-\!{1 \over \pt^2}\!\r)\!\l(\!1\!+\!{\ellp \over R}\r)^{\!\!2}\!-\pf^2}\,\r)\r] +}
\vspace{-0.15cm}
\end{array} $$
$$ +\,2 \l(\int_{1- \ellp/R}^{1- \ellm/R} + \int_{1+\ellm/R}^{1+\ellp/R} \r) d\rv\;
{(\af\!+\!\at\, \rv)\,\Hr(\rv)\,\big(1\!-\!\Hr(\rv)\big)
- \big[\,\big(1\!-\!\Hr(\rv)\big)^2 - \at^2\,\Hr(\rv)^2\,\big]\,\pf \over
\rv\,\big(1\!-\!\Hr(\rv)\big)^2\! + \af\,\at\,\Hr(\rv)^2}\; \times
\vspace{-0.15cm} $$
$$ \!\! \times\! \l[\big[\,\rv \big(1\!-\!\Hr(\rv)\big)\!+\!\at\,\Hr(\rv)\pf\,\big]^2
\!-\! \big[\,\af\,\Hr(\rv)\!-\!\big(1\!-\!\Hr(\rv)\big)\pf\,\big]^2
\!- {1 \over \pt^2}\,\big[\,\rv\big(1\!-\!\Hr(\rv)\big)^2\!+\!\af\,\at\,\Hr(\rv)^2\,\big]^2\r]^{\!-1/2}\!;
\vspace{0.25cm}$$}
\parn\vbox{
\begin{equation}
{\ta_2 \over R} \,=\, {4\,\af \over \pt}\;{\ellm \over R}
\l[\,\pf^2 - \,{\af^2 \over \at^2} \l(1 + {\at^2 \over \pt^2}\r) \r]^{-1/2} +
\label{eqta2} \vspace{-0.1cm}
\end{equation}
$$ + \;{2/\pt \over 1\!-\!1/\pt^2}\!
\l[\!\sqrt{\!\l(\!1\!-\!{1 \over \pt^2}\!\r)\!\l(\!1-{\ellp \over R}\r)^{\!\!2}\!\!-\pf^2}
+ \!\sqrt{\!\l(\!1\!-\!{1 \over \pt^2}\!\r)\!\l({\rho_0 \over R}\r)^{\!2}\!\!-\pf^2}
- \!\sqrt{\!\l(\!1\!-\!{1 \over \pt^2}\!\r)\!\l(\!1+{\ellp \over R}\r)^{\!\!2}\!\!-\pf^2}\, \r]\! +
\vspace{-0.2cm} $$
$$ + \;{2 \over \pt} \l(\int_{1- \ellp/R}^{1- \ellm/R} + \int_{1+\ellm/R}^{1+\ellp/R} \r) d\rv\;
\Big[\rv\,\big(1\!-\!\Hr(\rv)\big)^2\! + \af\,\at\,\Hr(\rv)^2 \Big] \; \times
\vspace{-0.15cm} $$
$$ \!\! \times\! \l[\big[\,\rv \big(1\!-\!\Hr(\rv)\big)\!+\!\at\,\Hr(\rv)\pf\,\big]^2
\!-\! \big[\,\af\,\Hr(\rv)\!-\!\big(1\!-\!\Hr(\rv)\big)\pf\,\big]^2
\!- {1 \over \pt^2}\,\big[\,\rv \big(1\!-\!\Hr(\rv)\big)^2\!+\!\af\,\at\,\Hr(\rv)^2\,\big]^2\r]^{\!-1/2}\!.$$}
In each one of Eq.s \eqref{eqt2}-\eqref{eqta2}, the terms in
the right-hand sides have the following meaning: \vspace{-0.15cm}
\begin{enumerate}[-]
\item the terms in the first line of each equation are the contributions to $t_2/R$,
$\vfi_2$ or $\ta_2/R$ from the geodesic motion in the spacetime region inside the smaller
torus $\Tom$, where $R - \ellm \leqs \rho \leqs R + \ellm$; \vspace{-0.2cm}
\item the terms in the second lines of Eq.s \eqref{eqt2} \eqref{eqta2} and in the second
and third lines of Eq. \eqref{eqfi2} are the contributions from the motion in the
Minkowskian region outside the larger torus $\Top$, where $\rho \leqs R - \ellp$ or $\rho \geqs R + \ellp$;\vspace{-0.2cm}
\vfill\eject\noindent
$\phantom{a}$
\vskip-1.5cm
\noindent
\item the integrals occupying the third and fourth lines of Eq.s \eqref{eqt2} \eqref{eqta2}
and the fourth and fifth lines of Eq. \eqref{eqfi2} are the contribution from the transition
region where $R - \ellp < \rho < R - \ellm$ or $R + \ellm < \rho < R + \ellp$.
Let us notice that all the contributions on the right-hand side of
Eq. \eqref{eqta2} are strictly positive. This indicates that the arrival proper time $\ta_2$
of the test particle is always positive, which correctly corresponds to the fact that
we are using a future-oriented parametrization of the geodesic.
\end{enumerate}

\subsection{How to fulfil the previous claims about the time travel}\label{subProof}
Let us focus on the expression \eqref{eqt2} for $t_2$.
The term in the first line of the cited equation (the contribution from the
region inside $\Tom$) is certainly negative, in agreement with the remarks made
after Eq. \eqref{ppWX1}; on the contrary, the term in the second line is positive,
while the sign of the integral in the third and fourth lines is not evident a priori.
The hope is to make the negative term very large and dominant on the others,
by an appropriate choice of the parameters; this choice should induce the claimed condition
$t_2 < 0$ of \eqref{ct2}, corresponding to a time travel into the past. This goal can
be attained choosing the rescaled momentum $\pf$ so as to make very small the expression
within the square brackets in the first line of Eq. \eqref{eqt2}; we will analyse this
strategy in greater detail in the forthcoming paragraph \ref{parLim} and show that
it allows to fulfil all claims of subsection \ref{subqual}.\parn
Let us remark that, besides fulfilling $t_2 < 0$, the parameters should
also be tuned properly so that $\vfi_2$ given by Eq. \eqref{eqfi2} is
an integer multiple of $2 \pi$ (see Eq. \eqref{cfi2}). When this condition is realized,
the test particle travelling along the geodesic returns exactly to the initial spatial position,
from which its journey had started.

\subsubsection{Setting up the previous strategy}\label{parLim}
Following the previous idea, let us fix the attention on the variable
\begin{equation}
\xf \,:=\, \sqrt{\pf^2 - \,{\af^2 \over \at^2} \l(1 + {\at^2 \over \pt^2}\r)}
~~\in ~~ \l(0\,,\,\sqrt{\l(\!1 - {\ellp \over R} \r)^{\!\!2}\! \l(\!1 - {1 \over \pt^2}\!\r)\!
- {\af^2 \over \at^2}\! \l(\!1 + {\at^2 \over \pt^2}\!\r)\!}\;\r) \label{defxf}
\end{equation}
which is (the square root of) the term between square brackets in the first line
of Eq. \eqref{eqt2} for $t_2$. Our strategy is to make $\xf$ small.
\parn
For definiteness, let us assume the shape function $\HH = \HH_{(k)}$ to have the form
\eqref{HkExp} \eqref{Fkdef} given in Appendix \ref{App1} for some finite integer $ k \geqs 2$.\parn
In Appendix \ref{App2} we illustrate a method for high precision calculation
of $t_2/R$, $\vfi_2$ and $\ta_2/R$ when $\xf$ is small. This method uses directly the
definitions \eqref{eqt2}-\eqref{eqta2} of these quantities; in particular, the integrals
appearing therein are re-expressed in a way which is more convenient for their numerical
evaluation.\parn
Section \ref{subapest} of the above mentioned Appendix \ref{App2} also considers the limit
\begin{equation}
\xf \,\to\, 0^+ \label{xfto0}
\end{equation}
and derives the expansions
\begin{equation}
{t_2 \over R} \,=\,- \l({4\,\af \over \at^2}\;{\ellm \over R} \r){1 \over \xf}\;
\Big(1\, +\,O\big(\xf^{2 \over k+1}\big)\Big) ~; \label{eqt2lim}
\end{equation}
\begin{equation}
\vfi_2 \,=\, \l({4 \over \at}\,{\ellm \over R}\,\sqrt{1 + {\at^2 \over \pt^2}}\;\r) {1 \over \xf}\;
\Big(1\, +\,O\big(\xf^{2 \over k+1}\big)\Big) ~; \label{eqfi2lim}
\end{equation}
\begin{equation}
{\ta_2 \over R} \,=\, \l({4\,\af \over \pt}\;{\ellm \over R}\r) {1 \over \xf}\;
\Big(1\, +\,O\big(\xf^{2 \over k+1}\big)\Big) ~. \label{eqta2lim} \vspace{0.2cm}
\end{equation}
\vfill\eject\noindent
Let us briefly comment the above results. \vspace{-0.2cm}
\begin{enumerate}[-]
\item The asymptotic expansion \eqref{eqt2lim} shows that $t_2$ can be made negative,
with $|t_2|$ arbitrarily large (one simply has to choose $\xf$ small enough). \vspace{-0.2cm}
\item  On the other hand, Eq. \eqref{eqfi2lim} shows that $\vfi_2$ varies rapidly when $\xf$
is small, meaning that little changes of $\xf$ correspond to
non-negligible deviations of $\vfi_2$. This indicates that
fulfilling the condition $\vfi_2 \,=\, 0~ (\mbox{mod}\, 2\pi)$ (see Eq. \eqref{cfi2})
is always possible in principle, but requires a fine tuning of the parameter $\xf$;
the asymptotic expression \eqref{eqfi2lim} for $\vfi_2$ is not sufficient
to determine this fine-tuned value of $\xf$ and it is necessary to use directly the
exact expression \eqref{eqfi2}.\vspace{-0.2cm}
\item Finally, let us remark that the leading order in the asymptotic expansion \eqref{eqta2lim}
for $\ta_2$ is inversely proportional to the Lorentz factor $\pt$; in particular,
by comparison with the expansion \eqref{eqt2lim} we see that $\ta_2/|t_2| = (\at^2/\pt)\,\big(1+O(\xf^{2 \over k+1})\big)$.
So, $\ta_2$ can be made small with respect to $|t_2|$ choosing a large $\pt$, exactly
as in special relativity.
\end{enumerate}
Summing up: with appropriate, small values of $\xf$ and sufficiently large $\pt$
we can fulfil all claims of subsection \ref{subqual}. In the next section we describe
this situation via fully quantitative examples.

\subsection{Some numerical examples}\label{antic}
In this subsection we fix as follows the parameters of the problem and the shape function:
\begin{equation}\begin{array}{c}
\dd{\ellm \,=\, {3 \over 5}\,R~, \qquad \ellp \,=\, {4 \over 5}\,R~, \qquad
\af \,=\, {9 \over 100}~, \qquad \at \,=\,10 ~;} \vspace{0.25cm}\\
\dd{\mbox{$\HH = \HH_{(k)}$ as in Eq.s \eqref{HkExp} \eqref{Fkdef} of Appendix \ref{App1}}~, \qquad
k \,=\, 3 ~.} \label{cho1}
\end{array}\end{equation}
These choices determine the time machine up to the scale factor $R$, for which we will
subsequently consider different choices. \parn
Concerning the parameters of the geodesic motion, we set
\begin{equation}
\rho_0 \,=\, (1 + 10^{-3})\, (R + \ellp) \label{cho2}
\end{equation}
(meaning that, at $\ta = 0$, the particle is outside but very close to the external torus $\Top$).
The other parameters describing the particle motion are $\pt$ and $\xf$, defined by
Eq. \eqref{defxf}; they are free for the moment, but $\xf$ is assumed to be small.\parn
Due to Eq.s \eqref{defxf} \eqref{cho1} and \eqref{cho2}, in the expressions
\eqref{eqt2}-\eqref{eqta2} for $t_2, \vfi_2, \ta_2$ (as well as in their small-$\xf$
asymptotic versions \eqref{eqt2lim}-\eqref{eqta2lim}) everything depends only
on the scale parameter $R$ of the time machine and on the kinematic parameters $\pt, \xf$.
In particular, the $\xf \to 0^+$ asymptotic expressions \eqref{eqt2lim} and \eqref{eqta2lim} become
\begin{equation}
{t_2 \over R} \,=\,-\,{27 \over 12500\,\xf}\;\Big(1 + O\big(\sqrt{\xf}\big)\Big) ~,
\label{eqt2Num}
\end{equation}
\begin{equation}
{\ta_2 \over R} \,=\, {27 \over 125\,\pt\,\xf}\;\Big(1 + O\big(\sqrt{\xf}\big)\Big) ~.
\label{eqta2Num} \vspace{0.2cm}
\end{equation}
We have checked that, ignoring the remainder terms $O(\sqrt{\xf})$, the above asymptotic
expressions agree up to 4 significant digits with the numerical values of the
exact expressions \eqref{eqt2}-\eqref{eqta2} for $\xf \simeq 10^{-5}$; the agreement
is even more accurate for smaller values of $\xf$. \vspace{0.1cm}\parn
As an example, let us choose $R = 100\,m$ ($\simeq 100/(2.99792458 \cdot 10^{8})\,s$
in our units with $c=1$); we want to determine the remaining parameters $\pt,\xf$ so that
$t_2 \simeq - 1\,y$ and $\ta_2 \simeq 1\,d$ ($y$ and $d$ stand, respectively, for ``year'' and ``day'').
To this purpose, we first use the asymptotic expressions \eqref{eqt2Num} \eqref{eqta2Num}
for a preliminary, rough estimate. From Eq. \eqref{eqt2Num} we infer
$t_2 \simeq - 1\,y$ if $\xf \simeq 2.28 \cdot 10^{-17}$; on the other hand, keeping this
choice of $\xf$, Eq. \eqref{eqta2Num} gives $\ta_2 \simeq 1\,d$ for $\pt \simeq 10^4$.
Then, we fix $\pt = 10^4$ and consider small variations of $\xf$ about the value $2.28 \cdot 10^{-17}$
to get $\vfi_2 \simeq 0\,(\mbox{mod}\, 2 \pi)$; using Eq. \eqref{eqfi2}, we find that
$|\vfi_2| < 10^{-6} \,(\mbox{mod}\, 2 \pi)$ if we use the fine-tuned value
$\xf = 2.280021827804094 \cdot 10^{-17}$.
From the previous values of $R,\pt,\xf$ and from the exact expressions \eqref{eqt2} \eqref{eqta2}
we get $t_2 = -1.002...\,y$ and $\ta_2 = 3.657...\,d$
(indeed, for these calculations Eq.s \eqref{eqt2}-\eqref{eqta2} are used in the reformulations
described in Appendix \ref{App2}, more suitable for precise numerical evaluation of the
integrals therein).
\parn
In the first five columns of Tables 1-3 we summarize the above results and many others,
obtained along the same lines using (Eq.s \eqref{eqt2Num} \eqref{eqta2Num} and) Eq.s \eqref{eqt2}-\eqref{eqta2};
the parts of the tables containing the symbols $\aa$, $\EEf$ and $\EEg$ refer to
the tidal accelerations and energy densities already mentioned in the introduction,
and will be explained in the forthcoming Sections \ref{secacc} and \ref{secen}.
Concerning the values of $R$ chosen in the said tables, we note that $10^{11} m$ is the
order of magnitude of the Earth-Sun distance,
while $10^{18} m \simeq 100$ light years: this is the choice yielding the smallest values
for the tidal accelerations and the energy densities.
\begin{table}[t!]\centering
\begin{tabular}{|c|c|c|c|c|c|c|}\hline
\multicolumn{7}{|c|}{\textbf{TABLE 1\,:} \qquad $\bm{R = 10^2\,m}$ \qquad\qquad $\bm{\min\EEf} = -1.2809... \cdot 10^{23}\,\gr/cm^3$} \\ \hline
$\bm{\pt}$  &   $\bm{\xf}$  &   $\bm{|\vfi_2|}$\,{\small$(\mbox{mod}\,2\pi)$}\! &   $\bm{t_2}$  &   $\bm{\ta_2}$    &   $\!\bm{\max\aa}$\,{\small$(\gt/m)$}\!\! &   $\!\bm{\min\EEg}$\,{\small $(\gr/cm^3)$}\!  \\ \hline
$1.1$           &   $7.2001167584668 \!\cdot\! 10^{-10}$        &   $2 \cdot 10^{-5}$   &   $-1\,s$         &   $\!90.97\,s\!\!$    &   $\!6.679 \cdot 10^{16}$ &   $-1.347 \cdot 10^{23}$      \\ \hline
$10^2$          &   $7.2001526829246 \!\cdot\! 10^{-10}$        &   $6 \cdot 10^{-7}$   &   $-1\,s$         &   $1\, s$                     &   $\!4.144 \cdot 10^{17}$ &   $-6.779 \cdot 10^{25}$      \\ \hline
$10^2$          &   $2.280028356416717 \!\cdot\! 10^{-17}$      &   $10^{-6}$           &   $-1\,y$         &   $1\, y$                     &   $\!4.144 \cdot 10^{17}$ &   $-6.779 \cdot 10^{25}$      \\ \hline
$10^4$          &   $2.280021827804094 \!\cdot\! 10^{-17}$      &   $10^{-6}$           &   $-1\,y$         &   $\!3.66\,d\!$       &   $\!4.143 \cdot 10^{21}$ &   $-6.768 \cdot 10^{29}$      \\ \hline
$10^5$          &   $2.280020685289079 \!\cdot\! 10^{-20}$      &   $3 \cdot 10^{-6}$   &   $\!-10^3\,y\!$  &   $1\, y$                     &   $\!4.143 \cdot 10^{23}$ &   $-6.768 \cdot 10^{31}$      \\ \hline
$10^7$          &   $2.280020673890116 \!\cdot\! 10^{-20}$      &   $5 \cdot 10^{-6}$   &   $\!-10^3\,y\!$  &   $\!3.66\,d\!$       &   $\!4.143 \cdot 10^{27}$ &   $-6.768 \cdot 10^{35}$      \\ \hline
$10^8$          &   $\!2.2800206734492706 \!\cdot\! 10^{-23}\!$ &   $10^{-6}$           &   $\!-10^6\,y\!$  &   $1\, y$                     &   $\!4.143 \cdot 10^{29}$ &   $-6.768 \cdot 10^{37}$      \\ \hline
$\!10^{10}\!\!$ &   $\!2.2800206734492592 \!\cdot\! 10^{-23}\!$ &   $10^{-6}$           &   $\!-10^6\,y\!$  &   $\!3.66\,d\!$       &   $\!4.143 \cdot 10^{33}$ &   $-6.768 \cdot 10^{41}$      \\ \hline
\end{tabular}
$\phantom{a}$\vspace{0.2cm}\\
\begin{tabular}{|c|c|c|c|c|c|c|}\hline
\multicolumn{7}{|c|}{\textbf{TABLE 2\,:}\qquad $\bm{R = 10^{11}\,m}$\qquad\qquad $\bm{\min\EEf} = -1.2809... \cdot 10^{5}\,\gr/cm^3$} \\ \hline
$\bm{\pt}$  &   $\bm{\xf}$  &   $\bm{|\vfi_2|}$\,{\small$(\mbox{mod}\,2\pi)$}\! &   $\bm{t_2}$  &   $\bm{\ta_2}$    &   $\!\bm{\max\aa}$\,{\small$(\gt/m)$}\!\! &   $\!\bm{\min\EEg}$\,{\small $(\gr/cm^3)$}\!  \\ \hline
$10^2$          &   $2.28000847482\!\cdot\!10^{-8}$                 &   $4 \cdot 10^{-6}$   &   $-1\,y$         &   $1\, y$         &   $0.4144$                &   $-6.779 \cdot 10^{7}$       \\ \hline
$10^4$          &   $2.280027538409\!\cdot\!10^{-8}$                &   $4 \cdot 10^{-7}$   &   $-1\,y$         &   $\!3.66\,d\!$  &    $4.143 \cdot 10^3$      &   $-6.768 \cdot 10^{11}$      \\ \hline
$10^5$          &   $2.2800350451561739\!\cdot\!10^{-11}$           &   $3 \cdot 10^{-6}$   &   $\!-10^3\,y\!$  &   $1\, y$         &   $4.143 \cdot 10^5$      &   $-6.768 \cdot 10^{13}$      \\ \hline
$10^7$          &   $2.2800078145725598\!\cdot\!10^{-11}$           &   $4 \cdot 10^{-6}$   &   $\!-10^3\,y\!$  &   $\!3.66\,d\!$   &   $4.143 \cdot 10^9$      &   $-6.768 \cdot 10^{17}$      \\ \hline
$10^8$          &   $2.280021127512103772\!\cdot\!10^{-14}$         &   $7 \cdot 10^{-6}$   &   $\!-10^6\,y\!$  &   $1\, y$         &   $4.143 \cdot 10^{11}$   &   $-6.768 \cdot 10^{19}$      \\ \hline
$\!10^{10}\!\!$ &   $\!2.2800211275120923732\!\cdot\!10^{-23}\!\!$  &   $8 \cdot 10^{-6}$   &   $\!-10^6\,y\!$  &   $\!3.66\,d\!$   &   $4.143 \cdot 10^{15}$   &   $-6.768 \cdot 10^{23}$      \\ \hline
\end{tabular}
$\phantom{a}$\vspace{0.2cm}\\
\begin{tabular}{|c|c|c|c|c|c|c|}\hline
\multicolumn{7}{|c|}{\textbf{TABLE 3\,:}\qquad $\bm{R = 10^{18}\,m}$\qquad\qquad $\bm{\min\EEf} = -1.2809... \cdot 10^{-9}\,\gr/cm^3$} \\ \hline
$\bm{\pt}$  &   $\bm{\xf}$  &   $\bm{|\vfi_2|}$\,{\small$(\mbox{mod}\,2\pi)$}\! &   $\bm{t_2}$  &   $\bm{\ta_2}$    &   $\!\bm{\max\aa}$\,{\small$(\gt/m)$}\!\! &   $\!\bm{\min\EEg}$\,{\small $(\gr/cm^3)$}\!  \\ \hline
$10^5$          &   $\!2.28157870976775\!\cdot\!10^{-4}\!\!$    &   $2 \cdot 10^{-12}$  &   $\!-925\,y\!$   &   $\!1.02\, y\!$  &   $4.143 \cdot 10^{-9}$                   &   $-0.6778$                   \\ \hline
$10^7$          &   $\!2.28157869832448\!\cdot\!10^{-4}\!\!$    &   $3 \cdot 10^{-12}$  &   $\!-925\,y\!$   &   $\!3.73\, d\!$  &   $4.143 \cdot 10^{-9}$                   &   $-6.778 \cdot 10^{3}$       \\ \hline
$10^8$          &   $\!2.280006782789627\!\cdot\!10^{-7}\!\!$   &   $2 \cdot 10^{-10}$  &   $\!-10^6\,y\!$  &   $\!1\, y\!$     &   $4.143 \cdot 10^{-3}$                   &   $-6.768 \cdot 10^{5}$       \\ \hline
$10^9$          &   $\!2.280006782789616\!\cdot\!10^{-7}\!\!$   &   $2 \cdot 10^{-10}$  &   $\!-10^6\,y\!$  &   $\!36.6\, d\!$  &   $0.4143$                                &   $-6.768 \cdot 10^{7}$       \\ \hline
$\!10^{10}\!\!$ &   $\!2.280006782789615\!\cdot\!10^{-7}\!\!$   &   $4 \cdot 10^{-10}$  &   $\!-10^6\,y\!$  &   $\!3.66\, d\!$  &   $41.43$                                 &   $-6.768 \cdot 10^{9}$       \\ \hline
\end{tabular}
\caption*{Tables 1-3: some numerical examples corresponding to the choices \eqref{cho1}
for $\af,\at,\ellp/R,\ellm/R$ and $\HH$. The values of $|\vfi_2|$, $t_2$ and $\ta_2$
are computed via Eq.s \eqref{eqt2}-\eqref{eqta2} (in the reformulations of Appendix \ref{App2});
the value of $\max\aa$ is obtained using Eq. \eqref{aaNum2}, while those of $\min\EEf$
and $\min\EEg$ descend from Eq.s \eqref{efgetes} and \eqref{espeegmin}.
($m :=$ meter, $cm :=$ centimeter, $\gr :=$ gram, $s:=$ second, $d := $ day, $y := $ year,
$\gt :=$ Earth's gravitational acceleration).}
\label{TABLES}
\end{table}

\section{Tidal accelerations}\label{secacc}
In this section we give a quantitative analysis of the tidal accelerations experienced
by small extended bodies during free fall time travels into the past; we assume that
during such trips, the particles constituting these bodies move along geodesics of
the type analysed in Section \ref{freefallSec}. \parn
To introduce the subject, it is convenient to start with some general facts.

\subsection{Basics on tidal effects}\label{basics}
Let us consider an arbitrary spacetime $\MM$ with metric $g$, and a timelike geodesic with
its proper time parametrization $\cu: I \subset \R \to \MM$, $\ta \mapsto \cu(\ta)$.
For each $\ta \in I$, we introduce the vector space
\begin{equation}
\SS_{\ta} \,:=\, \big\{ X \in T_{\cu(\ta)} \MM ~\big|~
g\big(X,\dcu(\ta)\big) = 0 \,\big\} \label{defss}
\end{equation}
and the linear operator
\begin{equation}
\AA_{\ta} \,:\, \SS_{\ta} \to \SS_{\ta} ~, \qquad
X \mapsto \AA_{\ta} X \,:=\,-\,\Riem\big(X, \dcu(\ta)\big)~\dcu(\ta) \label{defra}
\end{equation}
where $\Riem$ denotes the Riemann curvature tensor.
It can be easily checked that $\SS_{\ta}$ is a 3-dimensional, spacelike linear
subspace of the tangent space $T_{\cu(\ta)}\MM$; with the restriction
of $g \equiv g_{\cu(\ta)}$ as an inner product, $\SS_{\ta}$ is in fact a Euclidean space.
$\AA_{\ta}$ is a self-adjoint linear operator in this Euclidean space
(and thus it is diagonalizable, with real eigenvalues). \parn
We refer to $\AA_{\ta}$ as the \textsl{tidal operator} for the geodesic
$\cu$ at $\ta$. This name is due to the following fact: if we consider another
timelike geodesic ``infinitesimally close to $\cu$'' and $\decu(\ta) \in \SS_{\ta}$
is its infinitesimal separation vector from $\cu(\ta)$, the \textsl{tidal acceleration}
$(\nabla^2 \decu/ d \ta^2)(\ta)$ equals $\AA_{\ta}\, \decu(\ta)$. Of course, $\cu$
and $\cu + \decu$ could be the worldlines of two particles in a freely falling
extended body. \parn
For a justification of all the previous statements about $\SS_{\ta}$ and $\AA_{\ta}$,
we refer to Appendix \ref{App3}. In the sequel we consider the scalar quantity
\begin{equation}
\aa(\ta) \,:=\, \sup_{X \in \SS_{\ta} \setminus \{0\}}
{\sqrt{g(\AA_\ta X, \AA_{\ta} X)} \over \sqrt{g(X,X)}} ~, \label{defaa}
\end{equation}
which is just the operator norm of $\AA_{\ta}$ corresponding to the Euclidean norm
$\sqrt{g(\,\cdot\,,\,\cdot\,)} \equiv \sqrt{g_{\cu(\ta)}(\,\cdot\,,\,\cdot\,)}$ on $\SS_{\ta}$.
By the spectral theorem for self-adjoint operators, the above $\sup$ equals the maximum
of the absolute values of the eigenvalues of $\AA_{\ta}$, and it is attained when $X$
is an associated eigenvector. \parn
For obvious reasons, we shall call $\aa(\ta)$ the \textsl{maximal tidal acceleration
per unit length}. Let us remark that, if a timelike geodesic has a separation $\decu$
from $\cu$, the tidal acceleration at $\ta$ associated to it has norm
\begin{equation}
\sqrt{g\big(\AA_\ta \decu(\ta), \AA_{\ta} \decu(\ta)\big)}
\,\leqs\, \aa(\ta)\, \sqrt{g\big(\decu(\ta),\decu(\ta)\big)} ~;
\end{equation}
the above relation holds as an equality if $\decu(\ta)$ is an eigenvector associated
to an eigenvalue of $\AA_\ta$ with maximum absolute value.
\vspace{-0.1cm}

\subsection{Tidal effects during time travel}\label{tidaltime}
Let us return to the spacetime $\Tar$ and choose for $\cu$
a timelike geodesic of the type considered in Section \ref{freefallSec}, describing
a time travel by free fall. A sketch of the computation of $\AA_{\ta}$ and $\aa(\ta)$
for this case is given in Section \ref{secC2} of Appendix \ref{App3}; therein we write
\begin{equation}
\AA_{\ta} \,=\, {\pt^2 \over R^2}\; \AF_{\ta} \label{afrak}
\end{equation}
with $\AF_{\ta} : \SS_{\ta} \to \SS_{\ta}$ self-adjoint, and show that
\begin{equation}
\aa(\ta) \,=\, {\pt^2 \over R^2}\;\ac \big(\rho(\ta)/R\big) ~, \label{atau}
\end{equation}
where $\ac(r)$ is a dimensionless function of a variable $\rv \in (0,+\infty)$,
here set equal to $\rho(\ta)/R$; this function also depends, parametrically, on the
quantities $\ellm/R,\ellp/R,\af,\at$ and $\pt,\xf$ (related, respectively, to the metric
$g$ and to the motion $\cu$).
The function $\ac(\rv)$ can be computed explicitly and vanishes identically for $\rv$
outside the region $(1-\ellp/R,1-\ellm/R) \cup (1+\ellm/R,1+\ellp/R)$ (because the Riemannian
curvature is zero for $\rho/R$ outside this region). \parn
In the simultaneous limits $\pt \to +\infty$ and $\xf \to 0^+$, $\AF_{\ta}$
has a zero eigenvalue of multiplicity $2$ and a simple, non-zero eigenvalue depending only
on $\rv = \rho(\ta)/R$; the explicit expression of this eigenvalue is
reported in Appendix \ref{App3} (see Eq. \eqref{ar} therein) and the corresponding
eigenvector, giving the direction of the tidal acceleration, is $\de_z \big|_{\cu(\ta)}$\,.
Of course the function $\rv \mapsto \ac(\rv)$ has a limit for $\pt \to +\infty$
and $\xf \to 0^+$, coinciding with the absolute value of this eigenvalue. \parn
For a better quantitative appreciation, it is convenient to express $\aa(\ta)$ in terms
of the ratio $\gt/m$ where ($m:=$ meter and) we are considering the nominal Earth's
gravitational acceleration
\begin{equation}
\gt \,:=\, 9.8\;m/s^2 \,=\, 1.090397... \cdot 10^{-16}\, m^{-1} \label{defgt}
\end{equation}
(the last equality follows from our convention $c = 1$). \parn
Eq.s \eqref{atau} and \eqref{defgt} imply
\begin{equation}
\aa(\ta) \,=\, \big(9.170971... \cdot 10^{15}\big)\; {\pt^2 \over (R/m)^2}\;
\ac\big(\rho(\ta)/R\big)~{\gt \over m} ~.
\end{equation}
\begin{figure}[]
    \centering
        \begin{subfigure}[b]{0.47\textwidth}
                \includegraphics[width=\textwidth]{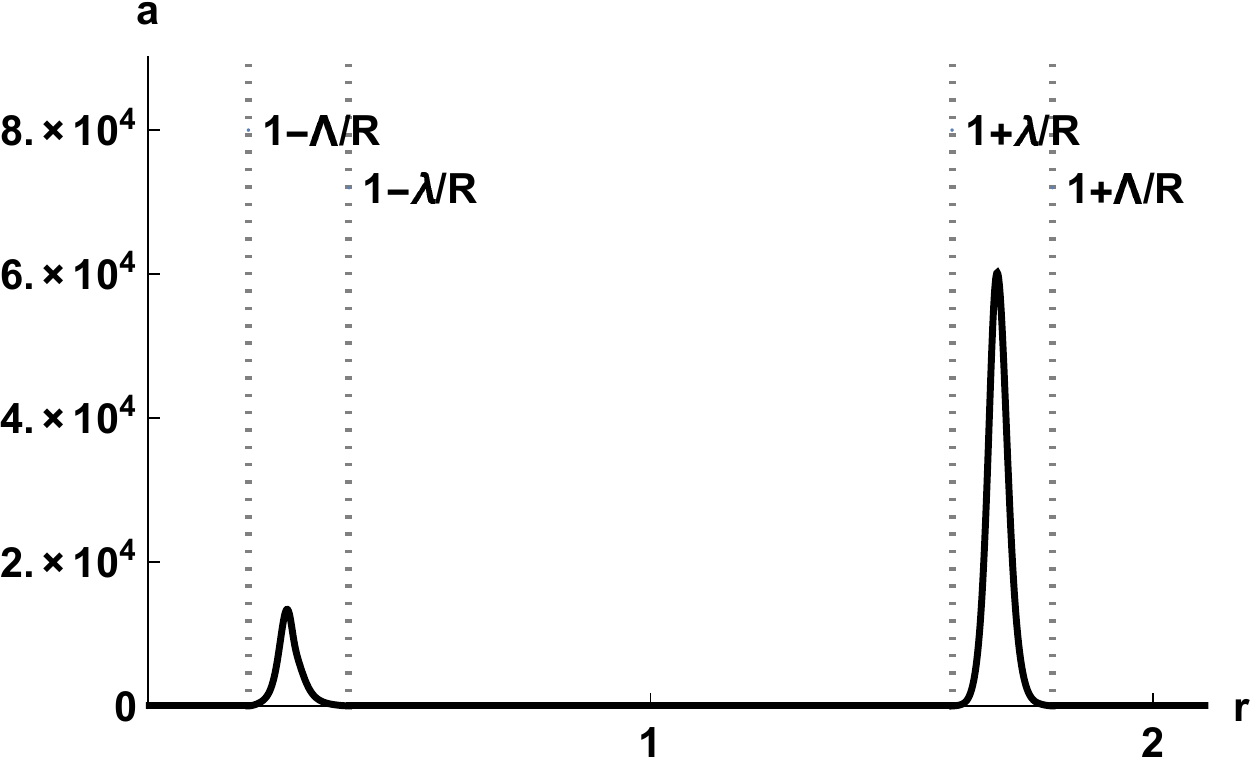}
                \caption*{Figure 4a} \label{A1}
                \vspace{0.cm}
        \end{subfigure}
        \hspace{0.3cm}
        \begin{subfigure}[b]{0.47\textwidth}
                \includegraphics[width=\textwidth]{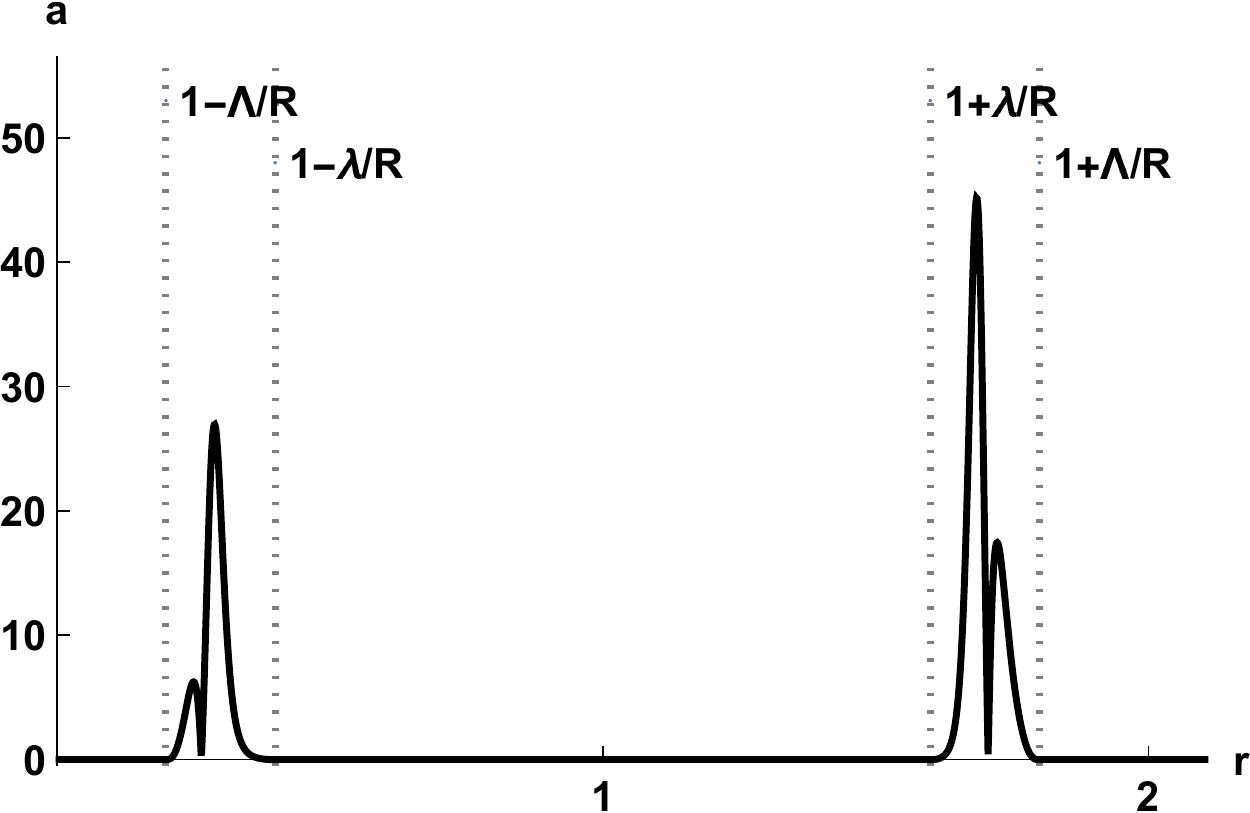}
                \caption*{Figure 4b} \label{A2}
                \vspace{0.cm}
        \end{subfigure}
        \caption*{{\small Figures 4a-4b: graphs of the function
        $\rv \mapsto \ac(\rv)$
        for $\ellm/R = 3/5$, $\ellp/R = 4/5$, $\af = 9/100$, $\at = 10$
        and for $\pt = 1.1$, $\xf = 7.2... \cdot 10^{-10}$ (case (a)) or
        in the limit $\pt \to +\infty$, $\xf \to 0^{+}$ (case (b)).
        Again, the shape function $\HH = \HH_{(k)}$ is the one given
        in Eq.s \eqref{HkExp} \eqref{Fkdef} of Appendix \ref{App1}, with $k = 3$.}}
        \label{fig:tidalacc}
\end{figure}
\parn
From here to the end of this subsection we fix $\ellm/R,\ellp/R,\af,\at$ and
$\HH = \HH_{(k)}$ as in Eq. \eqref{cho1}.
Fig.s 4a-4b show the graphs of the function $\rv \mapsto \ac(\rv)$ for a specific
choice of $(\pt,\xf)$ or for $\pt \to + \infty$, $\xf \to 0^{+}$.
The sixth columns of Tables 1-3 give for some choices of $R,\pt,\xf$ the maxima
of $\aa$ during the time travel, i.e.,
\begin{equation}
\max \aa \,:=\, \max_{\ta \in [0,\ta_2]}\,\aa(\ta)
\,=\, \big(9.170971... \cdot 10^{15}\big)\; {\pt^2 \over (R/m)^2}
\l(\max_{\rv \in [\rho_0/R,\rho_1/R]} \ac(\rv)\r) {\gt \over m} ~. \label{aaNum2}
\end{equation}
Finally, let us remark that Table 3 indicates
a fact already anticipated in paragraph \ref{antic}:
for $R = 10^{18} m \simeq 100$ light years, the tidal accelerations per unit length
are gentle (on a human scale) up to very large values of $\pt$ (say, up to $\pt = 10^{9}$).

\section{Energy density. Violation of the classical energy conditions}\label{secen}
A common drawback of spacetimes possessing CTCs is the violation
of the energy positivity conditions, which are fulfilled by the stress-energy
tensor of ordinary matter
({{\footnote{However, let us mention that the energy conditions are sometimes violated
by the expectation value of the stress-energy tensor of quantum systems; for example
this happens in Casimir configurations \cite{FPbook}, involving the vacuum states
of quantum fields in domains with boundaries. On the other hand, the appearance of
negative energy densities is typically constrained by some sort of averaged versions
of the energy inequalities \cite{Few,FoRo}; we will not consider such variations
of the energy conditions in the present work.}});
the same violations are known to occur in spacetimes describing wormholes \cite{MT1,MT2}
or warp drives for superluminal motions \cite{Alcu,Kras}. Another problematic
feature of time machines, wormholes and warp drives is that the (negative) energy densities
involved are enormous, unless the length scale of variations of the metric is gigantic.
The spacetime $\Tar$ that we are considering in this paper is no exception to the above trends.\parn
In the following, after a few preliminary considerations regarding the stress-energy tensor
(see subsection \ref{subsecTmn}), we consider two different classes of observers
and determine the energy densities which they measure.
More precisely: in subsection \ref{subsecET} we deal with the fundamental observers
introduced in Section \ref{SecOr}; in subsection \ref{subsecEG} we consider
the freely falling observers which perform a time travel into the past following
a geodesic of the type described in Section \ref{freefallSec}.
Our results show that, for both classes of observers, there are regions where the
measured energy densities becomes negative, thus violating the weak energy condition
(see \cite{Hawk}, page 89); this suffices to infer that the dominant
energy condition (see \cite{Hawk}, page 91) fails as well.
Similar arguments allow us to infer the violation of the strong energy condition
(see \cite{Hawk}, page 95). Concerning the size of the observed, negative
energy densities we refer to Tables 1-3 of page \pageref{TABLES}.

\subsection{Basics on the stress-energy tensor and the energy density}\label{subsecTmn}
Given any spacetime $\MM$ with metric $g$, we can define the associated stress-energy tensor
to be the symmetric bilinear form
\begin{equation}
\Te \,:=\, {1 \over 8 \pi G} \l(\Ric\,-\,{1 \over 2}\,\Rs\,g\r) \label{defT}
\end{equation}
where $G$, $\Ric$, $\Rs$ are the gravitational constant, the Ricci tensor and the
scalar curvature of $g$; this position automatically ensures that Einstein's
equations are fulfilled.
This approach somehow reverses the traditional viewpoint, according to which:
(i) the form of $\Te$ is prescribed on the grounds of a model for the matter content
of the system under analysis; (ii) Einstein's equations are solved to find $g$ and, possibly,
the few unknown functions appearing in $\Te$. The reversed viewpoint, in which
\eqref{defT} is a definition, is used when the metric has been constructed
\textsl{ad hoc} so as to exhibit some desired exotic features
(such as the features required by time machines, wormholes, warp drives and so on). \\
In the sequel we use the position \eqref{defT} with the following value for the
universal gravitational constant:
\begin{equation}
G \,=\, 6.67 \cdot 10^{-14} {m^3 \over gr \;s^2} \,=\,
7.421375... \cdot 10^{-31}\, {m/\gr} \label{defG}
\end{equation}
($\gr$ is the gram; the last equality follows from our convention $c = 1$). \parn
Let us consider any spacetime point $p \in \MM$ and a timelike vector $X \in T_p\MM$,
normalized so that
\begin{equation}
g(X,X) \,=\, -1~;
\end{equation}
then, the \textsl{energy density} measured at $p$ by an observer with instantaneous
$4$-velocity $X$ is
\begin{equation}
\EE(X) \,:=\, \Te(X,X) ~. \label{edens}
\end{equation}
Let us mention that, with the present notations, the weak energy condition
reads $\EE(X) \geqs 0$ for each $X$ as above. \parn
From now on, $\MM$ is the spacetime $\Tar$ of the present paper; Appendix \ref{App4}
gives some information on the calculation of $\Te$ in this case. We will consider
two choices for the vector $X$, corresponding to the observers already mentioned
at the beginning of the present section.
Using the fact that energies and masses are dimensionally equivalent in our setting
with $c=1$, we will measure energy densities in units of $\gr/cm^3$.

\subsection{The energy density measured by fundamental observers}\label{subsecET}
Let us consider the fundamental observer passing through a spacetime point $p \in \Tar$,
and remember that its four-velocity coincides with $E_{(0)}(p)$ (here and in the following
$E_{(0)}$ is the timelike vector field in the tetrad of Section \ref{SecOr}).
Consequently, the energy density measured by this observer is
\begin{equation}
\EEf(p) \,:=\, \EE\big(E_{(0)}(p)\big) ~. \label{defeef}
\end{equation}
The above function of $p$ depends on its coordinates $\rho,z$ in the following way:
\begin{equation}
\EEf \,=\, {1 \over 8 \pi G\,R^2}\; \EFf(\rho/R,z/R)
\,=\, {5.361369... \cdot 10^{22} \over (R/m)^2}\;\EFf\big(\rho/R,z/R\big)\,{\gr \over cm^3} ~, \label{espeef}
\end{equation}
for a suitable, dimensionless function $\EFf$ of two variables $\rv,\ze$ (here set equal
to $\rho/R,z/R$), which also depends on the parameters $\ellm/R,\ellp/R,\af,\at$
and on the shape function $\HH$.
The second equality in Eq. \eqref{espeef} follows from Eq. \eqref{defG} for $G$.
The function $\EFf$ can be computed analytically and vanishes identically for
$\sqrt{(\rv - 1)^2 + \ze^2} \in [0, \ellm/R] \cup [\ellp/R, + \infty)$
(i.e., where the curvature is zero).
Further details on this topic can be found in Appendix \ref{App4}.
\begin{figure}[t!]
    \centering
        \begin{subfigure}[b]{0.45\textwidth}
                \includegraphics[width=\textwidth]{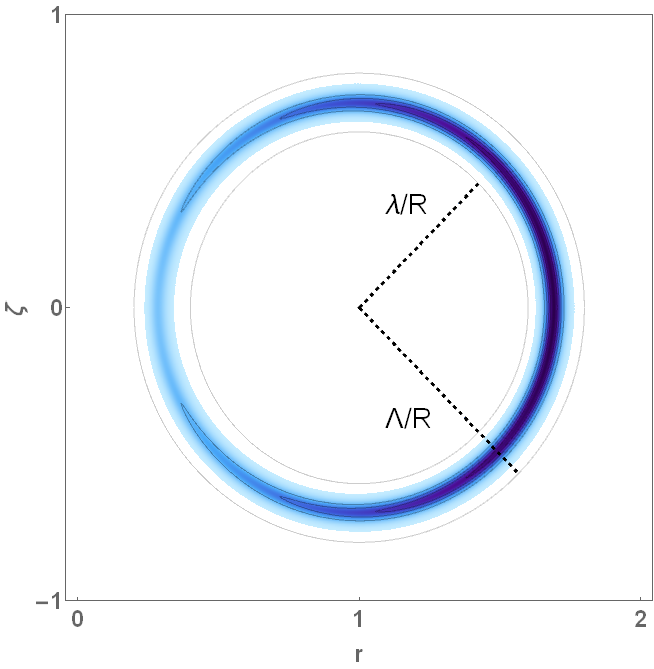}
                \caption*{Figure 5a} \label{E1}
                \vspace{0.cm}
        \end{subfigure}
        \hspace{0.5cm}
        \begin{subfigure}[b]{0.48\textwidth}
                \includegraphics[width=\textwidth]{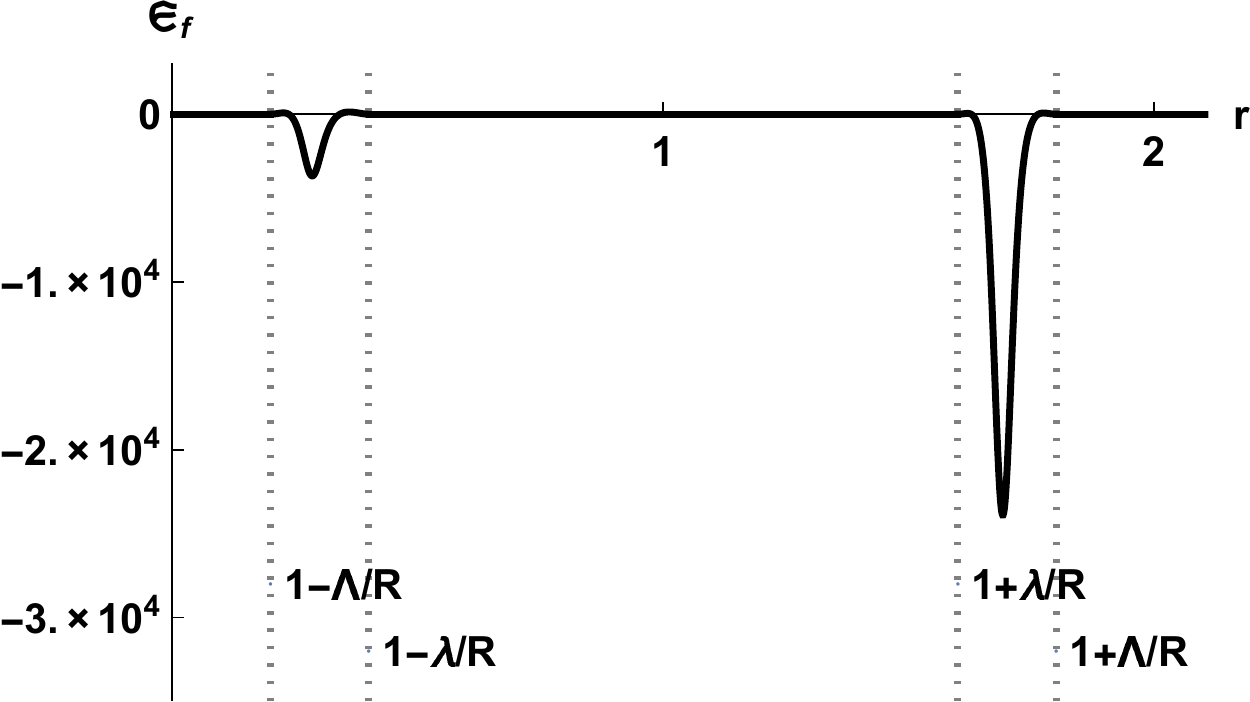}
                \caption*{$\phantom{a}$\vspace{1.2cm} \\ Figure 5b} \label{E2}
                \vspace{0.cm}
        \end{subfigure}
        \caption*{{\small Figures 5a-5b: density plot of the function
        $(\rv,\ze) \mapsto \EFf(\rv,\ze)$ and graph of the function
        $\rv \mapsto \EFf(\rv,0)$, for $\ellm/R = 3/5$, $\ellp/R = 4/5$, $\af = 9/100$, $\at = 10$.
        Again, the shape function $\HH = \HH_{(k)}$ is the one given
        in Eq.s \eqref{HkExp} \eqref{Fkdef} of Appendix \ref{App1}, with $k = 3$.}}
        \label{fig:EnDenT}
\end{figure}
\parn
For an appreciation of the above statements, let us fix $\ellm/R,\ellp/R,\af,\at$ and
$\HH = \HH_{(k)}$ as in Eq. \eqref{cho1}; for these choices, Figure 5a describes
$\EFf$ as a function of the variables $\rv=\rho/R$, $\ze = z/R$ and Figure 5b
is the graph of the function $\rv \mapsto \EFf(\rv,0)$.
\parn
A look at Figure 5b suffices to realize that $\EFf$ attains negative values;
so, the weak energy condition $\EE(X) \geqs 0$ fails for $X = E_{(0)}(p)$ at suitable
points $p \in \Tar$. With our choices \eqref{cho1}, $\EFf$ is a bounded function of $(\rv,\ze)$
with absolute minimum
\begin{equation}
\min \EFf \,=\, \EFf(1.691442...\,,\,0) \,=\, -\,2.389140... \cdot 10^4~.
\end{equation}
Using the above numerical value for the minimum, we infer from
\eqref{espeef} that
\begin{equation}
\min \EEf \,=\, -\,{1.280906... \cdot 10^{27} \over (R/m)^2}\; {\gr \over cm^3}
\label{efgetes}
\end{equation}
The outcomes of this formula for some values of $R$ are given in Tables 1-3 of page \pageref{TABLES}.
In passing, we remark that all the values of $|\min \EEf|$ arising from the cited tables
are considerably smaller than the Planck density $\rho_P := c^5/(\hbar\,G^2) = 5.155... \cdot 10^{93} gr/cm^3$
($\hbar$ is the reduced Planck constant); so, no quantum gravity effect seems to be
involved in the physical regimes described by the tables.
For the specific value $R = 10^{18} m \simeq 100$ light years, $|\min \EEf|$ is indeed
much smaller than $1 \gr/cm^3$.

\subsection{The energy density measured during time travel}\label{subsecEG}
Let us now pass to determine the energy density measured by a freely falling observer
who performs a time travel into the past, moving along a timelike geodesic $\cu$ of the
type analysed in Section \ref{freefallSec}. At proper time $\ta$, this observer has
four-velocity $\dcu(\ta) \in T_{\cu(\ta)}\Tar$ and measures the energy density
\begin{equation}
\EEg(\ta) \,:=\, \EE\big(\dcu(\ta)\big) ~; \label{defeeg}
\end{equation}
indicating with $\rho(\ta)$ the radial coordinate of $\cu(\ta)$, we find
\begin{equation}
\EEg(\ta) \,=\, {\pt^2 \over 8 \pi G\,R^2}\; \EFg\big(\rho(\ta)/R\big)
\,=\, {\big(5.361369... \cdot 10^{22}\big)\;\pt^2 \over (R/m)^2}\;
\EFg\big(\rho/R\big)\,{\gr \over cm^3} ~, \label{espeeg}
\end{equation}
where the second identity follows again from Eq. \eqref{defG}, and $\EFg$ is a suitable,
dimensionless function of the variable $\rv = \rho(\ta)/R$
which also depends on the parameters $\ellm/R,\ellp/R,\af,\at$
and $\pt,\xf$ (related, respectively, to the metric $g$ and to the geodesic $\cu$).
The function $\EFg(\rv)$ can be computed explicitly and vanishes identically for $\rv$
outside the region $(1-\ellp/R,1-\ellm/R) \cup (1+\ellm/R,1+\ellp/R)$ (where the curvature is zero);
moreover, it has a limit for $\pt \to +\infty$ and $\xf \to 0$, simultaneously.
\begin{figure}[t!]
    $\phantom{a}$ \vspace{-1cm}\parn
    \centering
        \begin{subfigure}[b]{0.45\textwidth}
                \includegraphics[width=\textwidth]{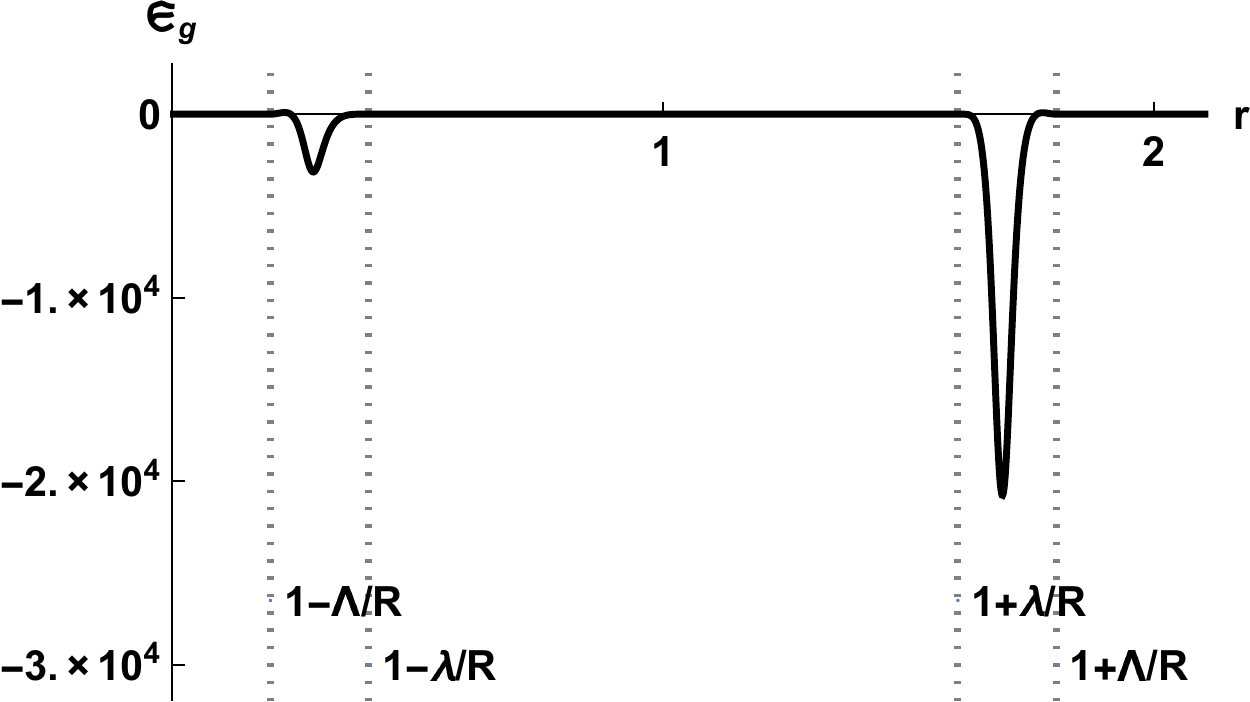}
                \caption*{Figure 6a} \label{EG1}
                \vspace{-0.1cm}
        \end{subfigure}
        \hspace{0.5cm}
        \begin{subfigure}[b]{0.48\textwidth}
                \includegraphics[width=\textwidth]{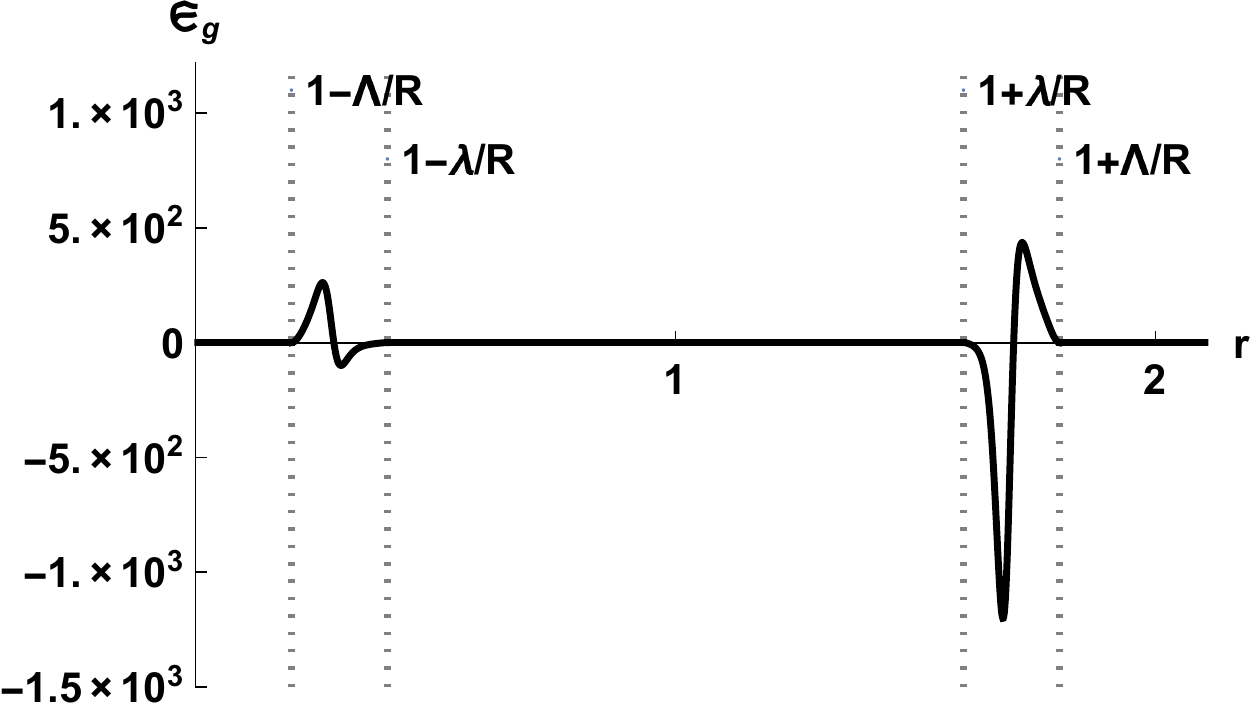}
                \caption*{Figure 6b} \label{EG2}
                \vspace{-0.1cm}
        \end{subfigure}
        \caption*{{\small Figures 6a-6b: graphs of the function
        $\rv \mapsto \EFg(\rv)$
        for $\ellm/R = 3/5$, $\ellp/R = 4/5$, $\af = 9/100$, $\at = 10$
        and for $\pt = 1.1$, $\xf = 7.2... \cdot 10^{-10}$ (Fig. 6a), or
        in the limit $\pt \to +\infty$, $\xf \to 0^{+}$ (Fig. 6b).
        Again, the shape function $\HH = \HH_{(k)}$ is the one given
        in Eq.s \eqref{HkExp} \eqref{Fkdef} of Appendix \ref{App1}, with $k = 3$.}}
        \label{fig:EnDenG}
\end{figure}
\parn
Again, let us fix $\ellm/R,\ellp/R,\af,\at$ and $\HH = \HH_{(k)}$ as in Eq. \eqref{cho1}.
Fig.s 6a-6b show the graphs of the function $\rv \mapsto \EFg(\rv)$ for a particular choice
of $(\pt,\xf)$ or for $\pt \to + \infty$, $\xf \to 0^+$; from these graphs, it can be
readily inferred that in general $\EFg$ is not positive, thus proving that the
weak energy condition $\EE(X) \geqs 0$ is violated for $X = \dcu(\ta)$ and for suitable
values of the proper time $\ta$.
Let us consider the absolute minimum of $\EEg$ along the travel, i.e.,
\begin{equation}
\min \EEg \,:=\, \min_{\ta \in [0,\ta_2]} \EEg(\ta) \,=\,
{\big(5.361369... \cdot 10^{22}\big)\;\pt^2 \over (R/m)^2}
\l(\min_{\rv \in [\rho_0/R,\rho_1/R]}\EFg(\rv)\r) {\gr \over cm^3}\;; \label{espeegmin}
\end{equation}
note that the minimum of $\EFg$ depends on the parameters $\pt,\xf$.
The last columns of Tables 1-3 on page \pageref{TABLES} give the numerical value of
$\min \EE_g$, descending from Eq. \eqref{espeegmin}, for some choices of $R,\pt,\xf$;
these minima appear to be much smaller than the Planck density (so we can repeat
the comments at the end of the previous subsection).
\parn
Finally, let us remark that for $R = 10^{18} m \simeq 100$ light years, the energy density
measured by a freely falling time traveller has an absolute value $\lesssim 1\,\gr/cm^3$
even for the quite large choice $\pt = 10^5$.
\parn
$\phantom{a}$\vspace{-0.1cm}
\parn
\textbf{Acknowledgments.} This work was supported by: INdAM, Gruppo Nazionale per la Fisica Matematica;
INFN; MIUR, PRIN 2010 Research Project ``Geometric and analytic theory of Hamiltonian systems in finite
and infinite dimensions''; Universit\`{a} degli Studi di Milano.\parn
We wish to thank Prof. Ori and the anonymous referees for valuable suggestions
and bibliographical indications which helped improving the quality of the manuscript.

\vfill \eject \noindent
\newpage

\appendix
\section{Appendix. On the shape function $\bm{\HH}$}\label{App1}
Let us make reference to the framework of Section \ref{secMod} and recall that the basic
function $\XX$, which determines the line element $\dsX^2$ of Eq. \eqref{dsXidef},
is defined in terms of a suitable, auxiliary shape function $\HH \in C^k([0,+\infty))$
($k \in\{2,3,...,\infty\}$).
In the present appendix we provide explicit examples $\HH_{(k)}$ of admissible choices
of $\HH$ for any $k \in \{0,1,...,\infty\}$
({\footnote{Notice that the stricter assumption $k \geqs 2$ made in Section \ref{secMod}
(and adhered to throughout all the paper) is motivated by the will to avoid the
appearance of curvature singularities; for the purposes of this appendix, that
assumption can be generalized requiring $k \geqs 0$.}}),
fulfilling the requirement \eqref{chidef} reported hereafter:
$$ \HH(\yr) = 1 ~~ \mbox{for $\yr \in [0,\ellm/R]$} ~, \qquad\quad
\HH(\yr) = 0 ~~ \mbox{for $\yr \in [\ellp/R,+\infty)$} ~. $$
In the two forthcoming paragraphs, we discuss separately the cases with $k \geqs 0$
finite and the case $k = +\infty$.
\vspace{-0.3cm}

\paragraph{An admissible choice $\bm{\HH = \HH_{(k)} \in C^k([0,+\infty))}$, for $\bm{k \geqs 0}$ finite.}
Let us first consider the function
\begin{equation}
\Jd_{(k)}(v) \,:=\, \l\{\!\begin{array}{ll}
\dd{v^{k}\,(1-v)^{k}} & ~~\mbox{for~ $v \in [0,1]$}~, \vspace{0.1cm} \\
\dd{0} & ~~\mbox{for~ $v \in (-\infty,0] \cup [1,+\infty)$}~.
\end{array} \r. \label{Jdef}
\end{equation}
Notice that the function $\Jd_{(0)}$, corresponding to $k = 0$, is piecewise constant
and discontinuous; on the other hand, we have $\Jd_{(k)} \in C^{k-1}(\R)$
for all $k \geqs 1$.\\
Next, we introduce the normalized primitive of $\Jd_{(k)}$ given by
\begin{equation}
\Hk_{(k)}(w) \,:= \,\l(\int_{-\infty}^{+\infty}\!\! dv\;\Jd_{(k)}(v) \r)^{\!\!-1}
\int_{-\infty}^w dv\;\Jd_{(k)}(v) \qquad (w \in \R) ~. \label{Hkdef}
\end{equation}
In view of the previous considerations on the regularity of $\Jd_{(k)}$, it can be
readily inferred that $\Hk_{(k)} \in C^k(\R)$.
Furthermore, let us stress that it is possible to derive an explicit, piecewise
polynomial expression for $\Hk_{(k)}$; more precisely,
taking into account the definition \eqref{Jdef} of $\Jd_{(k)}$ and using some known relations
for the incomplete beta function (see, e.g., Eq.s 8.17.1, 8.17.2 and 8.17.5 on page 183
of \cite{NIST}), it can be proved by simple computations that
\begin{equation}
\Hk_{(k)}(w) \,=\, \l\{\!\begin{array}{ll}
\dd{0} & ~\dd{\mbox{for~ $w \in (-\infty,0]$} ~,} \vspace{0.15cm} \\
\dd{\sum_{j = 0}^{k} \binom{2k\!+\!1}{j\!+\!k\!+\!1}\; w^{j+k+1}\, (1-w)^{k-j}} &
~\dd{\mbox{for~ $w \in [0,1]$} ~,} \vspace{0.15cm} \\
\dd{1} & ~\dd{\mbox{for~ $w \in [1,+\infty)$} ~.} \\
\end{array}\r. \label{HkExp}
\end{equation}
Finally, we set
\begin{equation}
\HH_{(k)}(\yr) \,:=\, \Hk_{(k)}\!\l({\ellp/R - \yr \over \ellp/R - \ellm/R}\r) \qquad
\mbox{for $\yr \in [0,+\infty)$} ~. \label{Fkdef}
\end{equation}
The previous considerations ensure that $\HH_{(k)}$ is of class $C^k$
and even fulfils the requirement \eqref{chidef}, thus determining
an admissible choice of $\HH$.
\vspace{-0.3cm}

\paragraph{An admissible choice $\bm{\HH = \HH_{(\infty)} \in C^\infty([0,+\infty))}$.}
In this case we can follow the same steps described in the previous paragraph,
starting with the smooth function
\begin{equation}
\Jd_{(\infty)}(v) \,:=\, \l\{\!\begin{array}{ll}
\dd{e^{-\,{1 \over v(1-v)}}} & ~~\mbox{for~ $v \in [0,1]$}~, \vspace{0.1cm} \\
\dd{0} & ~~\mbox{for~ $v \in (-\infty,0] \cup [1,+\infty)$}~.
\end{array} \r.
\label{Jinfdef}
\end{equation}
\newpage
$\phantom{a}$\vspace{-1.cm}\parn
Using the above function, we can proceed to introduce a map $\Hk_{(\infty)}$
by a definition analogous to \eqref{Hkdef}; more precisely, we set
\begin{equation}
\Hk_{(\infty)}(w) \,:= \,\l(\int_{-\infty}^{+\infty}\!\! dv\;\Jd_{(\infty)}(v) \r)^{\!\!-1}
\int_{-\infty}^w dv\;\Jd_{(\infty)}(v) \qquad (w \in \R) ~. \label{HInfdef}
\end{equation}
This function is itself smooth and fulfils, in addition,
({\footnote{Even though in this case it is not possible to derive a fully explicit
expression for $\Hk_{(\infty)}$ in terms of elementary (or special) functions,
one can use directly the integral representation \eqref{HInfdef} for the numerical
computations of interest in the applications.}})
\begin{equation}
\Hk_{(\infty)}(w) = 0 ~~ \mbox{for $w \in (-\infty,0]$} ~, \qquad\quad
\Hk_{(\infty)}(w) = 1 ~~ \mbox{for $w \in [1,+\infty)$} ~.
\end{equation}
In conclusion, similarly to what we did in the preceding paragraph for $k \geqs 0$
finite, we determine an admissible, smooth choice of $\HH$ fulfilling Eq. \eqref{chidef}
with the position
\begin{equation}
\HH_{(\infty)}(\yr) \,:=\, \Hk_{(\infty)}\!\l({\ellp/R - \yr \over \ellp/R - \ellm/R}\r)
\qquad \mbox{for $\yr \in [0,+\infty)$} ~.
\label{Finfdef}
\end{equation}
\textbf{A final remark.} The functions $\Jd_{(k)}$, $\Jd_{(\infty)}$ of Eq.s \eqref{Jdef},
\eqref{Jinfdef} are positive on $(0,1)$. Due to this, the normalized primitives $\Hk_{(k)}$,
$\Hk_{(\infty)}$ of Eq.s \eqref{Hkdef}, \eqref{HInfdef} are strictly increasing on $(0,1)$.
Therefore the functions $\HH_{(k)}$, $\HH_{(\infty)}$ of Eq.s \eqref{Fkdef}, \eqref{Finfdef}
are strictly decreasing on $(\ellm/R,\ellp/R)$, i.e., they fulfil condition \eqref{HHcond}.
\vspace{-0.1cm}

\section{Appendix. A class of integrals depending on a parameter. Applications to the integrals in Eq.s \eqref{eqt2}-\eqref{eqta2}}\label{App2}
In the present appendix we report a number of results about a class of integrals
depending on a parameter, focusing especially on their evaluation for small values
of the parameter. These results allow us, in particular, to treat the integrals
appearing in Eq.s \eqref{eqt2}-\eqref{eqta2} of subsection \ref{subProof},
in all the cases of interest for the applications discussed therein.\\
In the forthcoming subsection \ref{sub1App2} we show that, by means of simple changes
of variables, the study of all the above cited integrals can be reduced to the analysis
of a general class of integrals $\JJ(\ee)$, depending on a small parameter $\ee$.
In the following subsection \ref{subapest} we derive a preliminary bound for $\JJ(\ee)$
holding for all $\ee > 0$, which provides a control of its singular behaviour for
$\ee \to 0^+$. In subsection \ref{subalt} we determine an alternative representation
for $\JJ(\ee)$, allowing to cure some difficulties which arise in the numerical evaluation
for small $\ee$ (this is an essential result for the applications of subsections \ref{subProof}, \ref{antic}).
In the conclusive subsection \ref{subasym} we use this alternative representation of $\JJ(\ee)$
to determine its leading-order contributions for $\ee \to 0^+$.
\vspace{-0.3cm}

\subsection{The general structure for the integrals in Eq.s \eqref{eqt2}-\eqref{eqta2}}\label{sub1App2}
Let us consider the integral expressions appearing in Eq.s \eqref{eqt2}-\eqref{eqta2}
and recall that, for the applications discussed in subsections \ref{subProof} and \ref{antic},
it is of interest to evaluate them when (see Eq.s \eqref{imp3}, \eqref{defxf} and \eqref{xfto0})
$$ \pf \,=\, \sqrt{\,{\af^2 \over \at^2}\l(1+{\at^2 \over \pt^2}\r) + \xf^2}
\qquad \mbox{and} \qquad 0 < \xf \ll 1 ~. $$
Assuming the shape function $\HH = \HH_{(k)}$ has the form \eqref{HkExp} \eqref{Fkdef}
given in Appendix \ref{App1} for some finite $k \geqs 2$ and making the change of variables
\begin{equation}
\xr \,:=\, \l\{\!\begin{array}{ll}
\dd{{\rv - 1 + \ellp/R \over \ellp/R - \ellm/R}} &
\dd{\mbox{for}~~ \rv \in \l(1- {\ellp \over R}, 1 - {\ellm \over R}\r) ,} \vspace{0.1cm} \\
\dd{{\rv - 1 - \ellm/R \over \ellp/R - \ellm/R}} &
\dd{\mbox{for}~~ \rv \in \l(1+{\ellm \over R}, 1+ {\ellp \over R}\r)}
\end{array}\r. \qquad~
\ss \,:=\,b^2\,\xf^2 ~, \label{defxss}
\end{equation}
all the integrals mentioned above can be written in the following form:
\begin{equation}
\II(\ss) \,:= \int_{0}^{1} d\xr ~{\PP(\xr,\ss) \over \sqrt{\ss + \xr^h\,\QQ(\xr,\ss)}} ~, \label{IIdef}
\end{equation}
where $\ss \in (0,\ss_\ast)$ for some $\ss_\ast > 0$
(see Eq.s \eqref{defxf} \eqref{defxss}), $h := k+1 \geqs 3$,
$\PP,\QQ \in C^{\infty}([0,1]\times[0,\ss_\ast])$
({\footnote{As a matter of fact, for the analysis discussed in the following it
would suffice to assume that $\PP,\QQ \in C^{[(h+3)/2]}([0,1]\times[0,\ss_\ast])$,
where $[(h+3)/2]$ denotes the integer part of $(h+3)/2$. However, for the cases of
interest to us, actually both $\PP$ and $\QQ$ are smooth functions; more precisely,
they are polynomials with respect to the first variable $\xr$ (or ratios of polynomials
with non-vanishing denominators), with coefficients depending smoothly on $\ss$.}})
and
\begin{equation}\begin{array}{c}
\dd{\ss + \xr^h\,\QQ(\xr,\ss) > 0 \quad
\mbox{for all\; $(\xr,\ss) \in \big([0,1] \!\times\! [0,\ss_\ast]\big)\! \setminus\! \{(0,0)\}$} ~,} \vspace{0.2cm}\\
\dd{\q0 \,:=\, \QQ(0,0) \,>\, 0 ~. }
\label{ZZdef}
\end{array}\end{equation}
The rest of this appendix is devoted to the analysis of any integral of this form,
with $h \in \{3,4,5,...\}$
({\footnote{Apart from the fact that in all the applications of interest in
this work there holds $h \geqs 3$, let us point out for completeness that the
analysis of the cases where $h = 1,2$ would actually require a separate treatment,
involving slightly different computations.}}).
To this purpose, it is convenient to introduce the rescaled parameter
\begin{equation}
\ee \,:=\, (\ss/\q0)^{1/h} \,\in\, \big(0,\ee_\ast\big)~,
\qquad\; \ee_\ast := (\ss_\ast/\q0)^{1/h}
\end{equation}
and use it to re-express Eq. \eqref{IIdef} as
\begin{equation}
\II(\ss) \,=\, \JJ\big((\ss/\q0)^{1/h}\big) ~, \qquad
\JJ(\ee) \,:=\, {1 \over \sqrt{\q0}} \int_{0}^{1} d\xr ~ {\PP(\xr,\q0\,\ee^h) \over
\sqrt{\ee^h + \xr^h\,\QQ(\xr,\q0\,\ee^h)/\q0}} ~. \label{JJdef}
\end{equation}
As previously hinted at, our main concern is the investigation of the asymptotic
behaviour of $\JJ(\ee)$ for $\ee \to 0^+$. Let us notice that the integrand function
in Eq. \eqref{JJdef} evaluated at $\ee = 0$ diverges in a non-integrable way for
$\xr \to 0^+$; this suggests that $\JJ(\ee)$ should become singular for $\ee \to 0^+$.

\subsection{A preliminary bound for $\bm{\JJ(\ee)}$}\label{subapest}
Keeping in mind the definition \eqref{JJdef} of $\JJ(\ee)$, let us consider the function
\begin{equation}
\FF(\xr,\ee) \,:= \,{\PP(\xr,\q0\,\ee^h)\;\sqrt{\ee^h + \xr^h} \over
\sqrt{\q0}\,\sqrt{\ee^h + \xr^h\,\QQ(\xr,\q0\,\ee^h)/\q0}} ~. \label{FFdef}
\end{equation}
In view of the regularity features enjoyed by $\PP$ and $\QQ$, we can readily infer
that $\FF(\xr,\ee)$ is smooth on $\big([0,1] \!\times\! [0,\ee_\ast]\big)\! \setminus\! \{(0,0)\}$;
on the other hand, it can be proved that $\FF$ has a continuous extension to the
origin
({\footnote{To prove this claim, let us consider on $\R^2$ the norm
$\|(\xr,\ee)\|_h := (|\xr|^h\!+|\ee|^h)^{1/h}$ (recall that all norms on $\R^2$ are
equivalent, since we have a finite-dimensional vector space). Then, on account of
the fact that $\PP,\QQ \in C^\infty([0,1]\times[0,\ss_\ast])$ (so that, in particular,
$\PP,\QQ \in C^1([0,1]\times[0,\ss_\ast])\,$), we have $\PP(\xr,\ee) = \PP(0,0) + O(\|(\xr,\ee)\|_h)$
and $\QQ(\xr,\ee)/\q0 = 1 + O(\|(\xr,\ee)\|_h)$ for $(\xr,\ee) \to (0,0)$; in the same limit,
it is $\xr^h = O(\|(\xr,\ee)\|_h^h)$. Recalling the definition \eqref{FFdef} of $\FF$,
the previous considerations allow us to infer by simple computations that
$$ \FF(\xr,\ee) \,= \,{\PP(0,0) + O(\|(\xr,\ee)\|_h)\; \over
\sqrt{\q0}\, \sqrt{1 + \,O(\|(\xr,\ee)\|_h)}} \, = \, {\PP(0,0) \over
\sqrt{\q0}}\; + O(\|(\xr,\ee)\|_h) ~; $$
thus, we can extend continuously $\FF$ to the origin setting $\FF(0,0) := \PP(0,0)/\sqrt{\q0}$\,.}}).
Thus we can regard $\FF$ as a continuous function on $[0,1] \!\times\! [0,\ee_\ast]$, and
\begin{equation}
\|\FF\|_{C^0} \,:=\,
\sup_{(\xr,\ee) \,\in\, [0,1] \times [0,\ee_\ast]} \big|\FF(\xr,\ee)\big| \,<\, +\infty ~.
\end{equation}
Now, from the definition \eqref{JJdef} of $\JJ(\ee)$ we infer the bound
\begin{equation}
\big|\JJ(\ee)\big| \,=\, \l|\,\int_{0}^{1} d\xr ~ {\FF(\xr,\ee) \over \sqrt{\ee^h + \xr^h}}\,\r|
\,\leqs\, \|\FF\|_{C^0}\! \int_{0}^{1} d\xr ~ {1 \over \sqrt{\ee^h + \xr^h}} ~.
\end{equation}
On the other hand, for any $0 \!<\! \ee \!<\! 1$ we have the following
chain of elementary estimates (recall that we are assuming $h \geqs 3$):
\begin{equation}
\int_{0}^{1} d\xr ~ {1 \over \sqrt{\ee^h + \xr^h}} \;\leqs\,
\int_{0}^{\ee} d\xr\; {1 \over \sqrt{\ee^h}}\; + \int_{\ee}^{1}\! d\xr\; {1 \over \sqrt{\xr^h}} \;\leqs\;
{h \over h-2}\; {1 \over \ee^{h/2-1}} ~.
\label{estInt}
\end{equation}
Summing up Eq.s \eqref{FFdef}-\eqref{estInt} show that, if $\ee_\ast < 1$,
\begin{equation}
\big|\JJ(\ee)\big| \,\leqs\, \l({h \over h-2}\; \|\FF\|_{C^0}\!\r)\,
{1 \over \ee^{h/2-1}} \qquad \mbox{for all\; $\ee \in \big(0,\ee_\ast\big]$} ~.
\end{equation}

\subsection{An alternative representation of $\bm{\JJ(\ee)}$}\label{subalt}
In the applications considered in this work, it is often of interest to determine with
great accuracy the numerical value of integrals such as $\JJ(\ee)$
for small values of the parameter $\ee$. This task is plagued by unpleasant numerical
instabilities, which can be ascribed to the singular behaviour of $\JJ(\ee)$ for
$\ee \to 0^+$. In the following we shall derive an alternative
representation for $\JJ(\ee)$, isolating the leading order contributions which become
singular for $\ee \to 0^+$; this representation allows to perform the numerical evaluation
of $\JJ(\ee)$ with greater efficiency and better accuracy.
\vspace{0.1cm} \\
To begin with, let us consider the integral $\JJ(\ee)$ defined in Eq. \eqref{JJdef}; making
the change of variable $\tt := \xr/\ee \in (0,1/\ee)$, by simple manipulations we obtain
\begin{equation}
\JJ(\ee) \,=\, \int_{0}^{1/\ee} d\tt ~ \GG(\tt,\ee) ~,
\label{JJGG}
\end{equation}
where we have set
\begin{equation}
\dd{\GG(\tt,\ee) \,:=\,{1 \over \ee^{h/2-1}\;\sqrt{\q0}\; \sqrt{1 + \tt^h}}~
{\PP(\ee\,\tt,\q0\,\ee^h) \over  \sqrt{1 + {\tt^h \over 1 + \tt^h}
\l({\QQ(\ee\,\tt,\q0\,\ee^h) \over \q0} - 1\r)}}~.}
\label{GGdef}
\end{equation}
Recalling the regularity assumptions on $\PP$ and $\QQ$, we can proceed to compute
the Taylor expansion of $\GG(\tt,\ee)$ for $\ee \to 0^+$ up to order $\ee$.
To this purpose, it is useful to notice that
\begin{equation}
\PP(\ee\,\tt,\q0\,\ee^h) \;=\,
\sum_{n = 0}^{[(h+1)/2]} {1 \over n!}\,(\de_1^n \PP)(0,0)\;(\ee\,\tt)^n \,+ \,
O\big(\ee^{[(h+3)/2]}\big) \qquad \mbox{for $\ee \to 0^+$} ~;
\end{equation}
of course, an analogous relation holds as well for $\QQ$. By a few additional
computations, we obtain
\begin{equation}
\GG(\tt,\ee) \,=\, {1 \over \ee^{h/2-1}\; \sqrt{1 + \tt^h}} \,
\sum_{m,n = 0}^{[(h+1)/2]}\, \gmn_{m,n} \l({\tt^h \over 1+\tt^h}\r)^{\!\!m}\! (\ee\,\tt)^n
+\, O\big(\ee^{[(h+3)/2]}\big) ~ , \label{GGTay}
\end{equation}
where, the coefficients $\gmn_{m,n}$ are completely determined by the values at the
origin of $\PP,\QQ$ and of their derivatives with respect to the first variable up to $[(h+1)/2]$-th order.
For subsequent use, we define a reminder $\RR_{[(h+1)/2]} \in C^0([0,1/\ee]\times[0,(\ss_\ast/\q0)^{1/h}])$
via the equation
\begin{equation}
\RR_{[(h+1)/2]}(\tt,\ee) :=
\GG(\tt,\ee) \,-\, {1 \over \ee^{h/2-1}\; \sqrt{1 + \tt^h}} \, \sum_{m,n = 0}^{[(h+1)/2]}\,
\gmn_{m,n} \l({\tt^h \over 1+\tt^h}\r)^{\!\!m}\! (\ee\,\tt)^n ~ . \label{GGTay2}
\end{equation}
Thus, Eq.s \eqref{JJGG}, \eqref{GGTay} and \eqref{GGTay2} give
\begin{equation}
\JJ(\ee) \,=\, \sum_{m,n = 0}^{[(h+1)/2]} \gmn_{m,n}~ \jmn_{m,n}(\ee)~ + ~
\RJ_{[(h+1)/2]}(\ee) ~,
\label{JJAlt}
\end{equation}
where we introduced the functions
\begin{equation}
\jmn_{m,n}(\ee) \;:=\;
{1 \over \ee^{h/2-1-n}} \int_{0}^{1/\ee} d\tt ~ {\tt^{h\,m + n} \over (1+ \tt^h)^{m+1/2}} ~,
\end{equation}
\begin{equation}
\RJ_{[(h+1)/2]}(\ee) \;:=\, \int_{0}^{1/\ee} d\tt ~ \RR_{[(h+1)/2]}(\tt,\ee) ~.
\end{equation}
Concerning the terms $\jmn_{m,n}(\ee)$, making the change of variable $\uu := \ee^h\tt^h$
we obtain an integral coinciding with the well-known representation for a Gaussian
hypergeometric function $\2F1$ (see, e.g., Eq. (10) on page 59 of \cite{Erd});
taking this into account, we get
\begin{equation}
\jmn_{m,n}(\ee) \;=\; {1 \over (h\,m \!+\! n \!+\! 1)\;\ee^{h(m+1/2)}}~
\2F1\!\l(m\!+\!{1 \over 2}\,,\,m\!+\!{n\!+\!1 \over h}\,;
\,m\!+\!{n\!+\!1 \over h}\!+\!1\,;\,-\,{1 \over \ee^h}\r) .
\label{jmnHyp}
\end{equation}
In passing, let us point out that the above result can be re-expressed as follows,
for all $h \geqs 3$ and $n \in \{0,...,[(h+1)/2]\}$ such that $(n+1)/h-1/2 \notin \{0,1,2,...\}$
(see Eq. (18) on page 63 and Eq. (2) on page 56 of \cite{Erd}):
\begin{equation}\begin{array}{c}
\dd{\jmn_{m,n}(\ee) \;=} \vspace{0.1cm} \\
\dd{{\Ga(m\!+ \!{n + 1 \over h})\,\Ga({1 \over 2}\!-\!{n + 1 \over h})
\over h\,\Ga(m + {1 \over 2})}~ {1 \over \ee^{{h \over 2}-1 - n}} +
{1 \over n \!+\! 1\! -\! {h \over 2}}~
\2F1\!\l(m\! +\! {1 \over 2}\,,\,{1 \over 2}\! - \!{n\! +\! 1 \over h}\,;\,
{3 \over 2}\! -\! {n \!+\! 1 \over h}\,; -\,\ee^h\r).} \label{jmnHyp2}
\end{array}\end{equation}
As for the remainder $\RJ_{[(h+1)/2]}$, we can return to the integration variable
$\xr = \ee\,\tt \in (0,1)$; this gives
\begin{equation}
\RJ_{[(h+1)/2]}(\ee) \;=\, \int_{0}^{1} d\xr ~ {1 \over \ee}\;\RR_{[(h+1)/2]}(\xr/\ee,\ee) ~.
\label{RJInt}
\end{equation}
The map $(\xr,\ee) \mapsto (1/\ee)\,\RR_{[(h+1)/2]}(\xr/\ee,\ee)$ is defined in principle
on $[0,1] \times (0,\ee_\ast]$; however, in all cases of interest in this work it has a continuous
extension to $[0,1] \times [0,\ee_\ast]$
({\footnote{Moreover, in all cases of interest for us one has
$$ \lim_{(\xr,\ee) \to (0^+,0^+)} \l({1 \over \ee}\,\RR_{[(h+1)/2]}(\xr/\ee,\ee)\r) = 0 ~. $$}});
this allows us to infer by Lebesgue's dominated convergence that $\RJ_{[(h+1)/2]}(\ee)$
has a continuous extension to $\ee = 0$.
The computation of the reminder \eqref{RJInt} is numerically stable due to the
above mentioned continuity features.\\
In conclusion, Eq.s \eqref{GGTay}-\eqref{RJInt} give the alternative representation
of $\JJ(\ee)$ mentioned at the beginning of this subsection.
This has been used to compute the integrals in Eq.s \eqref{eqt2}-\eqref{eqta2} for the
values of the parameters reported in Tables 1-3; in this case, $\ee$ is proportional to
$\xf^{1/2}$.

\subsection{Asymptotic expansion of $\bm{\JJ(\ee)}$ for $\bm{\ee \to 0^+}$}\label{subasym}
The alternative representation \eqref{GGTay}-\eqref{RJInt} of $\JJ(\ee)$ discussed
in the previous subsection can be used to determine the leading-order, contributions
to $\JJ(\ee)$ for $\ee \to 0^+$; hereafter we shall give more details on this topic.\\
First of all, let us investigate the asymptotic behaviour of the functions
$\jmn_{m,n}(\ee)$ for $\ee \to 0^+$; this amounts to determine the leading order terms
in the Gaussian hypergeometric function $\2F1$ appearing in Eq. \eqref{jmnHyp}.
To this purpose, we follow the analysis reported in the paragraph 2.1.4 of \cite{Erd}.\\
Let us notice that, in principle, the cases where $(n+1)/h-1/2$ is either in $\{0,1,2,...\}$
or not must be treated separately. Since we have $h \geqs 3$ and $n \in \{0,...,[(h+1)/2]\}$
(see Eq. \eqref{JJAlt}), the condition $(n+1)/h-1/2 \in \{0,1,2,...\}$ is fulfilled only
if $h$ is even and $n = [(h-1)/2] = h/2-1$;
in this case, we can use Eq. (18) on page 63 of \cite{Erd} (along with
some known identities for the digamma function $\psi$; see, e.g., Eq.s 5.4.12 and 5.4.15
on page 137 of \cite{NIST}) to infer the identity
\begin{equation}
\jmn_{m,{h\over 2}-1}(\ee) \,=\, - \ln \ee
+ {2 \over h}\l(\ln 2 - \sum_{\ell = 1}^m {1 \over 2\ell - 1}\r) \qquad\big(\,\mbox{$h$ even}\,\big) \;.
\end{equation}
In all the remaining cases where $(n+1)/h-1/2 \notin \{0,1,2,...\}$ (i.e., for $h$ even and
$n \in \{0,...,h/2-2\}\cup\{h/2\}$, or for $h$ odd and $n \in \{0,...,(h+1)/2\}$),
we can use the representation \eqref{jmnHyp2} of $\jmn_{m,n}$ and the series representation
of the Gaussian hypergeometric function $\2F1$ (see, e.g., Eq. (2) on page 56 of \cite{Erd})
to obtain
\begin{equation}
\jmn_{m,n}(\ee) \;=\;
{\Ga(m+{n+1 \over h})\,\Ga({1 \over 2}-{n+1 \over h}) \over
h\,\Ga(m+{1\over 2})}\;{\ee^n \over \ee^{h/2-1}} \;+ \; {1 \over n+1-h}\;
+\;O\big(\ee^h\big) \qquad \mbox{for $\ee \to 0^+$} ~.
\end{equation}
Next, let us proceed to analyse the remainder term $\RJ_{[(h+1)/2]}(\ee)$.
Let us assume that the map $(\xr,\ee) \mapsto (1/\ee)\,\RR_{[(h+1)/2]}(\xr/\ee,\ee)$
has a continuous extension to $[0,1] \times [0,\ee_\ast]$
(see the comments on this after Eq. \eqref{RJInt}); then, we can introduce the continuous function
\begin{equation}
\rr_{[(h+1)/2]}(\xr) \,:=\, \lim_{\ee \to 0^+} \l({1 \over \ee}\;\RR_{[(h+1)/2]}(\xr/\ee,\ee) \r)
\qquad \big(\,\xr \in [0,1]\,\big) ~;
\end{equation}
and infer (again, by dominated convergence theorem) that
\begin{equation}
\RJ_{[(h+1)/2]}(0) \,:=\, \lim_{\ee \to 0^+} \RJ_{[(h+1)/2]}(\ee) \,=\,
\int_0^1 d\xr ~\rr_{[(h+1)/2]}(\xr) \,<\,+\infty ~.
\end{equation}
Summing up, the above arguments yield the following asymptotic expansions for $\ee \to 0^+$:\\
\textsl{i}) for \textsl{$h$ even}, there holds
\begin{equation}\begin{array}{c}
\dd{\JJ(\ee) \,=\, {1 \over \ee^{{h \over 2}-1}}
\sum_{n = 0}^{{h \over 2}-2}\! \l(\sum_{m = 0}^{{h \over 2}}
{\Ga(m\!+\!{n+1 \over h})\,\Ga({1 \over 2}\!-\!{n+1 \over h}) \over
h\,\Ga(m+{1\over 2})} ~ \gmn_{m,n}\!\r)\!\ee^n
- \!\l(\sum_{m = 0}^{{h \over 2}} \gmn_{m,{h \over 2}-1}\!\r)\! \ln \ee ~ +} \\
\dd{- \sum_{m = 0}^{{h \over 2}}\!\l[
\sum_{n = 0}^{{h \over 2}-2} {\gmn_{m,n} \over h\!-\!1\!-\!n}
+ {2 \over h}\!\l(\sum_{\ell = 1}^m {1 \over 2\ell\!-\!1} - \ln 2\!\r)\!\gmn_{m,{h \over 2}-1}
+ {2\,\gmn_{m,{h \over 2}} \over h\!-\!2}\r]\! +\,
\RJ_{h \over 2}(0)\, +\, o(1) ~;}
\end{array}\end{equation}
\textsl{ii}) for \textsl{$h$ odd}, there holds
\begin{equation}\begin{array}{c}
\dd{\JJ(\ee) \,=\, {1 \over \ee^{{h \over 2}-1}}\,
\sum_{n = 0}^{{h-3 \over 2}} \l(\sum_{m = 0}^{{h+1 \over 2}}
{\Ga(m+{n+1 \over h})\,\Ga({1 \over 2}-{n+1 \over h}) \over
h\,\Ga(m+{1\over 2})}~ \gmn_{m,n}\r)\ee^n ~ +} \\
\dd{- ~ \sum_{m,n = 0}^{{h+1 \over 2}} {\gmn_{m,n} \over h\!-\!1\!-\!n}~ + ~
\RJ_{{h+1 \over 2}}(0)\, +\, o(1) ~.}
\end{array}\end{equation}

\vfill\eject\noindent
$\phantom{a}$
\vskip-1.7cm
\noindent
\section{Appendix. On tidal accelerations}\label{App3}
\subsection{General facts} \label{App3gen}
Our present aim is to account for all statements of subsection
\ref{basics} on the grounds of well known facts. \parn Given an
arbitrary spacetime $\MM,$ with metric $g$, we consider a timelike geodesic
with a proper time parametrization $\cu: I \to \MM, \ta \mapsto
\cu(\ta)$ ($I\!\subset\!\R$ an interval). We also introduce a \textsl{variation}
of this parametrized geodesic: by this we mean a (regular) family
of maps $\cu_{\si} : I_\si \to \MM$ ($\si \in (-\eps,\eps)$,
$I_\si\!\subset\!\R$ an interval) such that, for each $\si$, the map
$\cu_{\si}$ is a timelike geodesic in a proper time
parametrization and $I_0 = I$, $\cu_0 = \cu$. The deviation vector
field associated to such a variation is
\begin{equation}
\pacu: I \to T \MM~, \qquad \ta \mapsto \pacu(\ta) \,:=\,
\l. {\de \over \de \si} \r|_{\si = 0} \cu_{\si}(\ta) \,\in\,
T_{\cu(\ta)} \MM~.
\end{equation}
By arguments very similar to those described
in \S 3.3 of \cite{Wald}, it can be proved that
$g\big(\pacu(\ta),\dcu(\ta)\big) = \mbox{const.} \equiv C$, and
that the latter constant can be set equal to zero after possibly
replacing $\cu_{\si}$ with the (proper-time) re-parametrization
$\tilde{\cu}_{\si}(\ta) := \cu_{\si}(\ta - C\,\si)$ ($\tau \in
\tilde{I}_\si := I_{\si} + C \si$). Therefore, with no loss of
generality we can assume that
\begin{equation}
g\big(\pacu(\ta),\dcu(\ta)\big) \,=\, 0 ~.
\end{equation}
It can be shown (see again \cite{Wald}) that the deviation vector
field fulfils the \textsl{Jacobi equation}
\begin{equation}
{\nabla^2 \pacu \over d \ta^2}(\ta) \,=\,
-\,\Riem\big(\pacu(\ta), \dcu(\ta)\big)\, \dcu(\ta)
\end{equation}
where $\nabla$ and $\Riem$ are the covariant derivative and the
Riemann curvature tensor associated to the metric $g$. In a more
customary formulation, one thinks about two ``infinitesimally
closed geodesics''
\begin{equation}
\ta \,\mapsto\, \cu(\ta)\,,\, (\cu + \decu)(\ta)
\end{equation}
where $\decu(\ta) := \si\,\pacu(\ta) \in T_{\cu(\ta)} \MM$ and
$\si$ is infinitesimally small; of course, we have formally
\begin{equation}
g\big(\decu(\ta),\dcu(\ta)\big) \,=\, 0 ~, \qquad\;
{\nabla^2 \decu \over d \ta^2}(\ta) \,=\,
-\,\Riem\big(\decu(\ta), \dcu(\ta)\big)\, \dcu(\ta)~. \label{eqjac}
\end{equation}
The last equation, that we call again the Jacobi equation,
describes the evolution equation for the ``infinitesimal
separation vector'' $\decu$; its left-hand side is the \textsl{tidal
acceleration} between the nearby geodesics. \parn
To go on, let us report from subsection \ref{basics} the definitions \eqref{defss}\eqref{defra}:
$$ \SS_{\ta} \,:=\, \big\{ X \in T_{\cu(\ta)} \MM~\big|~g\big(X,\dcu(\ta)\big) = 0 \big\}~; $$
$$ \AA_{\ta} \,:\, \SS_{\ta} \to \SS_{\ta}~, \qquad X \mapsto \AA_{\ta} X \,:=\,
-\,\Riem\big(X, \dcu(\ta)\big)\,\dcu(\ta)~. $$
We note the following facts.  \vspace{-0.2cm}
\begin{enumerate}[-]
\item Being the orthogonal complement of the non-zero timelike vector $\dcu(\ta)$,
$\SS_{\ta}$ is a 3-dimensional, spacelike linear subspace of $T_{\cu(\ta)} \MM$\,.
We can view $\SS_{\ta}$ a Euclidean space, with the inner product
given by the restriction of $g \equiv g_{\cu(\ta)}$\,. \vspace{-0.2cm}
\item For any $X \in \SS_{\ta}$ (and, more generally, for any $X \in T_{\cu(\ta)} \MM$),
from a well-known symmetry of the Riemann tensor (see, e.g., Eq. 3.2.15 on page 39 of \cite{Wald})
we infer that
$g\big(\Riem\big(X, \dcu(\ta)\big)\,\dcu(\ta),\dcu(\ta)\big)=0$; therefore,
$\AA_{\ta}$ is actually well defined as a map to $\SS_{\ta}$.
\vfill\eject\noindent
$\phantom{a}$
\vskip-1.7cm
\noindent
\item Again by the symmetries of the Riemann tensor (see, e.g., Eq.s 3.2.15, 3.2.20
on page 39 of \cite{Wald}), for all $X,Y \in \SS_{\ta}$ we have
\begin{equation}
g\big(\AA_{\ta} X, Y\big) \,=\, g\big(X,\AA_{\ta} Y\big)~;
\end{equation}
this means that $\AA_{\ta}$ is a self-adjoint operator on the 3-dimensional Euclidean
space $\SS_{\ta}$. \vspace{-0.2cm}
\item Of course, the Jacobi equation \eqref{eqjac} can be rephrased as
\begin{equation}
{\nabla^2 \decu \over d \ta^2}(\ta) \,=\, \AA_{\ta}\,\decu(\ta)~.
\end{equation}
\end{enumerate}
The above facts justify all statements made in subsection \ref{basics} about $\SS_{\ta}$,
$\AA_{\ta}$ and explain the denomination of \textsl{tidal operator} employed
therein for the latter. In the cited subsection, after writing the definition of $\AA_{\ta}$
we have introduced the \textsl{maximal tidal acceleration per unit length} which is,
by definition, the scalar
$$ \aa(\ta) \,:=\, \sup_{X \in \SS_{\ta} \setminus \{0\}}
{\sqrt{g(\AA_\ta X, \AA_{\ta} X)}
\over \sqrt{g(X,X)}} ~; $$
see Eq. \eqref{defaa} and the comments that follow it. Here we just repeat one
of these comments, namely, that $\aa(\ta)$ coincides with the maximum
of the absolute values of the eigenvalues of $\AA_{\ta}$.

\subsection{Computing $\aa(\ta)$ for a geodesic in the $\Tar$ spacetime.}\label{secC2}
Let us consider the spacetime $\Tar$ of Section \ref{secMod}. Hereafter we give
some indications on the calculation of the Riemann curvature tensor $\Riem$; moreover,
we choose a timelike geodesic $\cu$ of the type described in Section \ref{freefallSec}
and sketch the computation of the tidal operator $\AA_\ta$ and of its norm $\aa(\ta)$
in this case. \vspace{-0.2cm}
\begin{enumerate}[i)]
\item First of all, we compute the coefficients $\Rm^{\kappa}_{~\upsilon \mu \nu}$
of the Riemann tensor $\Riem$ in the coordinate system $(x^{\mu}) = (t,\vfi,\rho,z)$,
starting from the coefficients $g_{\mu \nu}$ of the metric.
It appears that the coefficients of the Riemann tensor depend
only on the coordinates $\rho$ and $z$ and possess the following structure:
\begin{equation}
\Rm^{\kappa}_{~\upsilon \mu \nu} \,=\, {1 \over R^2} ~
\Rsc^{\kappa}_{~\upsilon \mu \nu}\big(\rho/R,z/R\big)~, \label{rieco}
\end{equation}
for suitable functions $\Rsc^{\kappa}_{~\upsilon \mu \nu}$ of the variables $\rv = \rho/R$
and $\ze = z/R$, also depending on the parameters $\ellm/R,\ellp/R,\af,\at$;
these have been computed using \verb"Mathematica", but their
expressions are too lengthy to be reported here. The term $1/R^2$ is factored out
in the right-hand side of Eq. \eqref{rieco} for future convenience
(see the forthcoming items (iv)(v)\,).\\
We are interested in the evaluation of the above coefficients along the geodesic
$\cu$, where $\rho = \rho(\ta)$ and $z = 0$. Writing
$\RF^{\kappa}_{~\upsilon \mu \nu}(\rv) := \Rsc^{\kappa}_{~\upsilon \mu \nu}\big(\rv,0)$,
we get
\begin{equation}
\Rm^{\kappa}_{~\upsilon \mu \nu}\big(\cu(\ta)\big) \,=\,
{1 \over R^2} ~ \RF^{\kappa}_{~\upsilon \mu \nu}\big(\rho(\ta)/R\big) ~;
\end{equation}
$\RF^{\kappa}_{~\upsilon \mu \nu}$ are functions of $\rho/R$,
also depending on the parameters $\ellm/R,\ellp/R,\af,\at$.
Let us remark that, since the metric $g$ of Eq. \eqref{dsXidef}
is flat outside $\Top$ and inside $\Tom$, the functions $\RF^{\kappa}_{~\upsilon \mu \nu}(\rv)$
are non-zero only for $\rv \in (1-\ellp/R,1-\ellm/R) \cup (1+\ellm/R,1+\ellp/R)$. \vspace{-0.2cm}
\item Concerning the components $(\dcu^\mu) = (\dot{t},\dot{\vfi},\dot{\rho},\dot{z})$
of the velocity vector field $\dcu$, let us recall that $\dot{z} = 0$.
As for $\dot{t}$ and $\dot{\vfi}$, we have the explicit expressions
\eqref{tp} \eqref{fp}; similarly, $\dot{\rho}$ is given by Eq. \eqref{dacit},
with $V_{\pt,\pf}$ as in Eq. \eqref{Vdef}, $\En = -1/2$ and $\si = \pm 1$
(depending on the direction of motion). By direct inspection of the cited equations, we obtain
\begin{equation}
\dcu^\mu(\ta) \,=\, \pt^2\;\Xc^\mu\big(\rho(\ta)/R\big) ~, \label{dcuCom}
\end{equation}
for suitable functions $\Xc^\mu$, depending on the parameters $\ellm/R,\ellp/R,\af,\at$
(related to the metric) and $\pt,\xf,\si$ (related to the geodesic under analysis).
Our choice to single out the coefficient $\pt^2$ in Eq. \eqref{dcuCom}
is motivated by the fact that the functions $\Xc^\mu$ so determined (thought still
depending on $\pt$) possess a finite limit for $\pt \to +\infty$ and $\xf \to 0^+$ (simultaneously). \vspace{-0.2cm}
\item Items i) ii) suggest the representation
$$ \AA_{\ta} \,=\, {\pt^2 \over R^2}\; \AF_{\ta}~, $$
appearing in the main text as Eq. \eqref{afrak} and involving a self-adjoint operator
$\AF_{\ta}$ on $\SS_{\ta}$; this maps a vector $X \in \SS_{\ta} \subset T_{\cu(\ta)} \Tar$
of components $X^\mu$ into the vector $\AF_\ta X \in \SS_{\ta} \subset T_{\cu(\ta)} \Tar$
with components
\begin{equation}
(\AF_{\ta} X)^\kappa \,=\, \RF^{\kappa}_{~\upsilon \mu \nu}(\rv)\;
\Xc^\nu(\rv)\; \Xc^\upsilon(\rv)\;X^\mu \Big|_{\rv\, =\, \rho(\ta)/R}~. \vspace{-0.1cm}
\label{AFdef}
\end{equation}
The right-hand side of Eq. \eqref{AFdef} depends on the parameters $\ellm/R,\ellp/R,\af,\at$
and $\pt,\xf$, but not on $\si$.
\item The next step is again supported by \verb"Mathematica";
the related calculations involve very long expressions that we do not
report here, with the exception of the forthcoming Eq. \eqref{ar}. \\
First of all, for each $\ta$, we determine an orthonormal basis
$\big(B_{(i)}(\ta)\big)_{i \in \{1,2,3\}}$ of $\SS_{\ta}$
applying the Gram-Schmidt algorithm to the set of vectors obtained projecting onto
$\SS_{\ta}$ the spacelike elements $\big(E_{(i)}(\cu(\ta))\big)_{i \in \{1,2,3\}}$
of the tetrad \eqref{EComp}. We get
\begin{equation}
g(B_{(i)}(\ta),\AF_\ta B_{(j)}(\ta)) \,=\,
\AF_{ij}\big(\rho(\ta)/R\big) ~, \label{matel}
\end{equation}
for suitable functions $\AF_{ij}$, depending parametrically on $\ellm/R,\ellp/R,\af,\at$
and $\pt,\xf$. An elementary analysis shows that each function $\AF_{ij}$ is dimensionless;
this is an advantage of the decision to factor out the term $1/R^2$ in most of the previous computations.
From the considerations reported at the end of item i) it follows that $\AF_{ij}(\rv)$
vanishes identically for $\rv$ outside $(1-\ellp/R,1-\ellm/R) \cup (1+\ellm/R,1+\ellp/R)$. \\
From the matrix elements \eqref{matel} we obtain the eigenvalues of $\AF_{\ta}$
(and thus, of $\AA_{\ta}$). In the simultaneous limit $\pt \to + \infty$, $\xf \to 0^+$
we find that $\AF_{\ta}$ has a zero eigenvalue of multiplicity two, and a simple,
non-zero eigenvalue depending on $\rv = \rho(\ta)/R$ and admitting the (normalized)
eigenvector $\de_z \big|_{\cu(\ta)}$. The latter non-zero eigenvalue is
\begin{equation}\begin{array}{c}
\dd{\am(\rv) \,=\,-\,{(\af + \at\,\rv)\;\Hr'(\rv) \over \at^2\, (\rv-1)\,
\big(\rv\,(1\!-\!\Hr(\rv))^2\!+ \af\,\at\, \Hr(\rv)^2\big)^3}\;\cdot } \vspace{0.1cm} \\
\dd{\hspace{-1.5cm}\cdot \;\Big[
(\af\,\at\!+\!\rv)(\af\!-\!\at\,\rv)\big(1\!-\!\Hr(\rv)\big)^3 \!
- \af\,\at\,(\af\,\at\!+\!\rv)\big(1\!-\!\Hr(\rv)\big)^2\big(1\!+\!2\,\Hr(\rv)\big)\,+} \\
\dd{\hspace{7.5cm} -\,\af\,\at\,(\af \!-\!\at\,\rv)\big(1\!-\!\Hr(\rv)\big)+\af^2 \at^2\Big]}\label{ar}
\end{array}\end{equation}
(as usual, $\Hr$ denotes the restriction of the shape function $\HH$ defined in Eq. \eqref{Hrdef};
notice that $\am(\rv) = 0$ for $\rv$ outside $(1-\ellp/R,1-\ellm/R) \cup (1+\ellm/R,1+\ellp/R)$,
since $\Hr'(\rv) = 0$ in this region). \vspace{-0.2cm}
\item Finally, we recall that the maximal tidal acceleration per unit length $\aa(\ta)$
coincides with the maximum of the absolute values of the eigenvalues of $\AA_\ta$; therefore,
in view of the facts mentioned in item iv), we obtain
$$ \aa(\ta) \,=\, {\pt^2 \over R^2}\;\ac\big(\rho(\ta)/R\big) ~, $$
where $\ac(\rv)$ denotes the maximum absolute value for the eigenvalues of
the matrix $\AF_{ij}(\rv)$; this corresponds to Eq. \eqref{atau} in the main text.
Like the matrix elements $\AF_{ij}(\rv)$, $\ac(\rv)$ is a dimensionless quantity depending on the parameters
$\ellm/R,\ellp/R,\af,\at$ and $\pt,\xf$; furthermore, from (i)(ii)(iv) it follows that
$\ac(\rv)$ vanishes identically for $\rv$
outside $(1-\ellp/R,1-\ellm/R) \cup (1+\ellm/R,1+\ellp/R)$ and that it has a finite limit
for $\pt \to +\infty$ and $\xf \to 0$, coinciding with $|\am(\rv)|$. \\
The graphs and the numerical values for the maxima of $\aa(\ta)$ reported in subsection
\ref{tidaltime} were obtained from the explicit expressions of $\ac(\rv)$, $|\am(\rv)|$.
\end{enumerate}

\section{Appendix. On the stress-energy tensor and energy density}\label{App4}
Let us recall the definition \eqref{defT} of the stress-energy tensor $\Te$,
and specialize it to our spacetime $\Tar$. In any coordinate system, we have
\begin{equation}
\Tm_{\mu \nu} \,=\, {1 \over 8 \pi G} \l(\Rm_{\mu \nu}\,-\,{1 \over 2}\;\Rs\;g_{\mu \nu}\r)~;
\end{equation}
the coefficients $\Rm_{\mu \nu}$ of the Ricci tensor $\Ric$ and the scalar curvature $\Rs$
are obtained from the coefficients $\Rm^{\kappa}_{~\upsilon \mu \nu}$ of the
Riemann curvature tensor $\Riem$ by obvious contractions. \parn
As usually, we employ the coordinate system $(x^\mu) = (t,\vfi,\rho,z)$; due
to the structure \eqref{rieco} of the Riemann curvature coefficients, we have
\begin{equation}
\Tm_{\mu \nu} \,=\, {1 \over 8 \pi G\,R^2}\;\TT_{\mu \nu}\big(\rho/R,z/R\big)
\label{Tdim}
\end{equation}
for suitable functions $\TT_{\mu \nu}$, which also depend on the parameters
$\ellm/R,\ellp/R,\af,\at$ and on the shape function $\HH$. These functions have been
computed explicitly via \verb"Mathematica", but will not be reported here.
The factor $1/R^2$ is singled out in the right-hand side of Eq. \eqref{Tdim}
for future convenience.\parn
Of course, the coordinate expression for the energy density \eqref{edens} is
\begin{equation}
\EE(X) \,=\, \Tm_{\mu \nu}\,X^\mu\, X^\nu
\,=\, {1 \over 8 \pi G\,R^2}\; \TT_{\mu \nu}\big(\rho/R,z/R\big)\; X^\mu\, X^\nu~,
\label{enecomp}
\end{equation}
for each normalized timelike vector $X$, tangent at a spacetime
point $p$ of coordinates $(t,\vfi,\rho,z)$. \vspace{0.1cm}\parn
As already indicated in the main text, the energy density measured
by the fundamental observer passing through a spacetime point $p$ is (see Eq. \eqref{defeef})
$$ \EEf(p) \,:=\, \EE\big(E_{(0)}(p)\big) ~. $$
From Eq. \eqref{enecomp} with $X = E_{(0)}(p)$ and from the expressions \eqref{EComp}
for $E_{(0)}$ and \eqref{Xidef} for $\XX$, we obtain (as in Eq. \eqref{espeef})
$$ \EEf \,=\, {1 \over 8 \pi G\,R^2}\; \EFf\big(\rho/R,z/R\big) $$
for a suitable function $\EFf$ depending on the parameters $\ellm/R,\ellp/R,\af,\at$
and on the shape function $\HH$; again, this function
has been computed by \verb"Mathematica" but its analytic expression is too long
to be reported here. The function $\EEf$ is found to be dimensionless,
an advantage of our strategy to factor out systematically the term $1/R^2$.
Of course
\begin{equation}
\EFf(\rv,\ze) \,=\, 0 \qquad \mbox{for}~~
\sqrt{(\rv - 1)^2 + \ze^2} \in [0, \ellm/R] \cup [\ellp/R, + \infty)~,
\end{equation}
due to the vanishing of the curvature in the corresponding spacetime regions.
Figures 5a-5b about $\EFf$, appearing in the main text, are based on its analytic expression.
\vspace{0.1cm}\parn
We already pointed out in Eq. \eqref{defeeg} of subsection \ref{subsecEG}
that the energy density measured at proper time $\ta$ by an observer in free fall
along a timelike geodesic $\cu$ of the type considered in Section \ref{freefallSec}
is given by
$$ \EEg(\ta) \,:=\, \EE\big(\dcu(\ta)\big) ~. $$
Using Eq. \eqref{enecomp} for $\EE$ and expressing the components
$(\dcu) = (\dot{t},\dot{\vfi},\dot{\rho},\dot{z})$ via the relations \eqref{tp} \eqref{fp} \eqref{dacit}
(with $\En = -1/2$) and $\dot{z} = 0$ we obtain, as in Eq. \eqref{espeeg},
$$ \EEg(\ta) \,=\, {\pt^2 \over 8 \pi G\,R^2}\; \EFg\big(\rho(\ta)/R\big) ~, $$
where $\EFg$ is a suitable, dimensionless function, depending parametrically on
$\ellm/R,\ellp/R,\af,\at$ and $\pt,\xf$.
We have computed this function explicitly using \verb"Mathematica"; the resulting
expression has been used to plot the graphs reported in Figures 6a-6b. Let us mention that
\begin{equation}
\EFg(\rv) \,=\, 0 \qquad \mbox{for} ~~
\rv \in (0,1-\ellp/R] \cup [1-\ellm/R,1+\ellm/R] \cup [1+\ellp/R,+\infty) ~,
\end{equation}
a fact that follows again from the vanishing of the curvature in the associated
spacetime regions.

\end{document}